\documentclass[twocolumn]{aastex62}
\usepackage{fdsymbol}
\usepackage{wasysym}
\graphicspath{{./}{fig/}}

\shorttitle{Late-time UV emission of TDFs}
\shortauthors{S. van Velzen et al.}

\begin{document}

\title{\bf \Large Late-time UV observations of tidal disruption flares reveal \\ unobscured, compact accretion disks\footnote{Based on observations made with the NASA/ESA Hubble Space Telescope, obtained at the Space Telescope Science Institute, which is operated by the Association of Universities for Research in Astronomy Inc., under NASA contract NAS 5-26555. These observations are associated with program GO-14255.}}

\correspondingauthor{Sjoert van Velzen}
\email{sjoert@nyu.edu}

\author[0000-0002-3859-8074]{Sjoert van Velzen}\altaffiliation{James Arthur Fellow}
\affiliation{Center for Cosmology and Particle Physics, New York University, New York, NY 10003}
\affiliation{Department of Astronomy, University of Maryland, College Park, MD, 20742}

\author{Nicholas C. Stone}\altaffiliation{NASA Einstein Fellow}
\affiliation{Columbia Astrophysics Laboratory, Columbia University, New York, NY 10027}

\author{Brian D. Metzger}
\affiliation{Columbia Astrophysics Laboratory, Columbia University, New York, NY 10027}

\author{Suvi Gezari}
\affiliation{Department of Astronomy, University of Maryland, College Park, MD, 20742}

\author{Thomas M. Brown}
\affiliation{Space Telescope Science Institute, 3700 San Martin Drive, Baltimore, MD 2128}

\author{Andrew S. Fruchter}
\affiliation{Space Telescope Science Institute, 3700 San Martin Drive, Baltimore, MD 2128}


\begin{abstract}
The origin of thermal optical and UV emission from stellar tidal disruption flares (TDFs) remains an open question. 
We present {\it Hubble Space Telescope} far-UV (FUV) observations of eight optical/UV selected TDFs 5-10 years post-peak. Six sources are cleanly detected, showing point-like FUV emission ($10^{41.5-42.5}~{\rm erg~s}^{-1}$) from the centers of their host galaxies. We discover that the light curves of TDFs from low-mass black holes ($<10^{6.5}\,M_\odot$) show significant late-time flattening. Conversely, FUV light curves from high-mass black hole TDFs are generally consistent with an extrapolation from the early-time light curve.  The observed late-time emission cannot be explained by existing models for early-time TDF light curves (i.e. reprocessing or circularization shocks), but is instead consistent with a viscously spreading, unobscured accretion disk. These disk models can only reproduce the observed FUV luminosities, however, if they are assumed to be thermally and viscously stable, in contrast to the simplest predictions of $\alpha$-disk theory. For one TDF in our sample, we measure an upper limit to the UV luminosity that is significantly lower than expectations from theoretical modeling and an extrapolation of the early-time light curve. This dearth of late-time emission could be due to a disk instability/state change absent in the rest of the sample.  The disk models that explain the late-time UV detections solve the TDF ``missing energy problem'' by radiating a rest-mass energy of $\sim 0.1 M_\odot$ over a period of decades, primarily in extreme UV wavelengths.
\end{abstract}

\section{Introduction} 
In the current era, most tidal disruption flares (TDFs) are discovered by optical imaging surveys \citep{vanVelzen10,Gezari12,Arcavi14,Holoien14, Hung17, Blagorodnova17}; see \citet{vanVelzen18} for a recent compilation. TDFs have great potential as probes of quiescent supermassive black holes (SMBHs), but the origin of their optical and UV \citep{Gezari06} emission remains unclear. In the canonical model for TDFs, put forward by \citet{Rees88}, their electromagnetic emission is powered by a compact accretion disk of size comparable to the tidal radius, fed by the returning stellar debris streams. In this scenario, the bolometric disk luminosity is expected---after a brief phase of rising luminosity---to track the $t^{-5/3}$ debris fallback rate, while emission on the Rayleigh-Jeans tail of the disk spectrum will follow a more shallow decay rate of $t^{-5/12}$ \citep{Lodato11}. 

The observed decay of optical TDF emission is often consistent with the steep power law of the fallback rate. However, the blackbody radii fitted to these optical observations are $\sim 10-100$ times larger than naive estimates of the outer radius of the accretion disk, and correspondingly, observed optical luminosities are $\sim 100-1000$ times larger than expected  \citep[][]{Gezari09,vanVelzen10,Wevers17}. Two models have been proposed to explain this behavior. The optical/UV light could be produced by reprocessing of X-ray and extreme ultraviolet (EUV) radiation from the inner disk, after this harder component is absorbed by optically thick stellar debris or disk outflows \citep{LoebUlmer97, Guillochon14,Miller15,Metzger16}.  Alternatively, the optical/UV emission could be powered by energy released as the streams of stellar debris self-intersect and shock \citep{Lodato12,Piran15,Krolik16}. 
Stream-stream interactions provide a mechanism to dissipate orbital energy, which is required for accretion disk formation. 

Both theories of early-time TDF emission must contend with a ``missing energy problem.'' Single-color blackbody fits to observed TDF spectra find total radiated energies at least one order of magnitude below the theoretical expectation for radiatively efficient accretion \citep{Piran15,Stone16b}.  This is typically explained in the circularization paradigm with a very low radiative efficiency; in the reprocessing paradigm, most of the disk luminosity may escape as  unobservable EUV light.


If stream-stream shocks were the only source of TDF light, then the light curve should track the fallback rate even at late times. The contribution of additional emission from an accretion disk (either directly, or through reprocessed X-rays/EUV) would lead to a flattening of the late-time light curve. If the accretion rate through the disk is set quasi-viscously, the light curve will eventually flatten because at times $t$ far after peak, the viscous timescale $t_{\rm visc}\gg t$. This timescale hierarchy is expected because the viscous time depends on the radius ($R$) and scale height $(H)$ of the disk as $t_{\rm visc}\propto (H/R)^{-2}$ \citep{Cannizzo90}. At early times, when the accretion rate is super-Eddington, we expect $H/R \sim 1$, while for moderate accretion rates ($\sim 1-10\%$ of the Eddington limit) we expect $H/R\ll 1$ \citep[e.g.][]{Abramowicz13}, resulting in a long viscous time and a slow decay of the power emitted by the accretion disk.

Even if the feeding rate of the disk tracks the fallback rate ($\dot{M}_{\rm fb} \propto t^{-5/3}$), viscously-regulated accretion will result in a more gradual decline of the accretion rate, $\dot{M}\propto t^{-1.2}$ \citep{Cannizzo90, Shen14}.  Slopes that are more shallow than a $t^{-5/3}$ decay have been observed in the X-ray light curves of some TDF candidates \citep{Auchettl16}, although very few securely-classified X-ray TDFs \citep{Auchettl16} have both well-sampled X-ray light curve and spectra that are required to measure the slope of the bolometric luminosity.  Some evidence for flattening exists in the NUV light curves of TDF candidates GALEX-D1-9 and GALEX-D3-13 \citep{Gezari08}, but in general, very little is known about the late-time optical/UV properties of TDFs.  With one exception \citep{Gezari15}, optically-selected TDF candidates have not been detected more than a handful of years past peak.

Here we present {\it HST} observations of a sample of eight tidal disruption flares, each observed in the far UV (FUV) 5-10 years after peak light.  This is the first large sample of optical- and UV-selected TDFs observed at such late times, and the goal of this paper is to determine which, if any, emission mechanisms are still operating in these flares.  In Section~\ref{sec:obs}, we present the sample selection, host galaxy photometry, {\it HST} and {\it Swift} data reduction, and light curve fitting procedure.  In Section \ref{sec:discussion} we discuss the astrophysical implications of our observations for TDF disk evolution and emission mechanisms.  We close with a summary of the results in Section \ref{sec:conclusions}. Throughout this work, all magnitudes are in the AB system \citep{oke74} and we adopt a flat $\Lambda$CDM cosmology with $H_0=70 \, {\rm km}\,{\rm s}^{-1}{\rm Mpc}^{-1}$. 

%
\section{Analysis}\label{sec:obs}
%

\subsection{Source selection}\label{sec:select}
We obtained {\it HST} FUV (ACS/SBC; F125LP and F150LP) follow-up observations of eight strong TDF candidates, see Table~\ref{tab:UVmags}. These source were selected based their age (time of peak at least five years before the new {\it HST} observations) and expected FUV magnitude. The latter requirement removed PS1-11af \citep{Chornock14} due its high redshift and PTF-09axc \citep{Arcavi14} due to its low blackbody temperature. 

A few TDFs that have been discovered more recently have reasonably long ($\approx 2$~yr)  UV light curves from monitoring observations of the Neil Gehrels Swift Observatory \citep[{\it Swift};][]{Gehrels04} with the UVOT instrument \citep{Roming05}. We add these to our sample for analysis: ASASSN-14ae \citep{Holoien14}, ASASSN-14li \citep{Holoien16,Brown16b}, ASASSN-15oi \citep{Holoien16b}, iPTF-16fnl \citep{Blagorodnova17}. For completeness, we also include iPTF-16axa \citep{Hung17} and iPTF-15af \citep{Blagorodnova19}, even though these have no late-time {\it Swift} observations. 
\begin{figure*}
\begin{center}

\includegraphics[width=175pt, trim=5mm 16mm 6mm 5mm, clip]{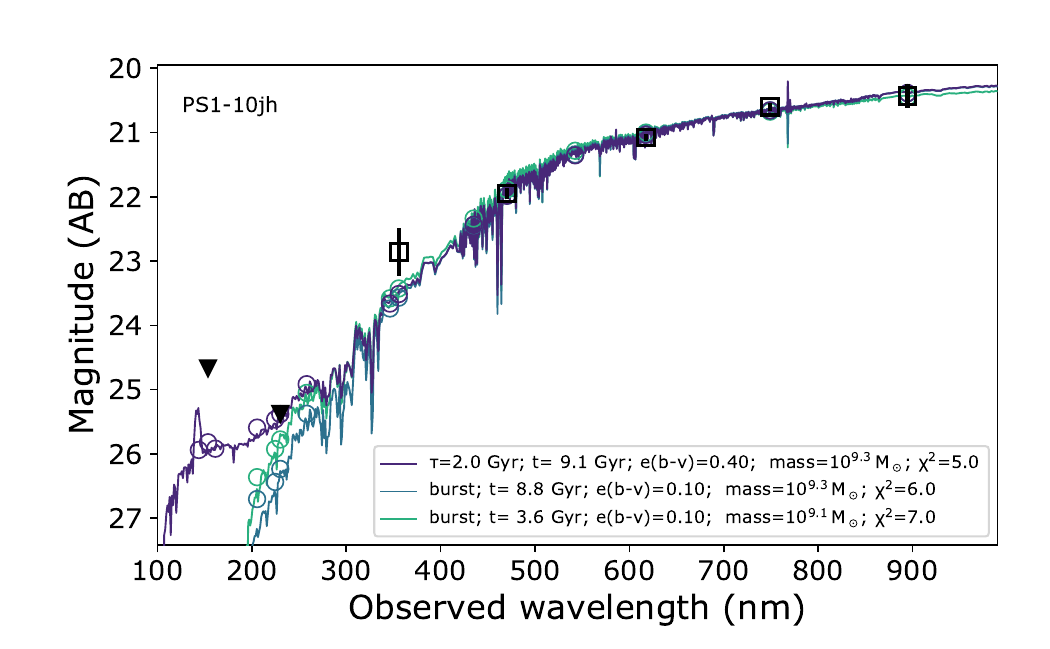} \qquad
\includegraphics[width=175pt, trim=5mm 16mm 6mm 5mm, clip]{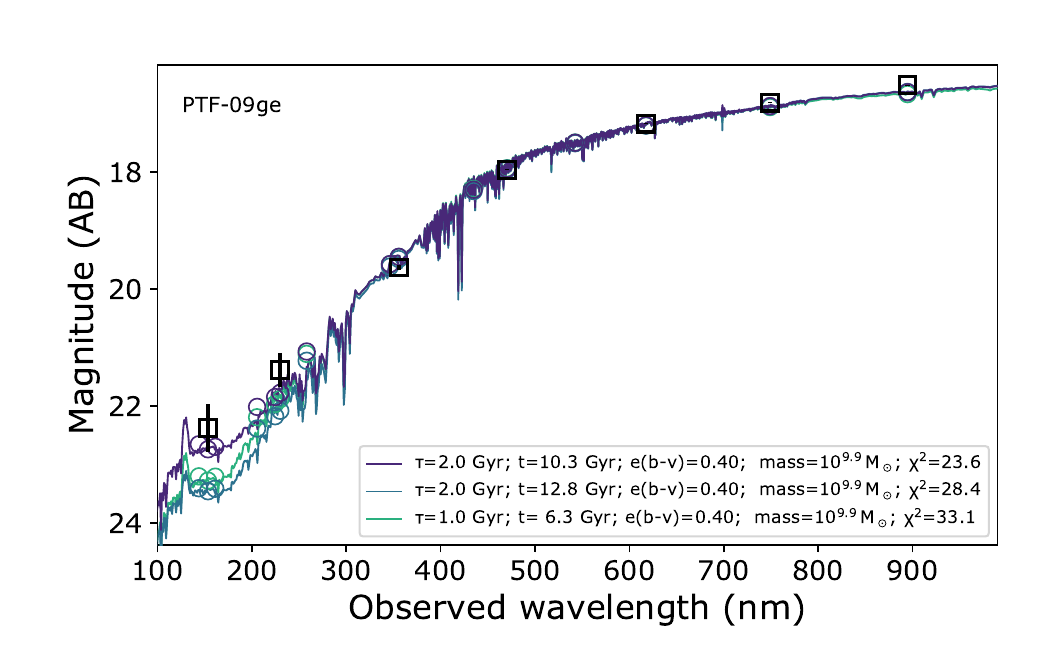} 

\includegraphics[width=175pt, trim=5mm 16mm 6mm 5mm, clip]{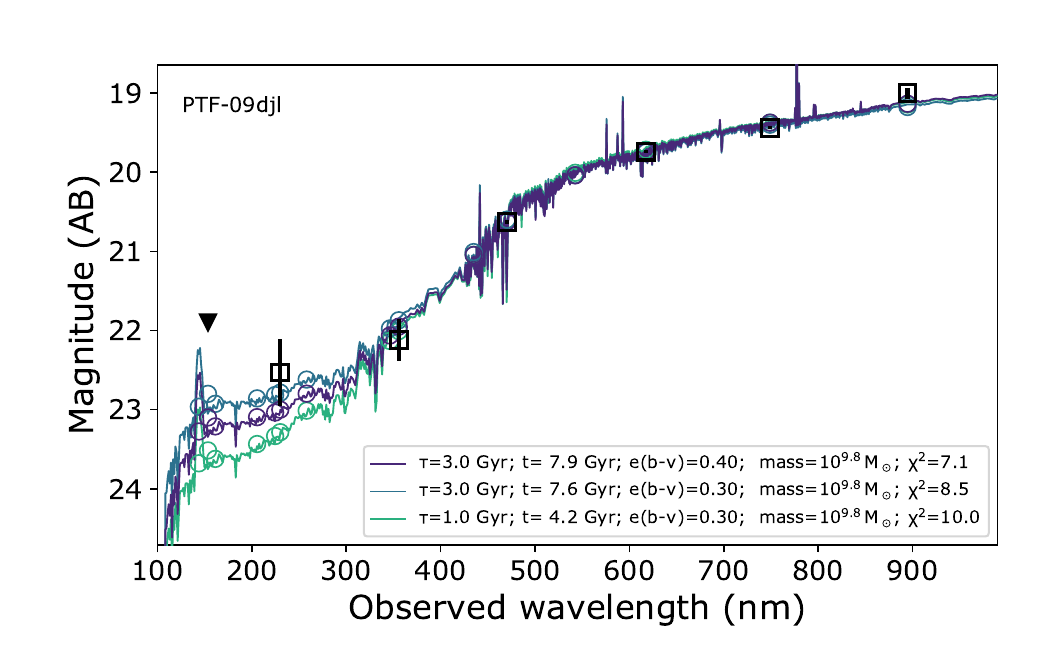} \qquad
\includegraphics[width=175pt, trim=5mm 16mm 6mm 5mm, clip]{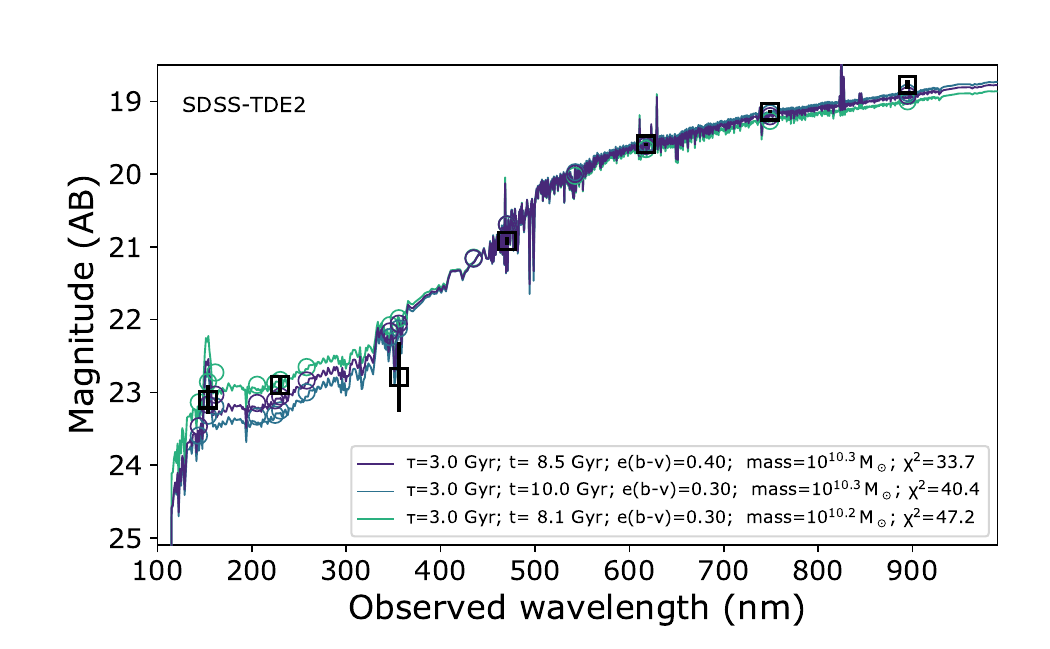}

\includegraphics[width=175pt, trim=5mm 16mm 6mm 5mm, clip]{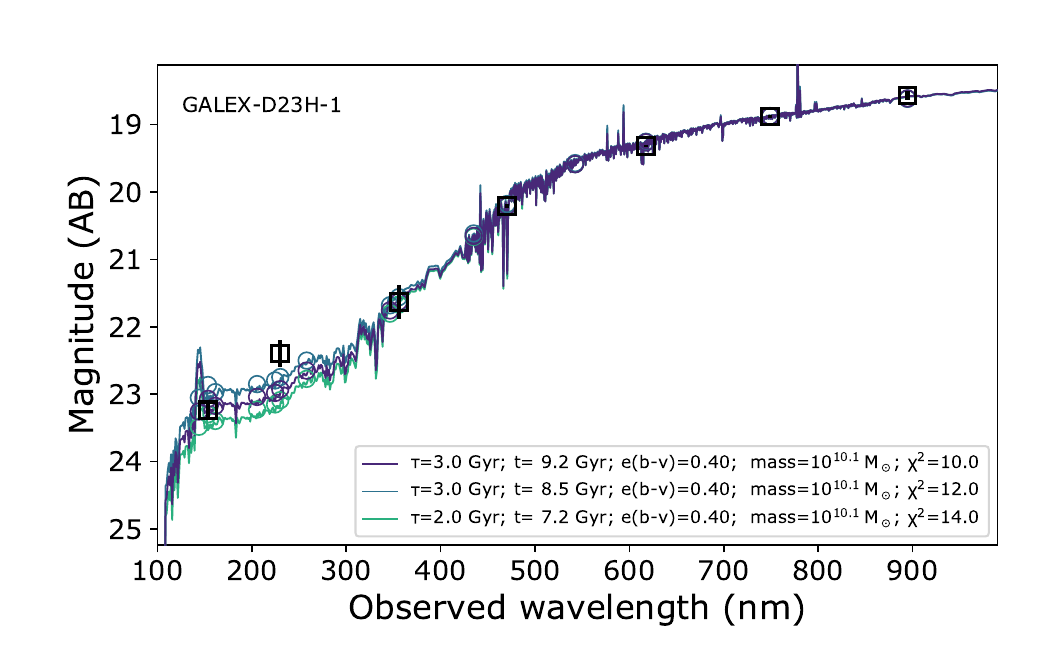} \qquad
\includegraphics[width=175pt, trim=5mm 16mm 6mm 5mm, clip]{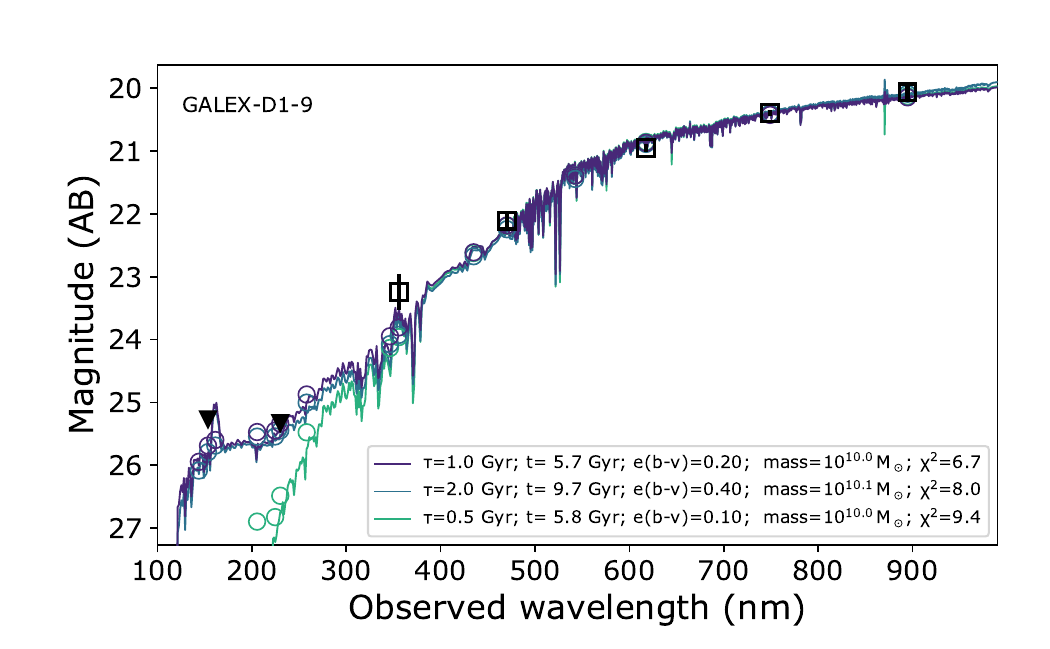} 

\includegraphics[width=175pt, trim=5mm 16mm 6mm 5mm, clip]{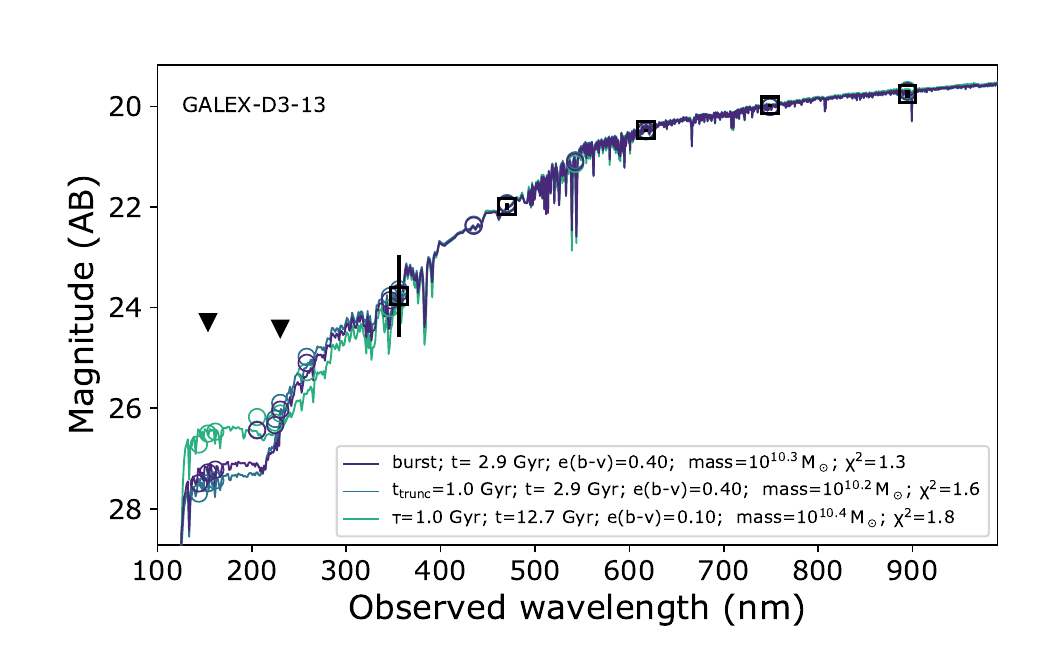} \qquad
\includegraphics[width=175pt, trim=5mm 16mm 6mm 5mm, clip]{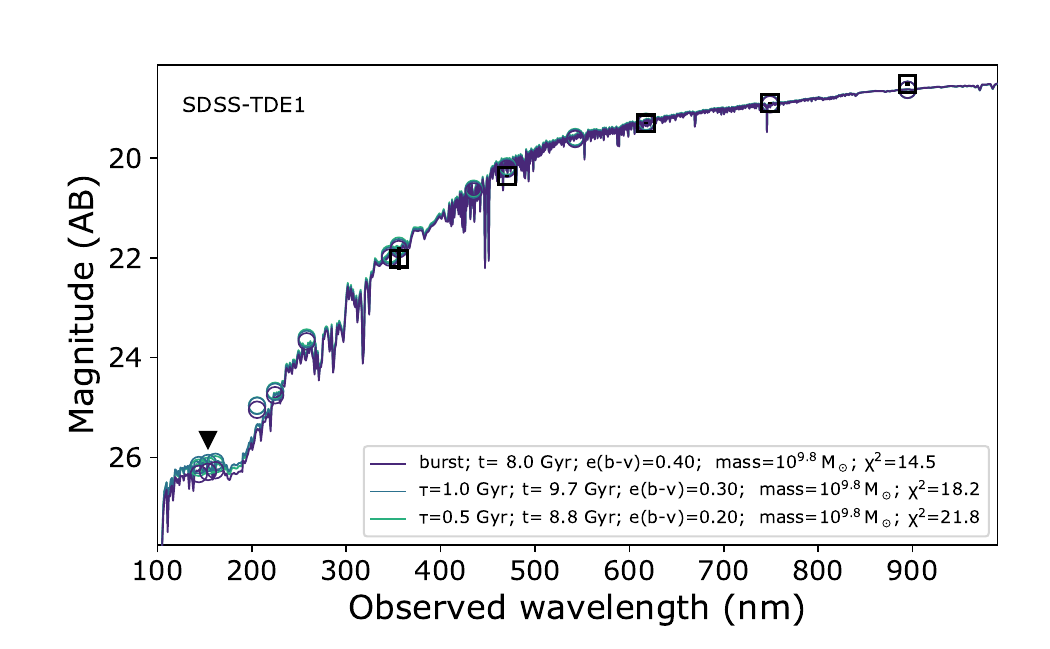}

\includegraphics[width=175pt, trim=5mm 16mm 6mm 5mm, clip]{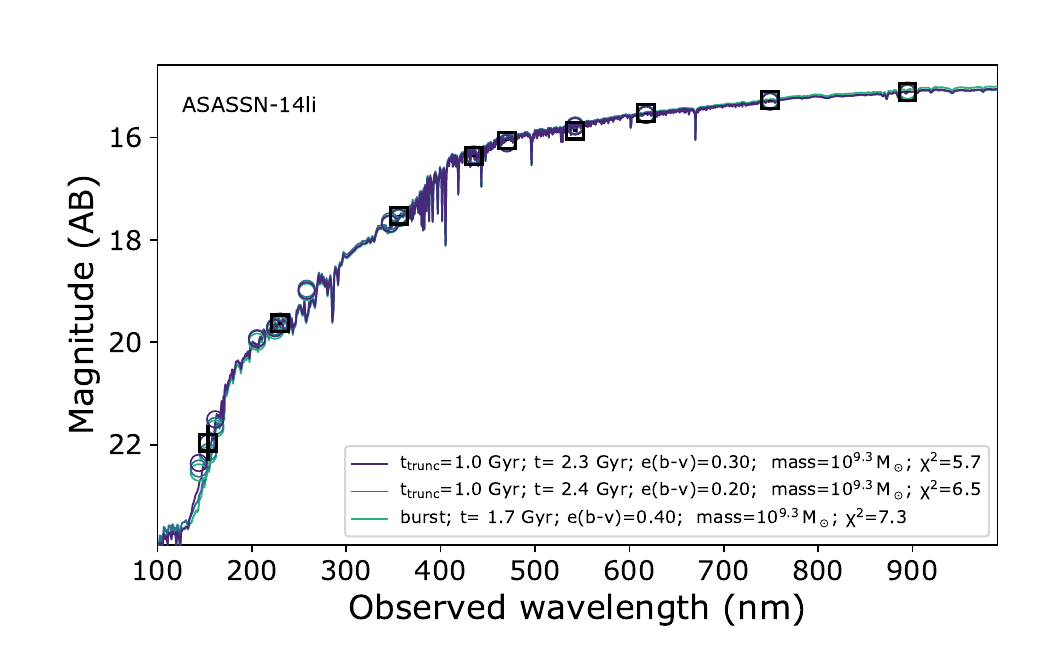} \qquad
\includegraphics[width=175pt, trim=5mm 16mm 6mm 5mm, clip]{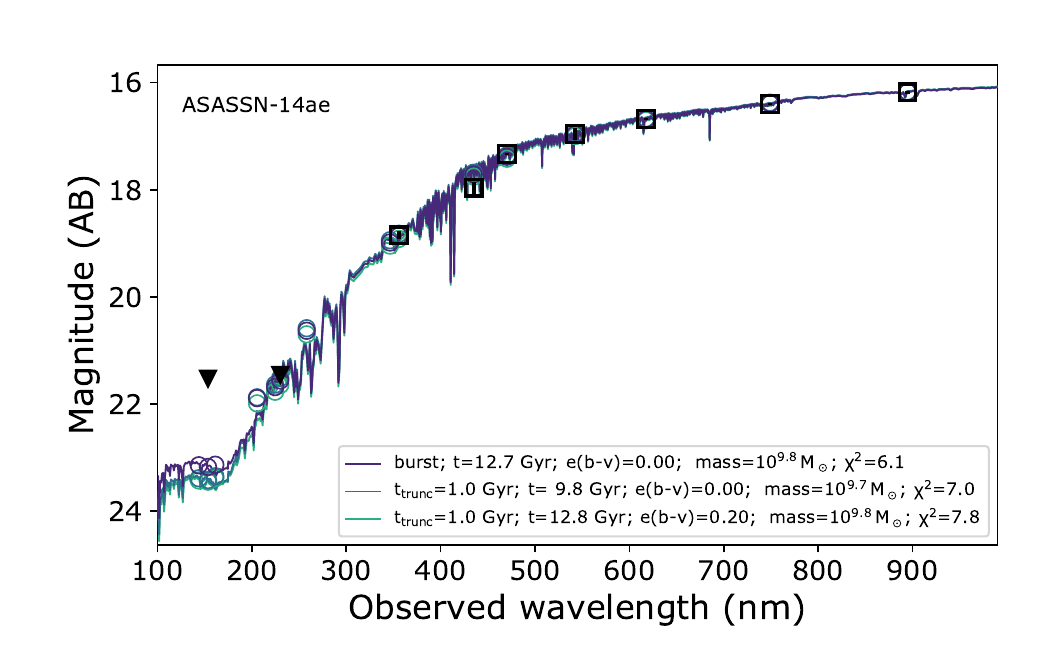}

\includegraphics[width=175pt, trim=5mm  2mm 6mm 5mm, clip]{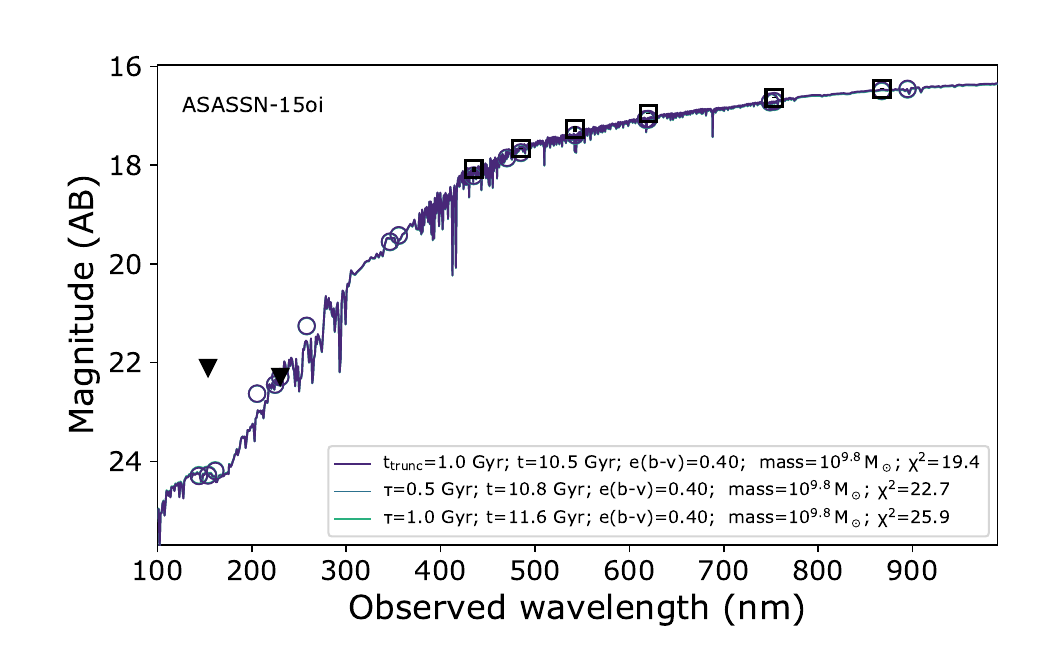}\qquad
\includegraphics[width=175pt, trim=5mm  2mm 6mm 5mm, clip]{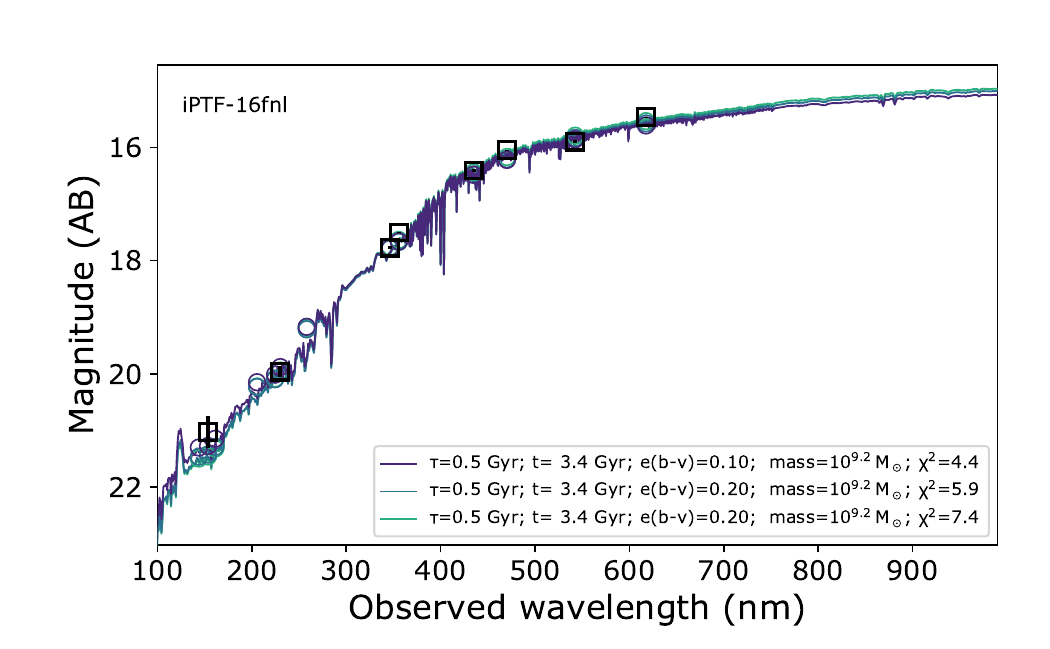} 

\caption{Comparison of synthetic galaxy spectral energy distributions to the observed magnitudes. Triangles indicate 5-$\sigma$ upper limits. 
The legend show the parameters of the galaxy model: time since the burst of star formation, the type of burst (instantaneous, burst lasting 1~Gyr, or exponentially decreasing rate with $e$-folding time $\tau$), and the amount of extinction applied to the host galaxy using the \citet{Calzetti00} attenuation law. 
The open circles show the synthetic magnitudes for each model (from blue to red, the bandpasses shown are: F125LP, FUV F150LP, UVW2, NUV, UVM2, UVW1, $U$ or $u$, $B$ or $g$, $V$ or $r$, $i$, $z$). }
\label{fig:popsyn}
\end{center}
\end{figure*}

%
\subsection{Host galaxy population synthesis}\label{sec:popsyn}
To estimate the contribution of the host galaxy to the total FUV flux of our sources we compared synthetic galaxy models to the observed UV and optical photometry of the host galaxies. 

All sources in our sample  have been observed by {\it GALEX} \citep{Martin05}. To estimate the host galaxy flux we use only the \textsl{GALEX} observations obtained well before the TDF. Five sources are detected pre-flare in at least one of the \textsl{GALEX} bands: D23H-1, TDE2, PTF-09ge, ASASSN-14li, iPTF-16fnl. For nondetections, we compute the upper limit on the flux, using the gPhoton software \citep{Million16} to estimate the noise in a 6" aperture. 

The optical flux of the host galaxy is taken from the SDSS \citep{york02} $u$, $g$, $r$, $i$ \citep{fukugita96} model magnitudes \citep{stoughton02}. The TDF ASASSN-15oi falls outside the SDSS footprint and we use Pan-STARRS \citep{Chambers16} Kron magnitudes \citep{Magnier16} for this source. For all source in our sample the flux of the host galaxy was measured before the TDF happened. 

For the sources that have been observed by {\it Swift}, we also include the mean late-time flux for the observations that are clearly dominated by the host galaxy. This assessment is made by comparing the light curves across the {\it Swift} UVOT bands. When a red band reaches a constant flux level while the higher-frequency bands continue to decay, we assume the red band is now dominated by the host galaxy. As expected, all four sources with {\it Swift} monitoring reach the host level in the $V$ and $B$ bands.  

The images of the low-redshift post-starburst host galaxy of iPTF-16fnl \citep{Blagorodnova17} clearly show two stellar populations, with a young central population inside an inclined disk that is dominated by a late-type population. The central component dominates the UV light of the host. To minimize the contribution from the late-type stellar population, we use a 4" aperture to measure the host galaxy photometry of this source. 

In order to extrapolate the observed host galaxy flux into other bands, we need to find the best-fit synthetic galaxy spectrum. For this task we used the Flexible Stellar Population Synthesis software \citep{Conroy09,Conroy10} to obtain stellar templates for different star formation histories. We adopt the default assumptions for the stellar parameters (\citealt{Kroupa01} initial mass function with stellar masses $0.08<M_{\rm star}/M_\odot<150$; Padova isochrones and MILES spectral library; \citealt{Vazdekis10}). We considered five different temporal shapes for the star formation history (SFH): an instantaneous burst, a burst truncated after 1~Gyr, and exponentially decreasing rates ($\propto e^{-t/\tau}$) with half-times of $\tau=0.5$~Gyr, $\tau=1$~Gyr, or $\tau=1.5$~Gyr. We allow the time since the burst to vary from 0.1 to 10~Gyr, using steps of 0.1~Gyr. We also include the effect of dust obscuration using a \citet{Calzetti00} extinction law to modify the spectrum to a maximum $E(B-V)$ of 0.4.

We place each galaxy in our grid of models at the redshift of the TDF host galaxy and apply Galactic extinction \citep{cardelli89} using the extinction at the celestial location of the TDF host galaxy \citep{Schlegel98}. After convolving the resulting synthetic spectrum with the bandpass we obtain synthetic magnitudes and the $\chi^2$ of each model. The convolution with the bandpass is important since it accounts for asymmetries in the wavelength dependence of the response (e.g. the ``red leak" of the UVW1 filter; see Appendix \ref{app:bandpass} and Fig.~\ref{fig:bandpass}).  

In Fig.~\ref{fig:popsyn} we show the best-fit galaxy models and, to visualize the  uncertainty in this fit, we also show models with higher $\chi^2$ values (selected to have log likelihood that is smaller than the best-fit by a factor of 0.5 and 1). 
For nearly all galaxies in our sample, the synthetic magnitudes in the near-UV bands can be determined with an accuracy of 0.1~mag or better. When only FUV upper limits from \textsl{GALEX} are available, the synthetic magnitudes at this wavelength are more uncertain (although often the best-fit model appears to provide an upper limit to the FUV flux; this happens because the FUV flux can become vanishingly small a few Gyr after the burst of star formation).

\begin{deluxetable*}{l c c c c c c c c}

\tablewidth{0pt}
\tablecolumns{9}

\tablecaption{Ultraviolet photometry.}

\tablehead{& \multicolumn{2}{c}{\textsl{GALEX} --- pre-flare} & \multicolumn{2}{c}{HST --- host predicted} & \multicolumn{2}{c}{HST --- 2.5" aperture}	& \multicolumn{2}{c}{HST --- 0.10" aperture } \\[4pt]
		name 	& FUV & NUV 			& F125LP 	& F150LP 				& F125LP 	& F150LP 		& F125LP	& F150LP \\}
\startdata 
PS1-10jh   & $>24.5$ & $>25.2$ & 25.8 & 25.8 &  $22.76\pm0.17$ & $22.82\pm0.06$ & $23.12\pm0.04$ & $23.11\pm0.06$\\
PTF-09ge   & $22.1\pm 0.4$ & $21.1\pm 0.3$ & 22.5 & 22.6 &  $21.64\pm0.06$ & $21.75\pm0.03$ & $23.09\pm0.04$ & $23.12\pm0.07$\\
PTF-09djl  & $>21.6$ & $22.3\pm 0.4$ & 23.1 & 23.1 &  $>23.18$ & $23.00\pm0.07$ & $23.10\pm0.04$ & $23.15\pm0.06$\\
SDSS-TDE2  & $22.6\pm 0.2$ & $22.4\pm 0.1$ & 23.2 & 22.8 &  $22.95\pm0.03$ & $22.51\pm0.05$ & $24.21\pm0.06$ & $23.89\pm0.09$\\
GALEX-D1-9 & $>24.9$ & $>25.0$ & 25.8 & 25.4 &  $>24.08$ & $>24.65$ & $25.05\pm0.09$ & $24.85\pm0.13$\\
GALEX-D23H-1 & $22.7\pm 0.1$ & $21.9\pm 0.2$ & 23.0 & 22.9 &  $22.80\pm0.05$ & $22.80\pm0.06$ & $24.98\pm0.10$ & $24.95\pm0.17$\\
GALEX-D3-13 & $>24.2$ & $>24.3$ & 27.4 & 27.2 &  $>23.37$ & $>24.94$ & $25.69\pm0.17$ & $25.62\pm0.17$\\
SDSS-TDE1  & &  & 26.1 & 26.1 &  $>25.65$ & $>24.56$ & $>27.39$ & $>26.31$\\
\enddata
\tablecomments{Observed and predicted UV flux of the eight TDF with late-time {\it HST} observations. The {\it HST} F125LP/F150LP filters probe a typical wavelength that is slightly lower/higher than the effective wavelength of the \textsl{GALEX} FUV filter (see Fig.~\ref{fig:bandpass}). The pre-flare \textsl{GALEX} observations measure the total galaxy flux. Together with optical data from the host galaxy, the \textsl{GALEX} observations of the host can be used to predict the galaxy magnitude in the two {\it HST} FUV bands (see Sec.~\ref{sec:popsyn}). The last two columns show our {\it HST} measurements in two apertures. The large 2.5" aperture includes both the TDF flux and total host galaxy flux, while the small 0.1" aperture is dominated by the flux of the TDF (see Fig.~\ref{fig:countshist}). All magnitudes are corrected for Galactic extinction.}\label{tab:UVmags}
\tablerefs{$^1$\citealt{Gezari12}, $^2$\citealt{Arcavi14}, $^3$\citealt{vanVelzen10}, $^4$\citealt{Gezari08}, $^5$\citealt{Gezari09}, $^5$\citealt{Gezari06}.}

\end{deluxetable*}

\subsection{HST photometry}\label{sec:HSTphot}
\begin{figure*}
\begin{center}
\includegraphics[width=115pt, trim=56mm 3mm 35mm 5mm, clip]{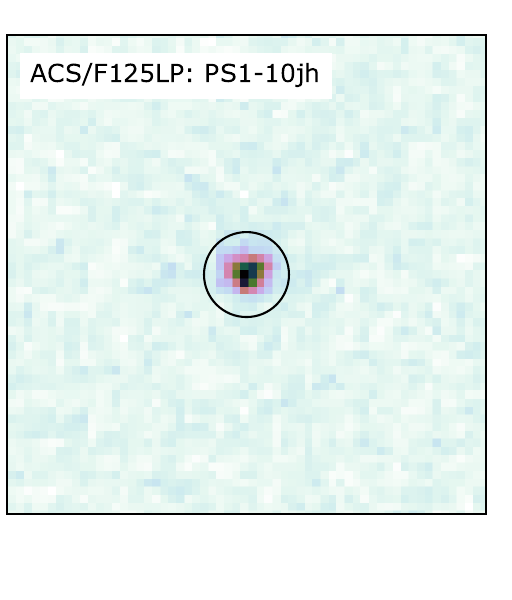}
\includegraphics[width=115pt, trim=56mm 3mm 35mm 5mm, clip]{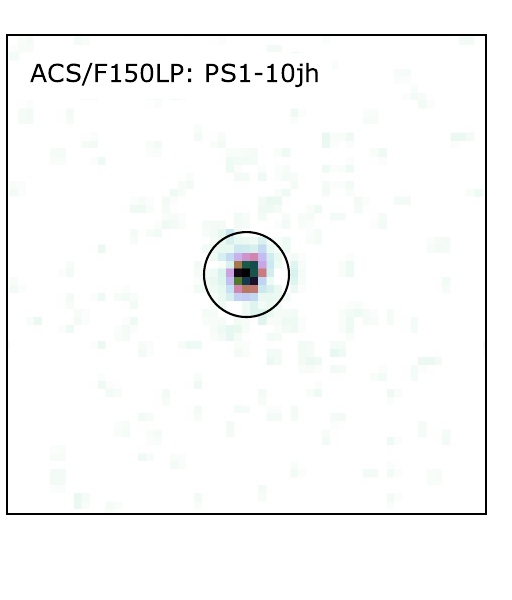} \qquad
\includegraphics[width=115pt, trim=56mm 3mm 35mm 5mm, clip]{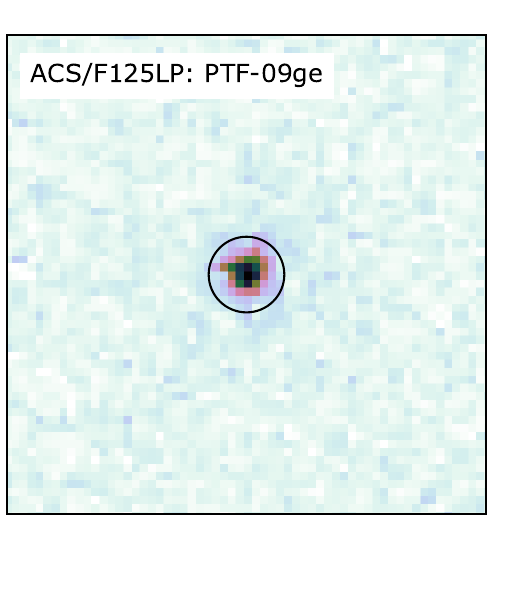}
\includegraphics[width=115pt, trim=56mm 3mm 35mm 5mm, clip]{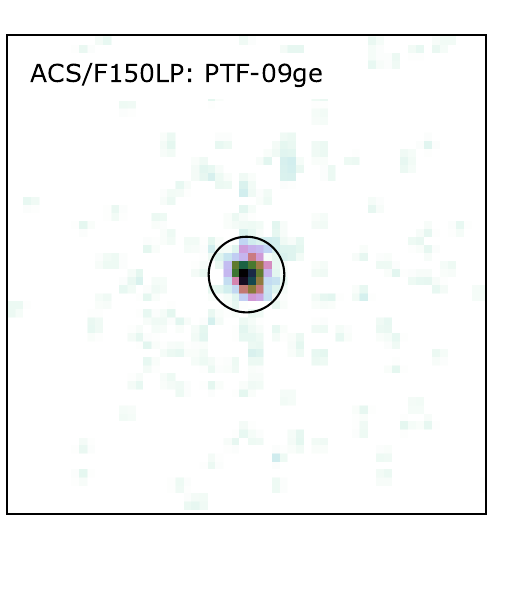} 

\includegraphics[width=115pt, trim=56mm 3mm 35mm 5mm, clip]{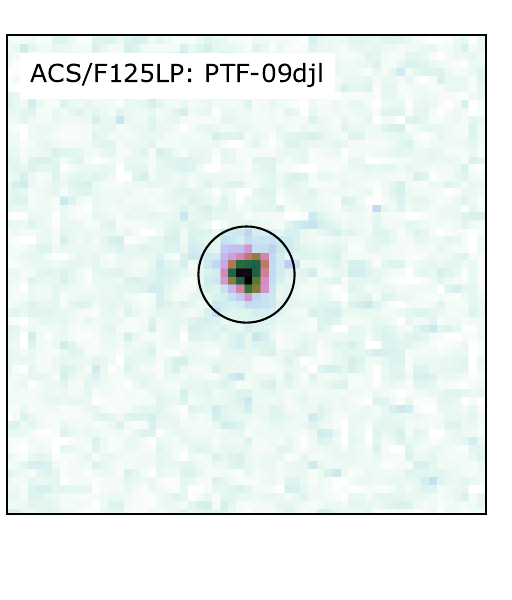} 
\includegraphics[width=115pt, trim=56mm 3mm 35mm 5mm, clip]{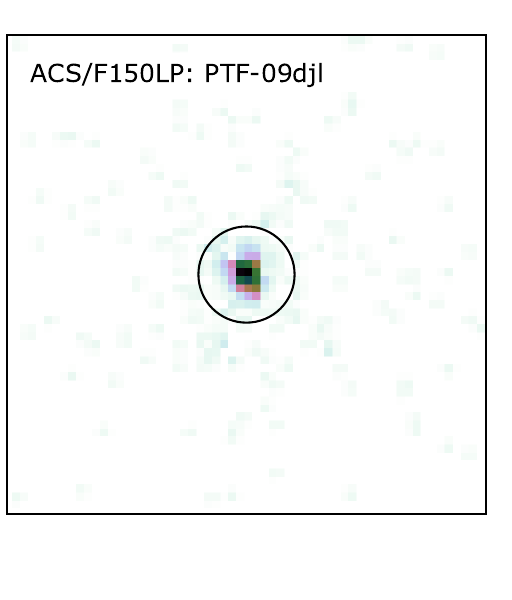} \qquad
\includegraphics[width=115pt, trim=56mm 3mm 35mm 5mm, clip]{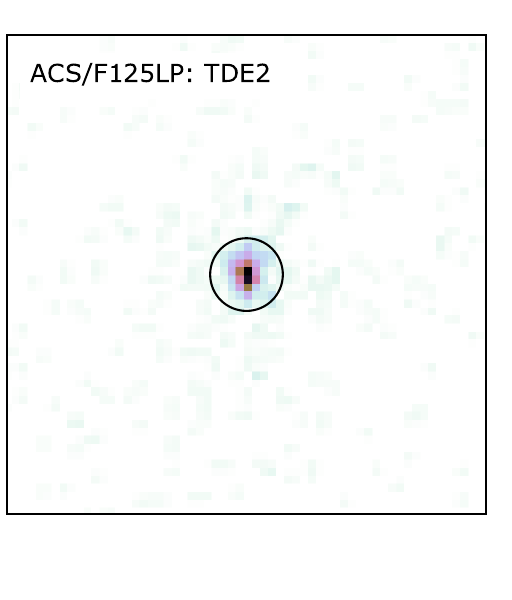}
\includegraphics[width=115pt, trim=56mm 3mm 35mm 5mm, clip]{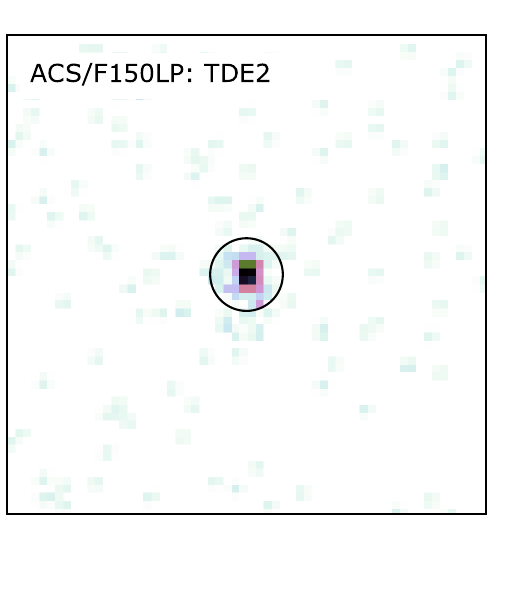} 

\includegraphics[width=115pt, trim=56mm 3mm 35mm 5mm, clip]{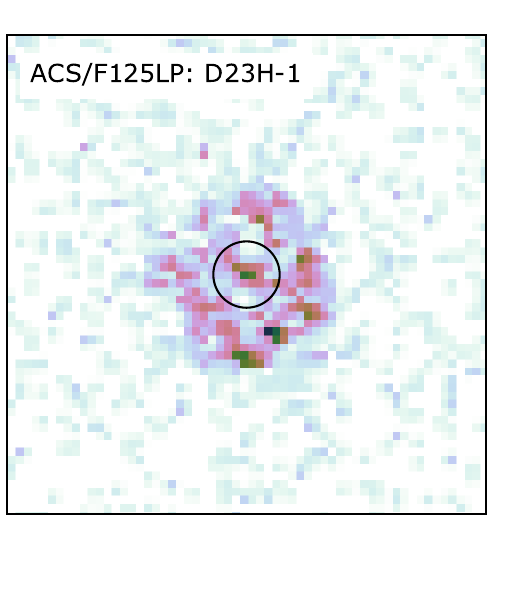}
\includegraphics[width=115pt, trim=56mm 3mm 35mm 5mm, clip]{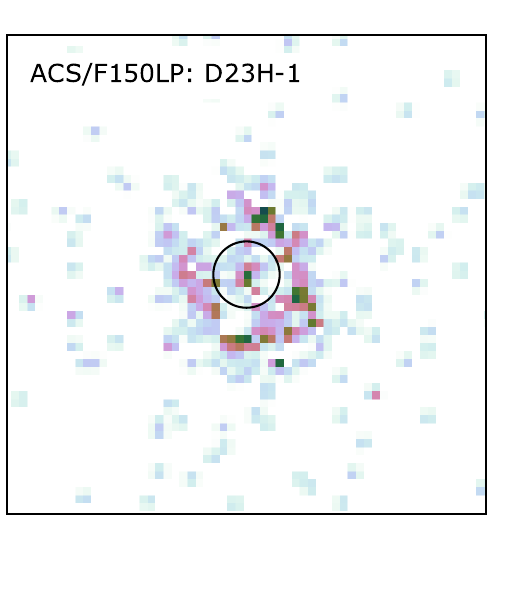} \qquad
\includegraphics[width=115pt, trim=56mm 3mm 35mm 5mm, clip]{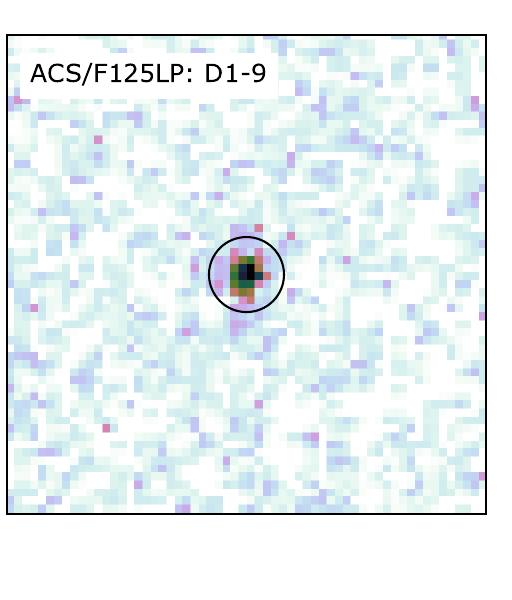}
\includegraphics[width=115pt, trim=56mm 3mm 35mm 5mm, clip]{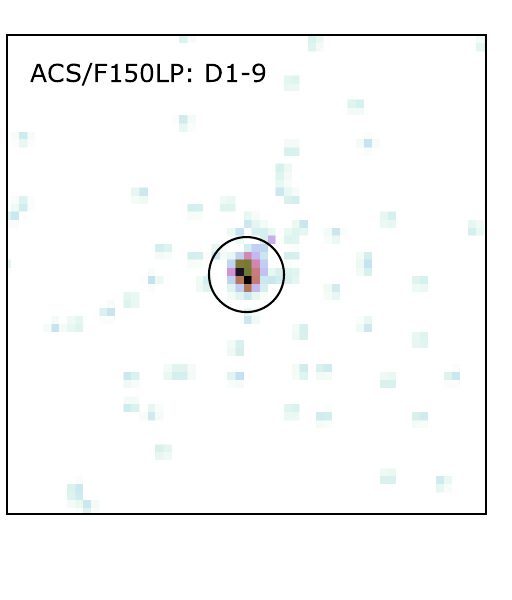} 

\includegraphics[width=115pt, trim=56mm 3mm 35mm 5mm, clip]{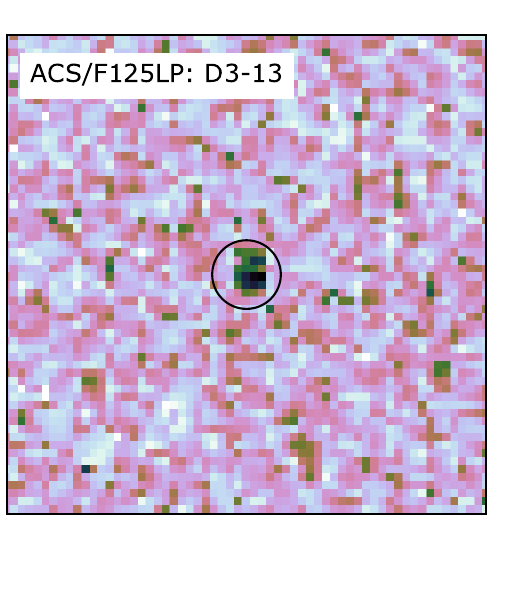}
\includegraphics[width=115pt, trim=56mm 3mm 35mm 5mm, clip]{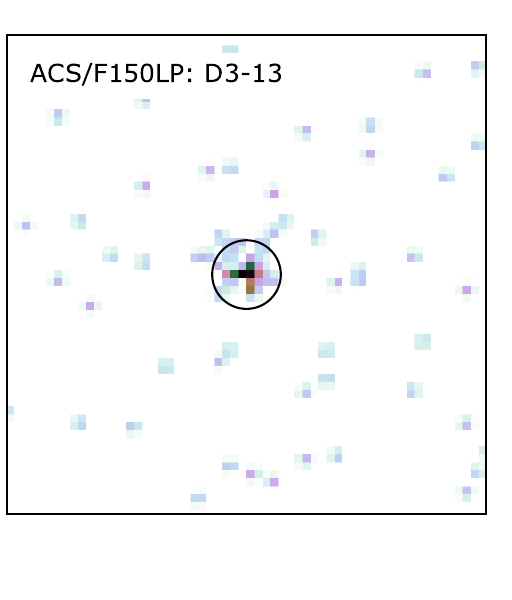} \qquad
\includegraphics[width=115pt, trim=56mm 3mm 35mm 5mm, clip]{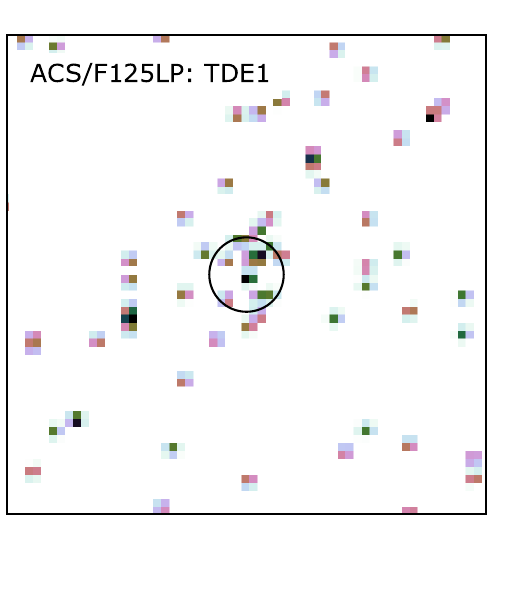}
\includegraphics[width=115pt, trim=56mm 3mm 35mm 5mm, clip]{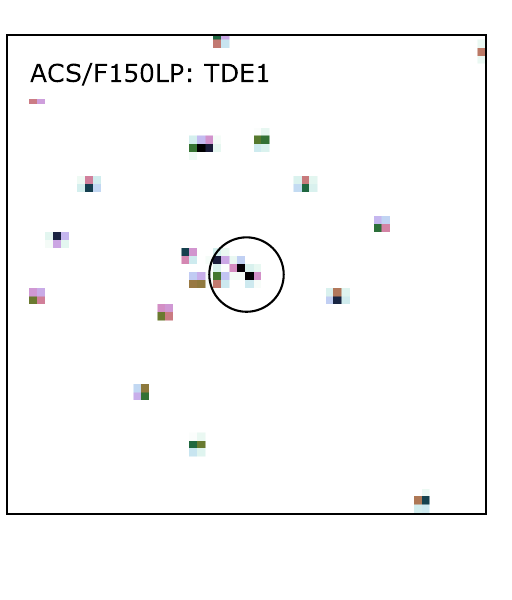}

\caption{Cutouts of the {\it HST} ACS FUV images, $1.3\times1.3$~arcsec, on a linear scale. The circle indicates our 0.10" aperture radius that is used to extract the flux. }
\label{fig:cutouts}
\end{center}
\end{figure*}

The {\it HST} ACS/SCB imaging observations were obtain with a two-point dither and we combined these exposures using the default ``drizzle" \citep{Fruchter02} parameters of the {\it HST} ACS reduction pipeline. Cutouts of the images are shown in Fig.~\ref{fig:cutouts}.
We measured the flux in the F125LP and F150LP images using a circular aperture of 0.1" (4.5 pixel). The aperture correction for this radius is 0.70 and 0.60~mag for F125LP and F150LP, respectively. This small radius was chosen to minimize any contribution of the host galaxy light. Seven source are detected with a signal-to-noise ratio of at least 6, independently in both the F125LP and F150LP images; only for TDE1 do we find a nondetection in both bands.

\begin{figure*}
\begin{center}

\includegraphics[width=240pt, trim=5mm 10mm 4mm 2mm, clip]{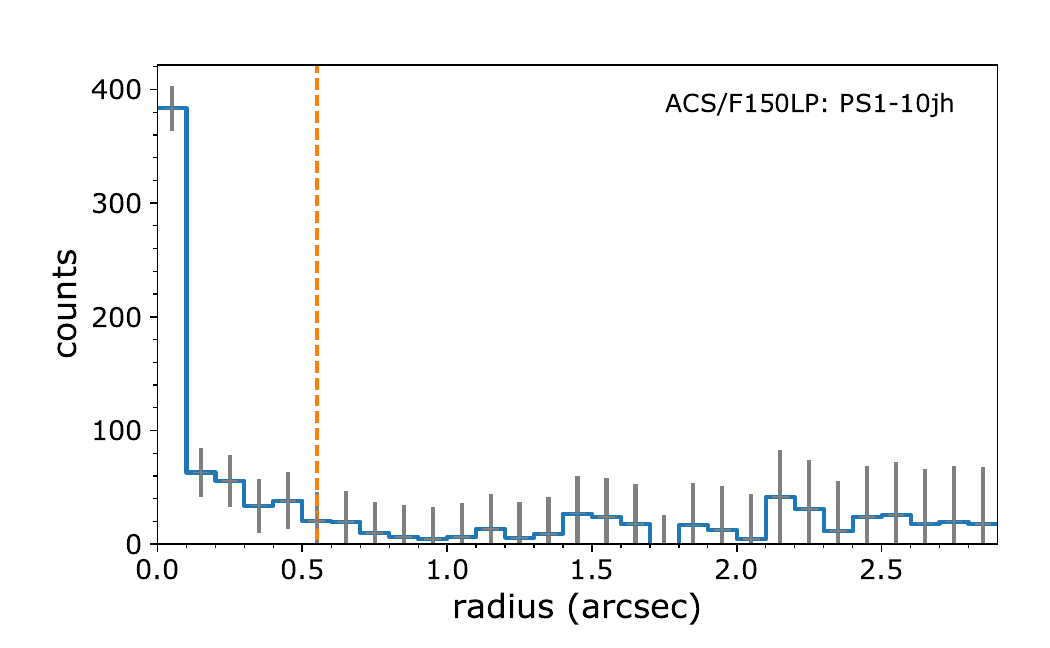} \quad
\includegraphics[width=240pt, trim=5mm 10mm 4mm 2mm, clip]{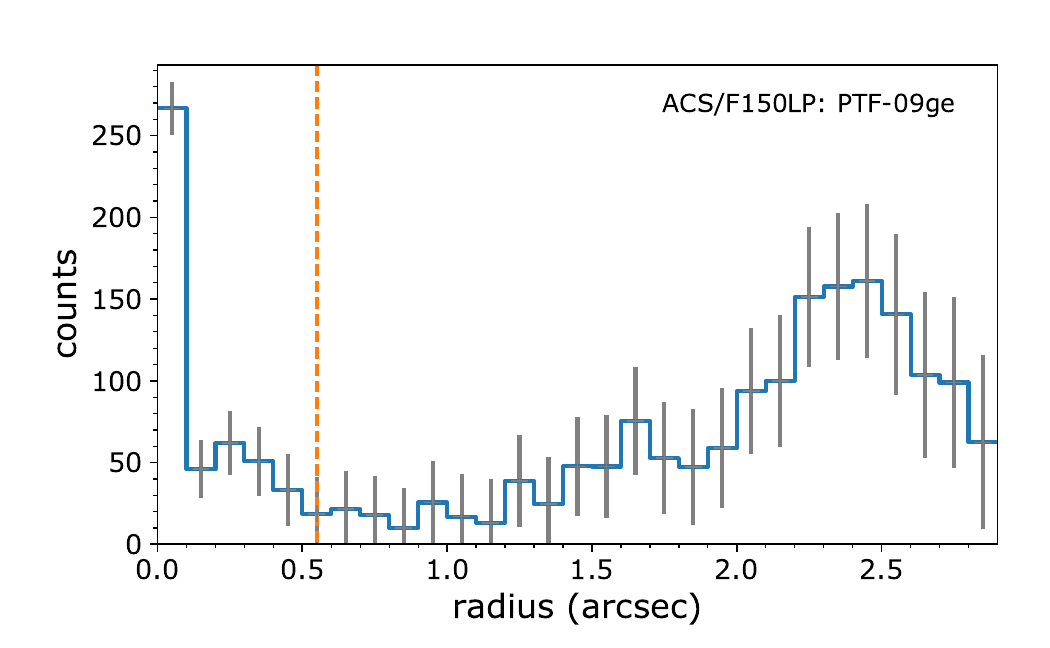} 

\includegraphics[width=240pt, trim=5mm 10mm 4mm 2mm, clip]{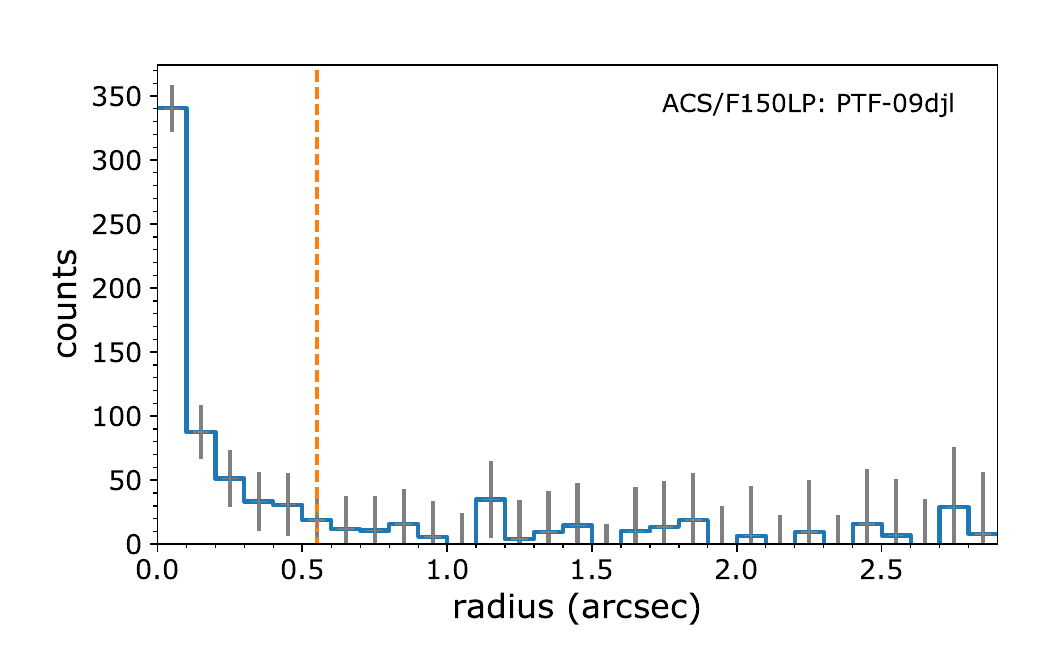} \quad
\includegraphics[width=240pt, trim=5mm 10mm 4mm 2mm, clip]{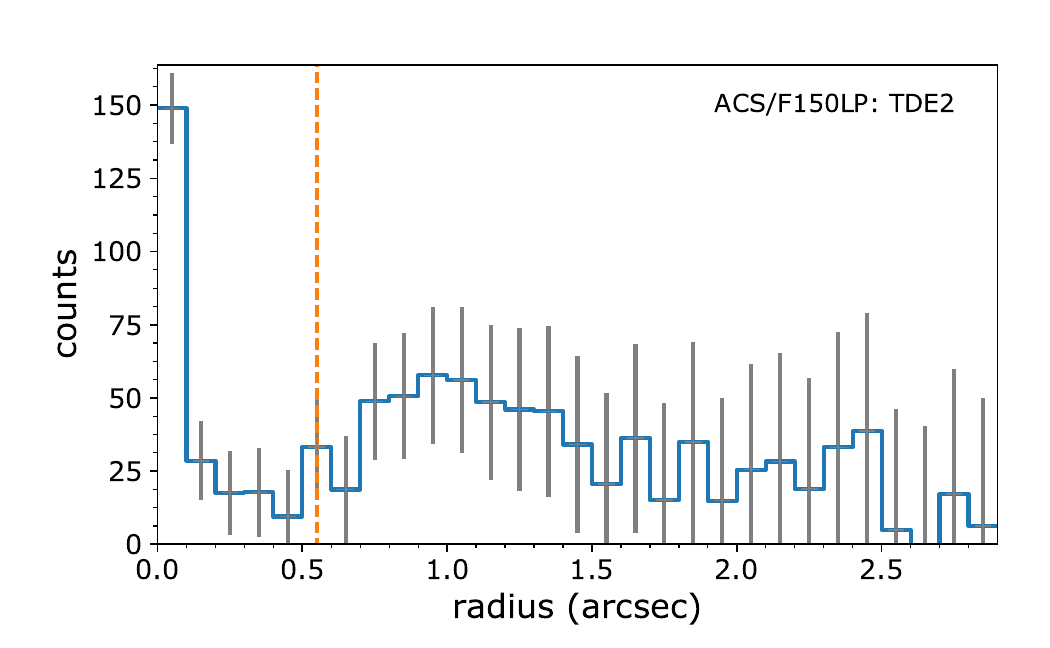}

\includegraphics[width=240pt, trim=5mm 10mm 4mm 2mm, clip]{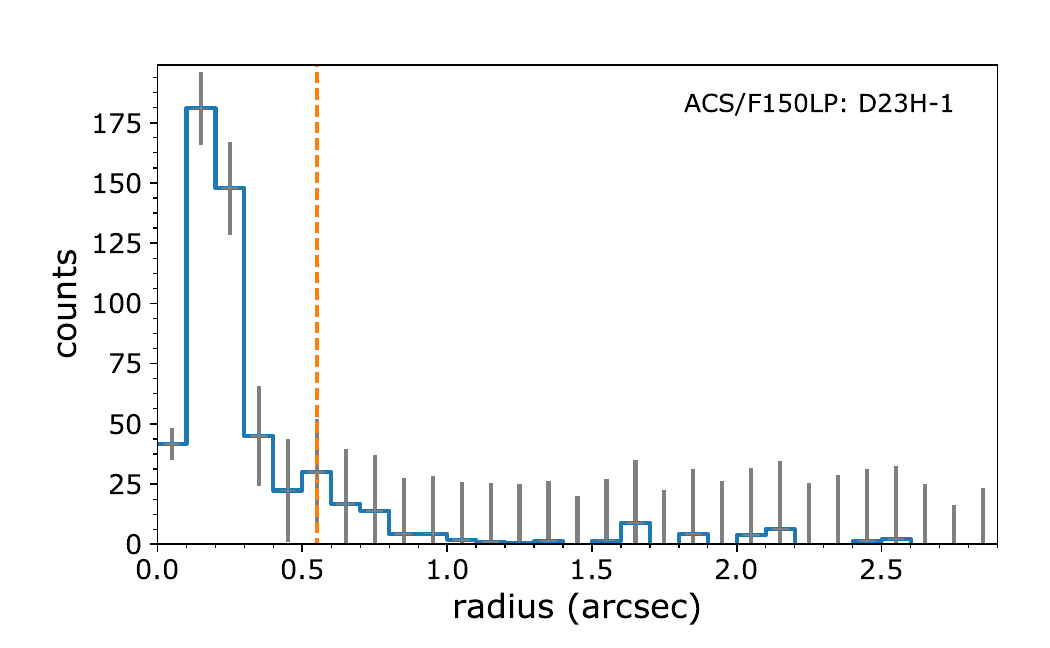} \quad
\includegraphics[width=240pt, trim=10mm 10mm 4mm 2mm, clip]{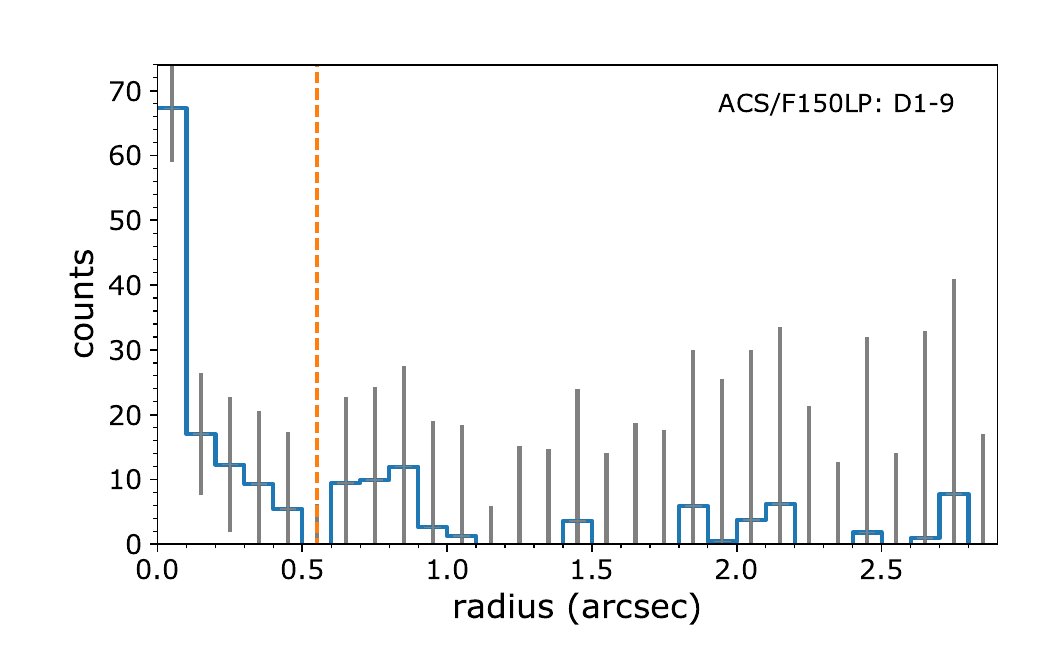} 

\includegraphics[width=240pt, trim=5mm 0mm 4mm 2mm, clip]{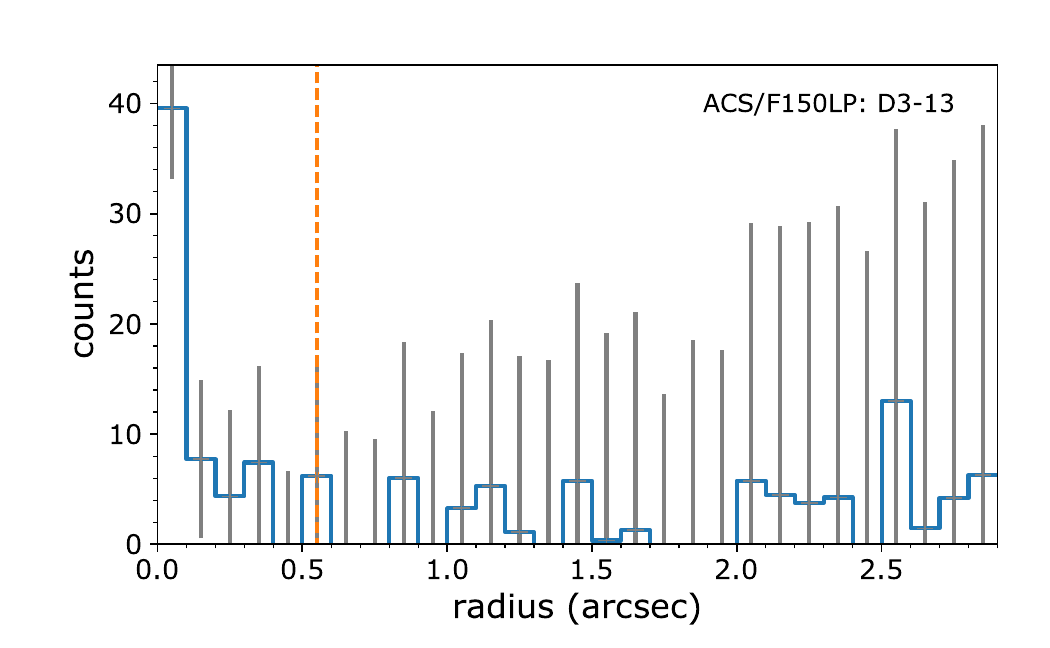} \quad
\includegraphics[width=240pt, trim=5mm 0mm 4mm 2mm, clip]{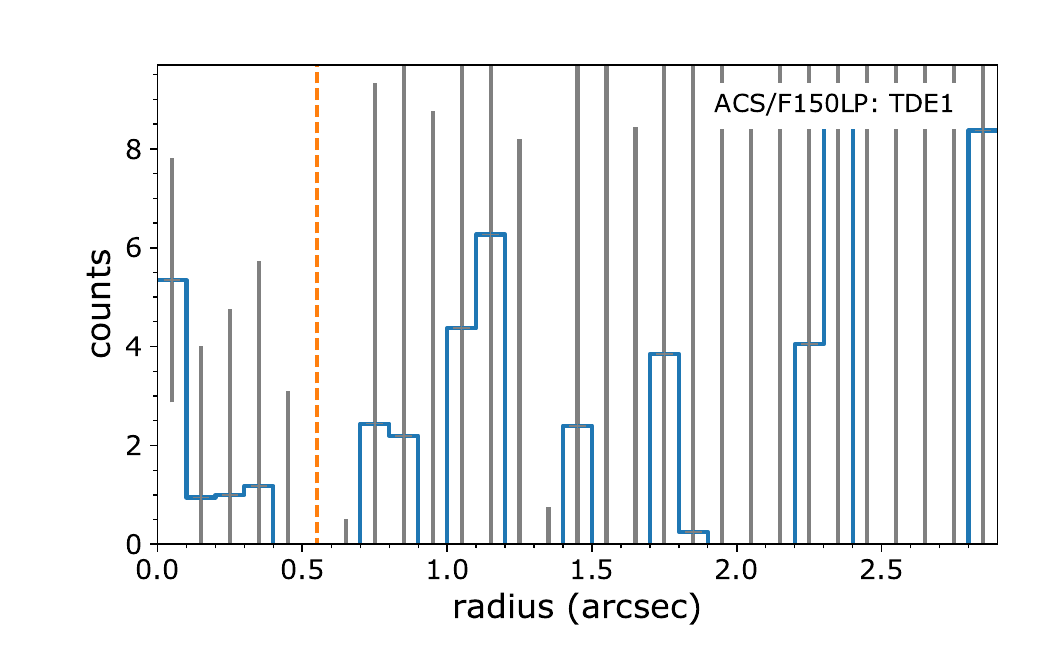}

\caption{Histogram of total counts in an annulus centered on the central point source. The innermost bin contains the counts within the 0.10" aperture that is used to measure the flux of the central source. The dashed orange line indicates the radius that contains 90\% of the flux from a point source.
We see that for three of the sources, PTF-09ge, SDSS-TDE2, and GALEX-D23H-1, the hosts show extended FUV emission surrounding a central unresolved source. Except for GALEX-D23H-1, the flux of the point-source dominates over the extended emission.}
\label{fig:countshist}
\end{center}
\end{figure*}

For six of the seven detections, a point source dominates the total FUV flux.  (Fig.~\ref{fig:countshist}). Since the galaxies in our sample have an effective radius that is $\sim 10$ times larger than the angular resolution of the {\it HST} observations ($\approx 0.1$~kpc at $z=0.1$), we expect that these six central point sources are of non-stellar origin (i.e. late-time TDF emission). We quantify this statement using additional analysis presented in section~\ref{sec:fuv_compare}.

In some cases, additional observations allow us to directly reject the hypothesis that a compact population of young stars is the origin of the FUV points sources in our sample. Below we discuss the evidence for non-stellar emission for each source separately: 

\begin{itemize}
	\item For PS1-10jh, the {\it HST} FUV flux in our nuclear aperture (0.1") is an order of magnitude larger than the upper limit on the FUV flux based on {\it GALEX} data obtained before the TDF. This confirms our late-time UV detections are transient. We also note that {\it HST} WFC3 observations in the F625W ($r$) band \citep{Gezari15} show no evidence for the strong central enhancement that would be needed to explain the post-flare nuclear FUV emission with a central population of young stars.  
	
	\item The {\it HST} FUV images of PTF-09ge show both extended emission and a nuclear point source. The FUV flux of the nuclear point source is contained within 0.1" or 0.1~kpc and is only a factor of 3.5 smaller than the total FUV flux of the galaxy. If this nuclear emission is due to the same population of stars that dominate the total galaxy light, one-third of this galaxy's mass would be contained within an aperture that is a factor of 20 smaller than the effective radius of the galaxy. A concentration of young stars on 0.1~kpc scales is inconsistent with post-flare {\it HST} imaging that shows a constant optical/IR color (F438W$-$F814W) for radii smaller than 1~kpc (K.\,D. French et al. 2019 in prep).
	
	
	\item For SDSS-TDE2, we detect the expected host galaxy FUV flux from our fit to the {\it GALEX} and SDSS data (Fig.~\ref{fig:popsyn}) only inside a large ($>$2") aperture. The FUV flux of the nuclear component is contained within a projected radius of 0.4~kpc and is a factor of 3 smaller compared to the total FUV of the galaxy. Based on the $r$-band De Vaucouleurs's radius, the expected light contained within is radius is only 5\%. 
	
	\item For PTF-09djl, the expected total host galaxy FUV flux from our fit to the SDSS and {\it GALEX} photometry is similar to the observed flux in our small {\it HST} aperture. In other words, if the nuclear FUV would be due to the same population of stars that are observed in the SDSS and {\it GALEX} photometry, most of the galaxy should have a size of 0.1" or 0.2~kpc, which is clearly not consistent with the SDSS observations. 
	
	\item Likewise, for the TDF GALEX-D1-9, the upper limit on the total host galaxy FUV flux is similar to the observed flux in our small {\it HST} aperture (0.5~kpc). This again  suggests the nuclear FUV flux is of a non-stellar origin. 
	
	\item For the GALEX-D3-13, the expected total FUV flux of this relatively high-redshift galaxy ($z=0.37$) is fainter than the observed {\it HST} flux within our 0.1" aperture (0.5~kpc). Most importantly,  no evidence for a strong central concentration of stars is seen in an {\it HST} $I$-band image that is available for this galaxy \citep{Gezari06}.
	
	
	\item The source GALEX-D23H-1 is an outlier in our sample of {\it HST} observations. We see a ring-like structure of FUV emission around what appears to be a central point source (Fig.~\ref{fig:cutouts}). The circumnuclear emission is qualitatively consistent with the host galaxy's star-forming classification from emission line observations \citep{Gezari09} and the {\it GALEX} observations.  For this source, we cannot be certain the nuclear flux is of non-stellar origin, and we therefore consider the flux in the 0.1" aperture to be an upper limit on the late-time TDF emission. 
\end{itemize}

\subsection{Swift photometry}\label{sec:Swiftphot}
The flux in the {\it Swift} UVOT images is measured using the \verb uvotsource  task. To capture the entire flux of the galaxy we use a 5" aperture (except, as motivated in Sec.~\ref{sec:popsyn}, for iPTF-16fnl, where we use a 4" aperture). The host galaxy flux is subtracted using the best-fit synthetic galaxy model (see Section~\ref{sec:popsyn}). To compute the uncertainty on the host-subtracted flux, we conservatively adopt an uncertainty of 0.1~mag on our estimate of the host flux level.   

\begin{figure*}
\begin{center}

\includegraphics[width=240pt, trim=6mm 12mm 4mm 6mm, clip]{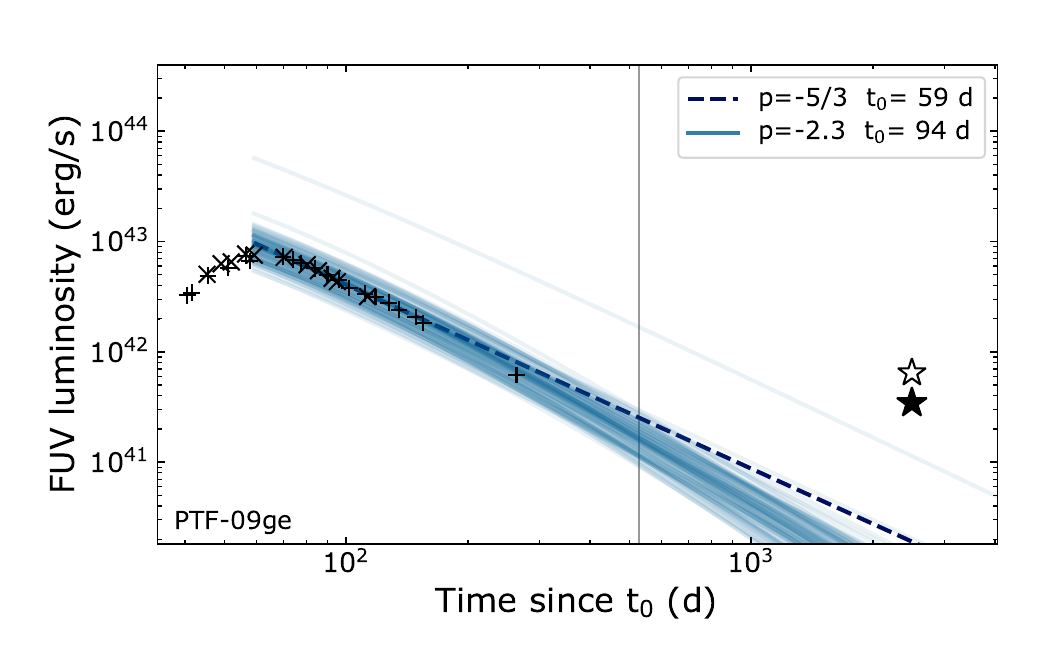}  \quad 
\includegraphics[width=240pt, trim=6mm 12mm 4mm 6mm, clip]{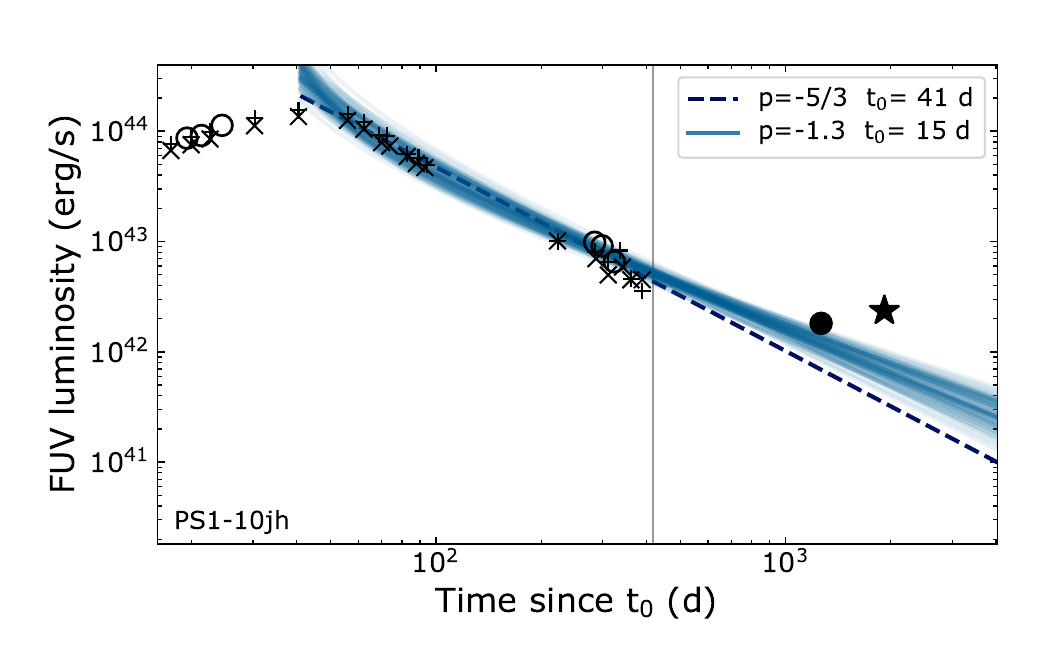}

\includegraphics[width=240pt, trim=6mm 12mm 4mm 6mm, clip]{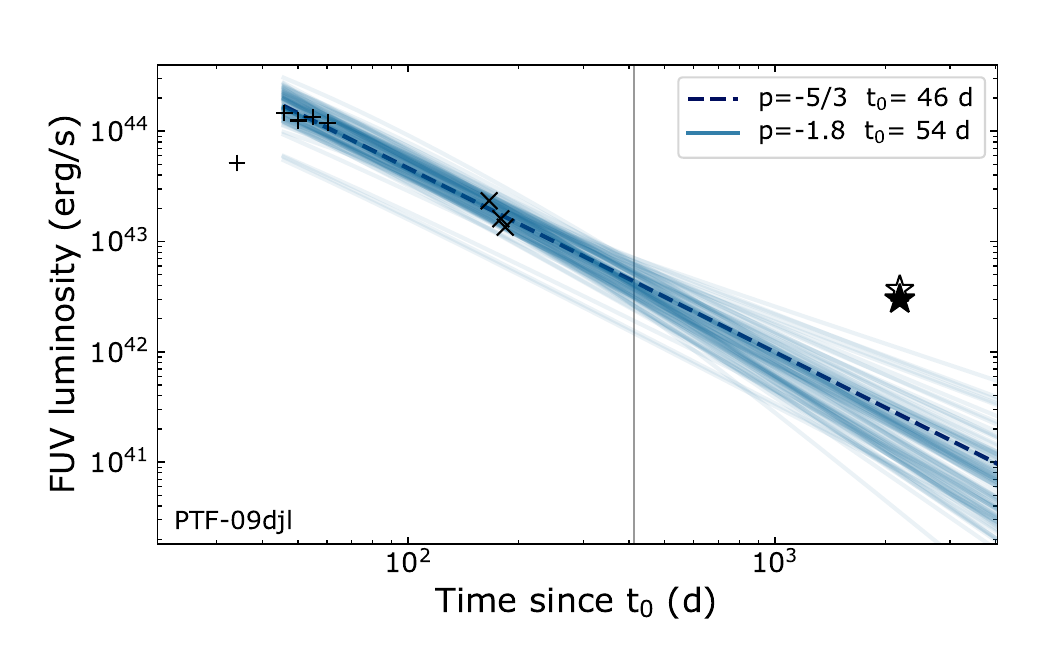} \quad
\includegraphics[width=240pt, trim=6mm 12mm 4mm 6mm, clip]{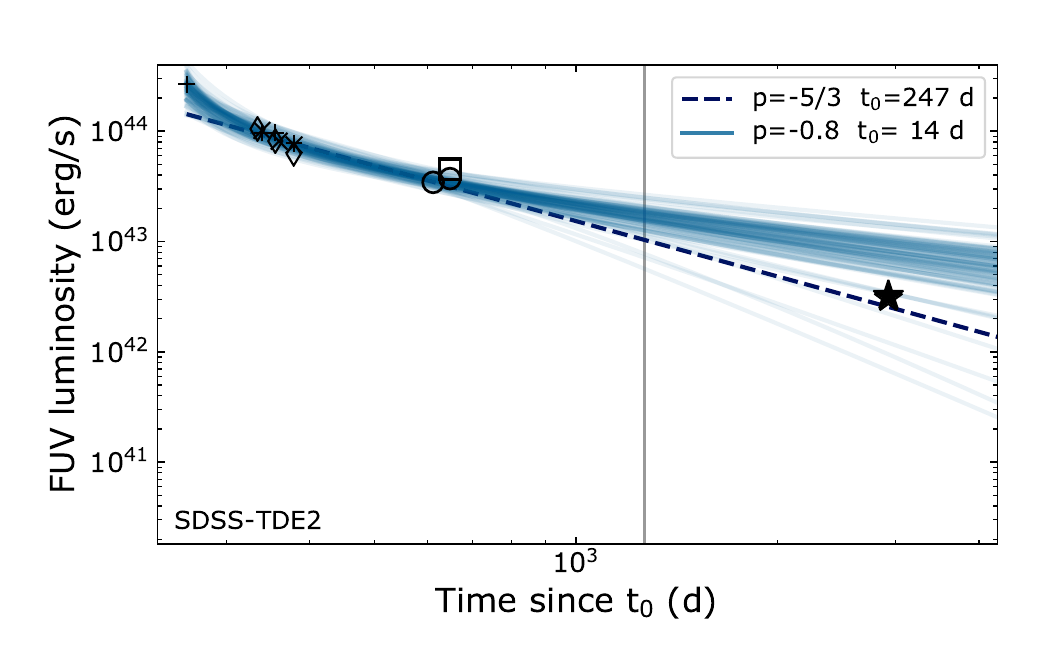}

\includegraphics[width=240pt, trim=6mm 12mm 4mm 6mm, clip]{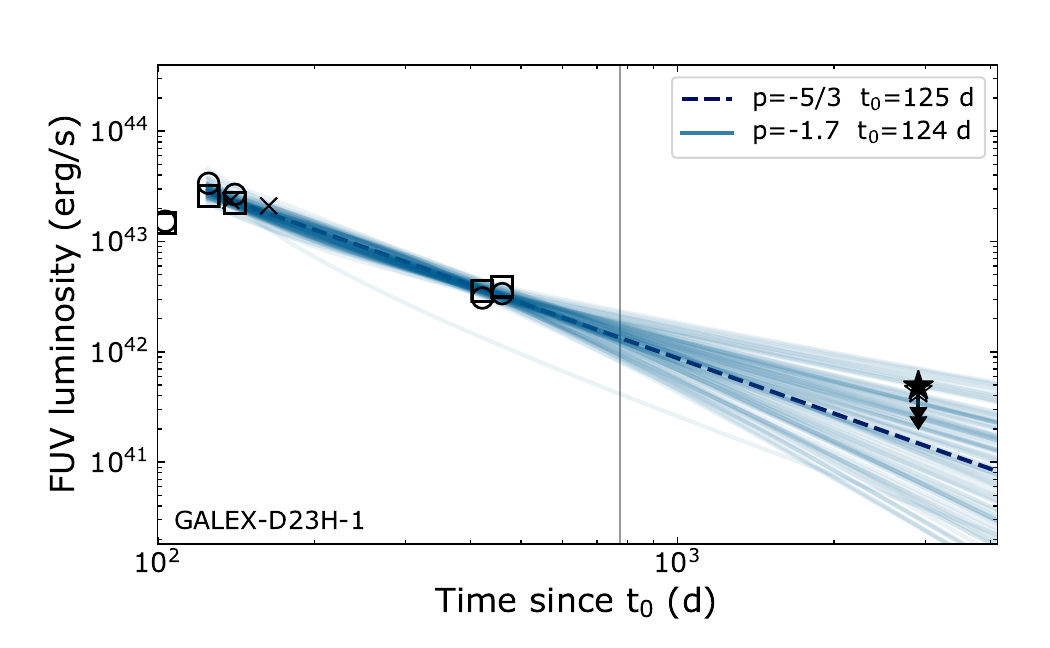} \quad
\includegraphics[width=240pt, trim=6mm 12mm 4mm 6mm, clip]{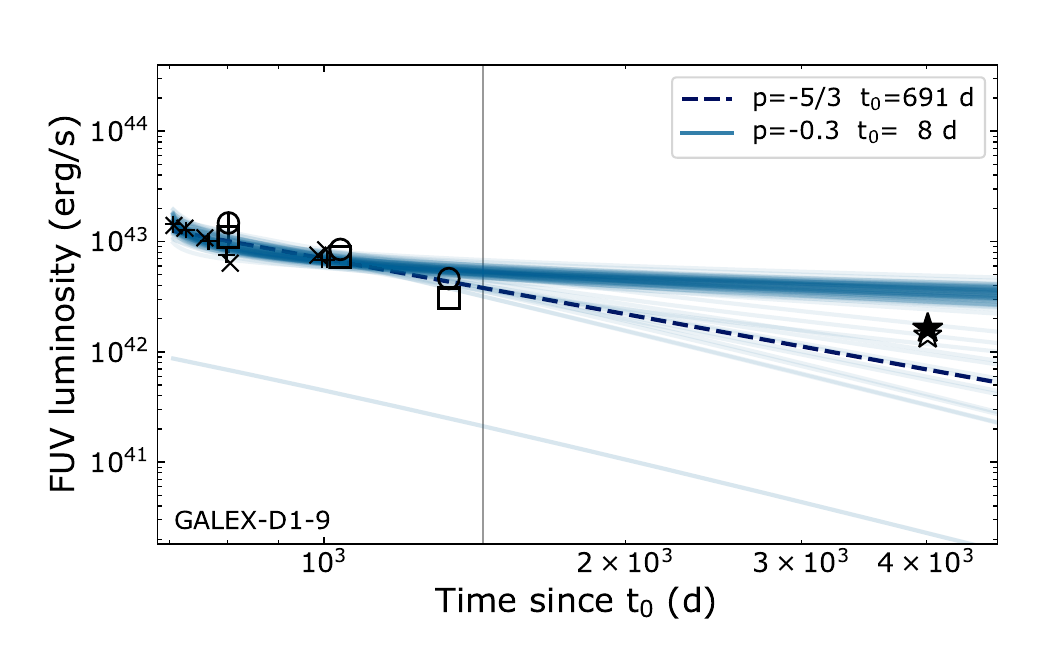} 

\includegraphics[width=240pt, trim=6mm  0mm 4mm 6mm, clip]{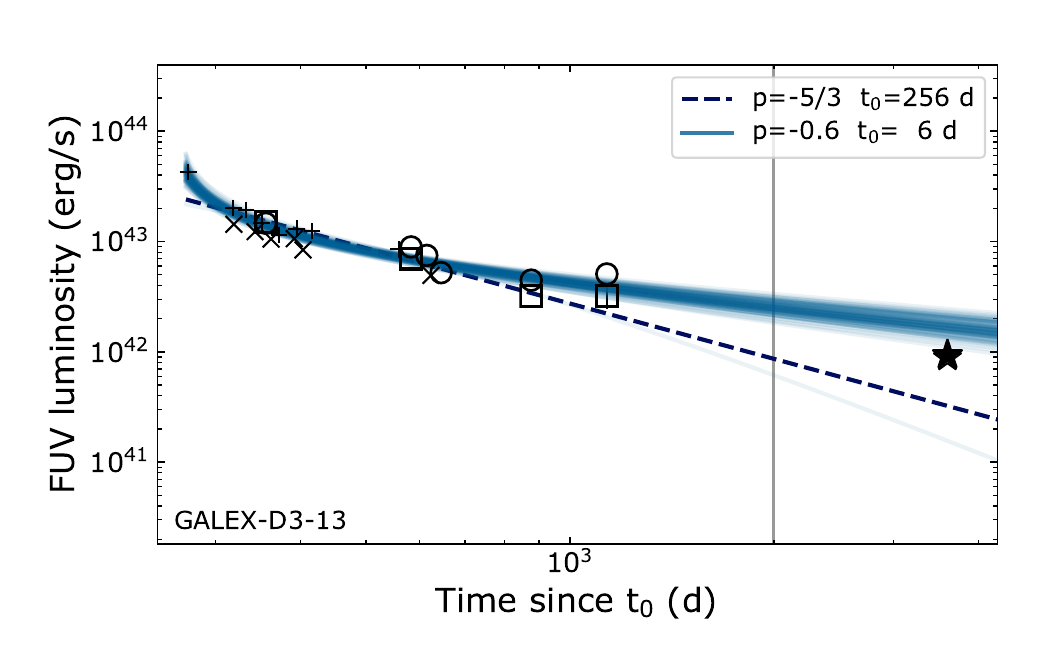} \quad
\includegraphics[width=240pt, trim=6mm  0mm 4mm 6mm, clip]{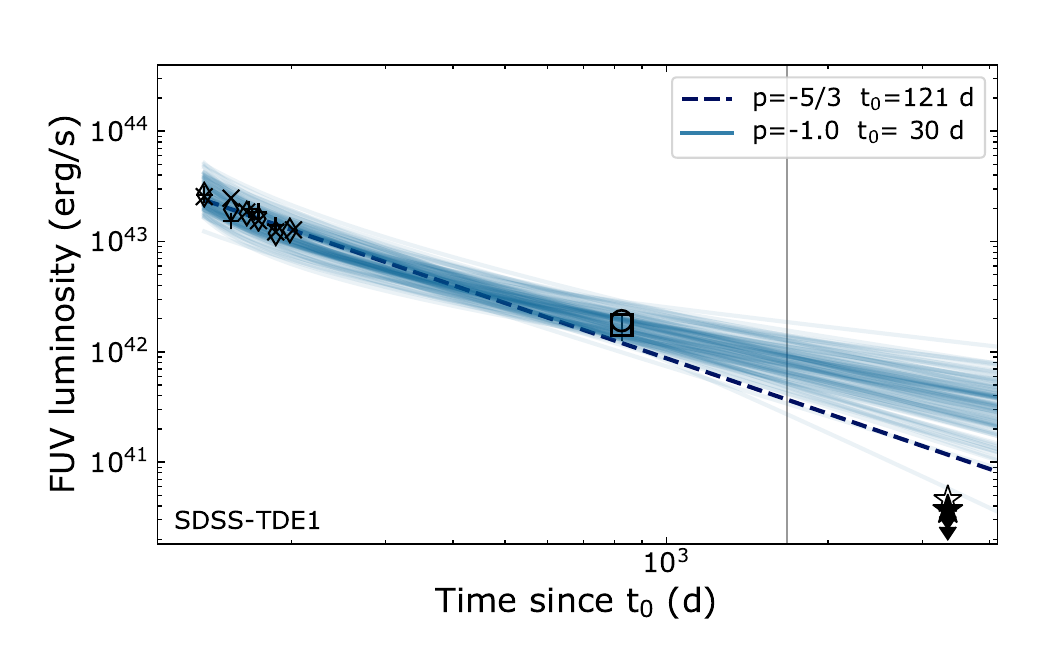}

\caption{Light curves of the eight TDFs with late-time {\it HST} FUV observations.  The dashed lines indicated the best-fit $ (t/t_{0})^{-5/3}$ power-law decay. The solid lines show the result for a fit with the power-law index ($p$) as a free parameter (the normalization of the x-axis uses $t_0$ as obtained from the fit with $p=-5/3$). We show the FUV luminosity, $\lambda L_{\lambda}$ at $\lambda=150$~nm in the rest frame, k-corrected using the mean blackbody temperature. The frequency of the observations is indicated by the plot symbol: {\it HST} F125LP $\largewhitestar$, {\it HST} F150LP $\bigstar$, FUV $\square$, NUV  $\ocircle$, {\it HST} F225W $\CIRCLE$, $u$-band $\lozenge$, $g$-band $\times$, $r$-band $+$. All power-law fits are to early-time data (observations before $10 t_{\rm fb}$), a demarcation indicated by the vertical gray line in each panel.}
\label{fig:lcfit}
\end{center}
\end{figure*}

%
\subsection{Light curve inference}\label{sec:lightcurves}
%
To compare the late-time UV luminosity to the flux expected based on the observations near peak, we describe the early-time (but post-peak) light curve with a power law, 
\begin{equation}\label{eq:lc}
L_{\nu}(t,\nu)  =  L_{\rm peak} ( [t+t_{\rm peak}+t_0] / t_0)^p ~ B_\nu(T,\nu)/B_\nu(T,\nu_0).
\end{equation}
Here $B_\nu$ is the Planck function and $t$ indicates the time relative to the light curve maximum, as measured in the source rest frame. Some sources in our sample are only detected post-peak; hence we include a nuisance parameter $t_{\rm peak}$ that counts the time between between the true peak of the light curve and the first detection. The other four free parameters of our model are: the blackbody temperature ($T$), the peak luminosity ($L_{\rm peak}$) at the reference frequency $\nu_0$ , the characteristic decay time ($t_0$), and the power-law index ($p$). 

We use a Markov Chain Monte Carlo (MCMC) method \citep{Foreman-Mackey13} to obtain the posterior distribution of the free parameters in our model. We use 200 walkers and 2000 steps; we confirm for each case that the walkers converge and only use the last 1000 steps in the computation of the posterior distributions. To estimate the likelihood of the data we use assume Gaussian statistics, but we allow the variance of the data to be overestimated by a factor $f$ \citep[e.g.][]{Guillochon18}. We use a flat prior for all free parameters: 
\begin{itemize}
	\item $4<\log\left( \frac{T}{\rm K}\right)<5$;
	\item $-3<p<3$;
	\item $0.5<\log \left( \frac{t_0}{\rm d}\right)  <3$;
	\item  $35<\log \left( \frac{L_{\rm peak}}{\rm erg~s^{-1}} \right)<50$;
	\item $-5<\ln(f)<-1$.
\end{itemize}
When the peak of the light curve is not resolved, we allow the true peak to be up to 30 days before the first observation in the light curve: $0<t_{\rm peak}<30$. When the peak is observed, this nuisance parameter is not needed, and we simply set $t_{\rm peak}=0$. The only role of the nuisance parameter is to propagate our uncertainty on the time of the true peak into the posterior distribution of $t_0$. However, because $t/t_0\gg t_{\rm peak}/t_0$ for the majority of the observations, the uncertainty about the true time of the peak has virtually no influence on the inferred values of $t_0$ or $p$.

For two sources in our sample we have to make an exception and use more informative priors to obtain convergent solutions for the model light curve (Eq.~\ref{eq:lc}). To avoid solutions with a very steep power-law index, we use a Gaussian prior at $p=-5/3 \pm 0.3$ for PTF-09ge and PTF-09djl. For PTF-09djl the light curve has limited frequency coverage (almost exclusively $r$-band observations), so for this source we therefore also use a Gaussian prior for $\log T$, centered on the blackbody temperature measured from  spectroscopic observations \citep{Arcavi14}.

\begin{deluxetable*}{l c c c c c c c c c c}
\tablewidth{0pt}
\tablecolumns{11}
\tablecaption{Observed properties: early time versus late time}
\tablehead{name & redshift & $M_\bullet$ & $T_{\rm early}$\tablenotemark{a}  & $\nu L_{\nu}$$_{\rm max}$\tablenotemark{b} & $R_{\rm max}$ &$t_{\rm late}$\tablenotemark{c} & $\Delta t$\tablenotemark{d} & $T_{\rm late}$\tablenotemark{e} & $\nu L_{\nu}$$_{\rm late}$ & $R_{\rm late}$\\
                    &   & $(\log M_\odot)$     &($\log$ K) & ($\log$\,erg/s)&  ($\log$ cm) &(MJD) & (days) & ($\log$ K) & ($\log$ erg/s) & ($\log$ cm)}
\startdata
PS1-10jh   & 0.170 & 6.06 & $4.59_{-0.09}^{+0.15 }$ & 44.2 & 14.6 & 57585.6 & 1877 &$4.7_{-0.2}^{+0.5}$ & $42.4\pm0.01$ & $13.5_{-0.2}^{+0.1}$ \\
PTF-09ge   & 0.064 & 6.25 & $4.08_{-0.03}^{+0.03 }$ & 42.9 & 15.2 & 57577.8 & 2436 &$4.7_{-0.3}^{+0.6}$ & $41.5\pm0.02$ & $13.2_{-0.3}^{+0.3}$ \\
PTF-09djl  & 0.184 & 6.03 & $4.41_{-0.08}^{+0.08 }$ & 44.2 & 14.9 & 57588.5 & 2144 &$4.8_{-0.3}^{+0.6}$ & $42.5\pm0.01$ & $13.6_{-0.3}^{+0.2}$ \\
TDE2       & 0.256 & 6.91\tablenotemark{f} & $4.37_{-0.04}^{+0.05 }$ & 44.4 & 15.1 & 57617.8 & 2666 &$4.4_{-0.2}^{+0.7}$ & $42.5\pm0.02$ & $14.2_{-0.4}^{+0.2}$ \\
D1-9       & 0.326 & 6.64\tablenotemark{f} & $4.59_{-0.05}^{+0.07 }$ & 43.2 & 14.2 & 57607.2 & 3302 &$4.5_{-0.3}^{+0.7}$ & $42.2\pm0.04$ & $13.7_{-0.4}^{+0.3}$ \\
D23H-1\tablenotemark{g}     & 0.185 & 6.53 & $4.70_{-0.09}^{+0.14 }$ & 43.5 & 14.2 & 57668.8 & 2780 &$4.6_{-0.2}^{+0.7}$ & $<41.7$ & $<13.3$ \\
D3-13      & 0.370 & 7.38 & $4.66_{-0.05}^{+0.06 }$ & 43.6 & 14.3 & 57583.8 & 3332 &$4.8_{-0.5}^{+0.6}$ & $42.0\pm0.07$ & $13.3_{-0.4}^{+0.4}$ \\
TDE1       & 0.136 & 7.28 & $4.42_{-0.10}^{+0.15 }$ & 43.5 & 14.6 & 57615.9 & 3206 &-- &$<40.6$ & -- \\
ASASSN-14ae & 0.044 & 5.68 & $4.29_{-0.03}^{+0.03 }$ & 43.8 & 15.0 & 57434.7 & 718 &$4.5_{-0.5}^{+0.8}$ & $41.5\pm0.18$ & $13.4_{-0.6}^{+0.8}$ \\
ASASSN-14li & 0.021 & 6.40 & $4.52_{-0.04}^{+0.04 }$ & 43.6 & 14.4 & 58224.5 & 1208 &$5.0_{-0.4}^{+0.5}$ & $42.1\pm0.03$ & $13.2_{-0.2}^{+0.3}$ \\
iPTF-16fnl\tablenotemark{h} & 0.016 & 5.75 & $4.47_{-0.02}^{+0.03 }$ & 43.2 & 14.3 & 57932.1 & 295 &$4.5$ &$41.7\pm0.12$ & $13.6$ \\
ASASSN-15oi\tablenotemark{h} & 0.048 & 5.94 & $4.60_{-0.07}^{+0.08 }$ & 44.5 & 14.8 & 57841.9 & 557 &$4.6$ & $41.6\pm0.16$ & $13.3$ \\
iPTF-16axa\tablenotemark{i} & 0.108 & 6.49 & $4.46_{-0.02}^{+0.02 }$ & 43.8 & 14.6 & -- & -- & -- & -- & -- \\
iPTF-15af\tablenotemark{i}  & 0.079 & 6.96 & $4.85_{-0.19}^{+0.18 }$ & 43.8 & 14.2 & -- & -- & -- & -- & -- \\
\enddata
\tablenotetext{a}{The mean early-time temperature. Obtained from the best-fit power law (Eq.~\ref{eq:lc}) using observations with $t<10 f_{\rm fb}$ (see Sec.~\ref{sec:lightcurves}).}
\tablenotetext{b}{The maximum value of the luminosity.}
\tablenotetext{c}{The time of the most recent observation (either from {\it HST} or {\it Swift}/UVOT).}
\tablenotetext{d}{The time difference (in the rest frame of the transient) between $t_{\rm late}$ and the time of maximum light.}
\tablenotetext{e}{The mean temperature inferred for the late-time observations, i.e. $t>10 t_{\rm fb}$.}
\tablenotemark{f}{TDE2 and D1-9 have a large systematic uncertainty on the measurement of the velocity dispersion, see discussion in \citet{Wevers19}. }
\tablenotetext{g}{For the source D23H-1 the late-time {\it HST} observations are reported as upper limits since the observed nuclear FUV flux is likely dominated by the host galaxy, see Sec.~\ref{sec:HSTphot}}
\tablenotetext{h}{Only a single late-time detection, hence $T_{\rm late}$ is unconstrained. To compute $R_{\rm late}$ we assume the $T_{\rm late}=T_{\rm early}$.}
\tablenotetext{i}{No late-time observations.}

\tablecomments{All luminosities are corrected for Galactic extinction and reported at 150~nm in the rest frame of the transient. Upper/lower bounds computed from the 90\% credible interval.}\label{tab:prop}
\end{deluxetable*}

\begin{figure*}
\begin{center}
\includegraphics[width=240pt, trim=6mm 12mm 4mm 6mm, clip]{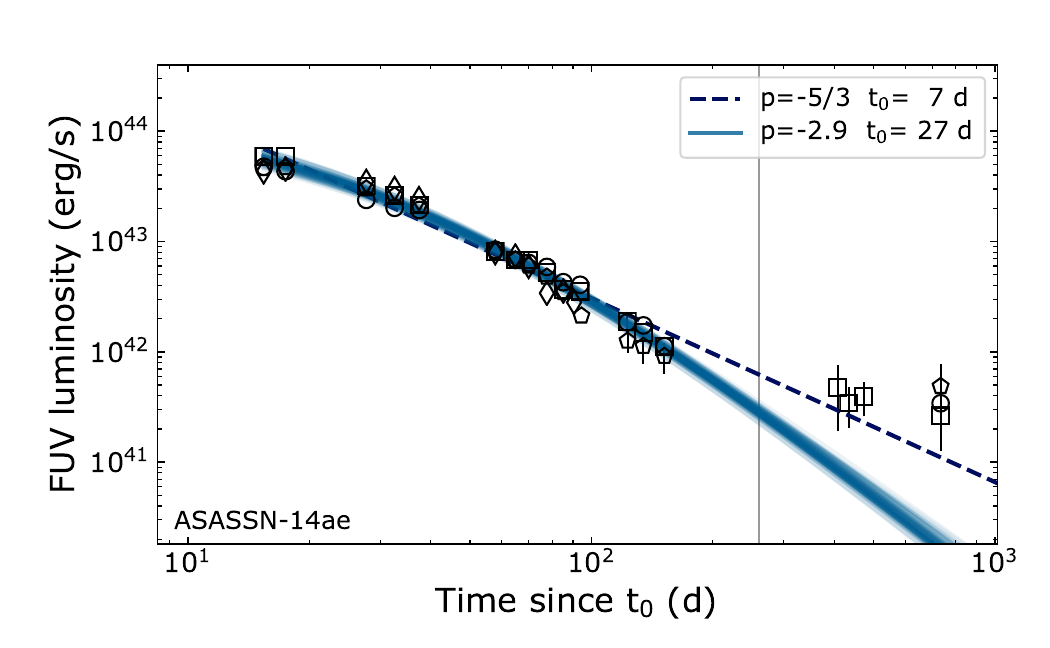}  \quad 
\includegraphics[width=240pt, trim=6mm 12mm 4mm 6mm, clip]{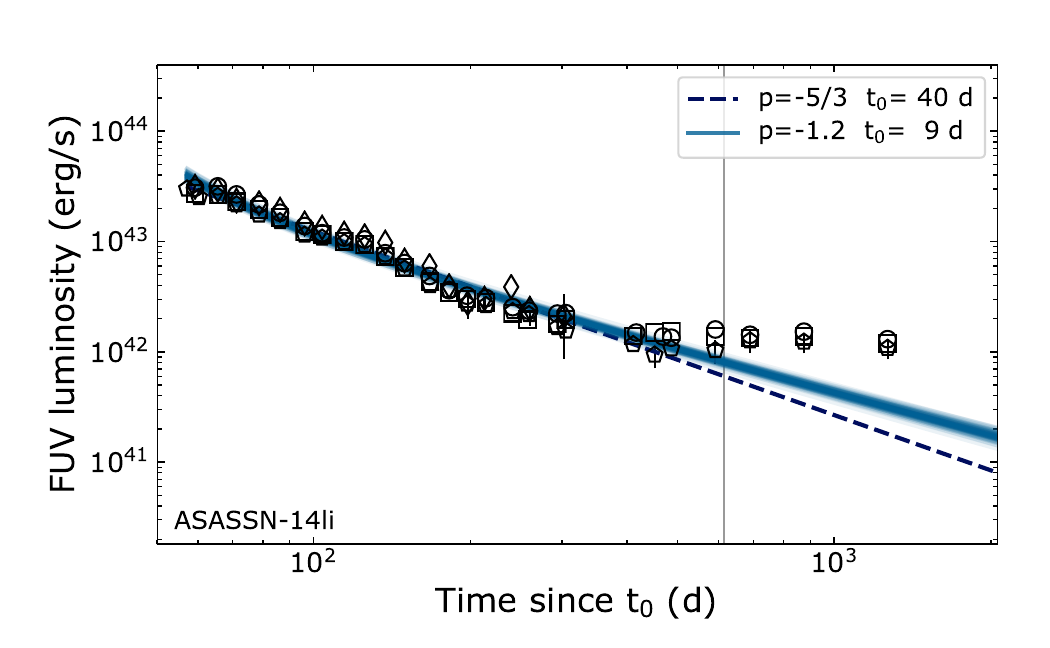} 

\includegraphics[width=240pt, trim=6mm 0mm 4mm 6mm, clip]{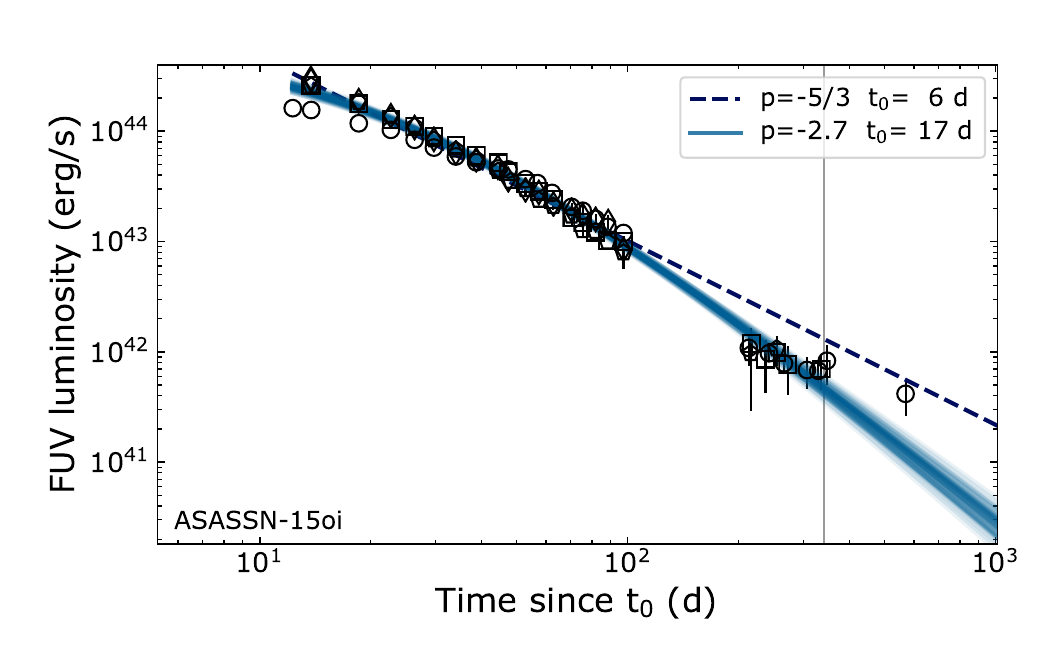}  \quad 
\includegraphics[width=240pt, trim=6mm 0mm 4mm 6mm, clip]{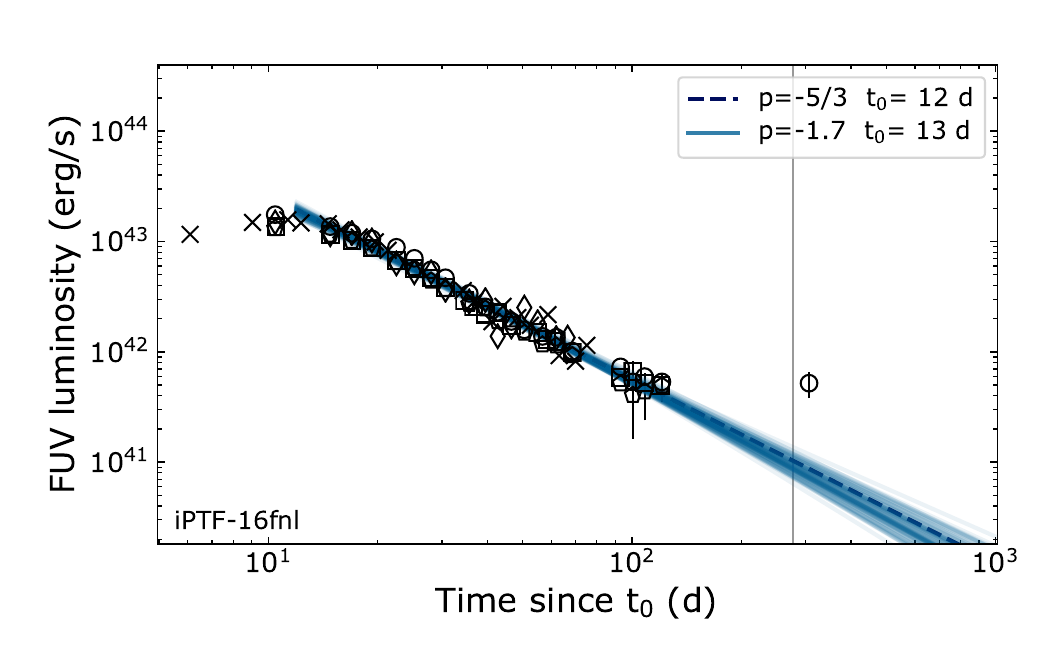}  

\caption{Light curves of more recent TDFs with {\it Swift} photometry. Identical to Fig.~\ref{fig:lcfit} except for the scale of the rest frame time axis. The symbols used for the {\it Swift} UVOT observations are: UVW2 $\ocircle$, UVW2 $\square$, UVW1  $\pentagon$, $U$ $\lozenge$. The overlapping light curves of these different bands is due to the limited temperature evolution. } 
\label{fig:lcfitswift}
\end{center}
\end{figure*}

We fit Eq. \ref{eq:lc} only to the early-time portion of the light curve. To differentiate between early-time and late-time emission, we set $t_{\rm late}=10t_{\rm fb}$, with $t_{\rm fb}$ the fallback time estimated for a star with a mass of 0.5~$M_\odot$. To compute the fallback time as a function of mass we adopt \citep[][]{Stone13}  
\begin{equation}
	t_{\rm fb} = 3.5\times 10^6~{\rm s} \left(\frac{M_\bullet} {10^6 M_\odot}\right)^{1/2}   
	\left(\frac{M_*} {M_\odot}\right)^{-1}
	\left(\frac{R_*} {M_\odot}\right)^{3/2}
	\quad.
\end{equation}
Here $M_*$ and $R_*$ are the stellar mass and radius, respectively.  
We use the black hole mass estimated from the $M_\bullet$-$\sigma$ relation using the \citet{Gultekin09} calibration and the velocity dispersion measurements reported in \citet{Wevers17,Wevers19}. For GALEX-D1-9 we adopt the measurement of the velocity dispersion of the entire galaxy (which is 24 km\,s$^{-1}$ larger than the dispersion obtained for the central extraction, indicating a large systematic uncertainty, likely due to the small angular size of this galaxy, see \citealt{Wevers19}). We indicate $t_{\rm late}$ with a vertical line in Figs.~\ref{fig:lcfit} \& \ref{fig:lcfitswift}. In most cases, the early-time data (i.e. $t<t_{\rm late}$) includes all points in the light curve prior to the {\it HST} observations. 
Besides our fits to the early-time data with the power-law index $p$ as a free parameter, we also estimate the parameters of our light curve model using the index fixed at $p=-5/3$. 

Since we use a single temperature to describe the TDF spectrum at any point in time, we can k-correct the multi-band light curves to a single (rest frame) wavelength, $\lambda_0$. In Figs.~\ref{fig:lcfit} and \ref{fig:lcfitswift} we show the results for $\lambda_0$=150~nm, using the mean temperature measured from the early-time data. This reference frequency is close to the effective wavelength of the{\it HST} FUV bands, thus minimizing the k-correction for the late-time observations. 
We also use our light curve model (Eq.~\ref{eq:lc}) to measure the temperature from the late-time data only. 

The very small dispersion between the multi-band light curves k-corrected to 150~nm (Figs.~\ref{fig:lcfit} and \ref{fig:lcfitswift}) testifies to the limited temperature evolution in TDF light curves. The flares  have a well-defined mean temperature that can be measured accurately by combining multi-epoch observations. The typical uncertainty on the mean temperature is only 0.05~dex or smaller (Table~\ref{tab:prop}). For a blackbody spectrum with $T=5\times 10^4$~K observed at $z=0.1$, the spectral index\footnote{Defined as $F_\nu \propto \nu^\alpha$.} is (1.5, 1.1, 0.6) for the ($u$, NUV, FUV) bands, respectively. We thus see that our observations are not on the Rayleigh-Jeans tail (i.e. $\alpha = 2$) of the SED, and therefore the near-constant temperature we observe is not due to limited spectral evolution on this tail.

\section{Discussion}\label{sec:discussion}

%

All HST-detected TDFs in our sample (Fig.~\ref{fig:lcfit}), as well as more recently-discovered TDFs that have late-time coverage (Fig.~\ref{fig:lcfitswift}), show, after a few years, a flattening of the UV light curve relative to a $t^{-5/3}$ fit to early-time data. For the high-mass black holes ($M_\bullet >10^{6.5}\, M_\odot$), the flattening is consistent with a single power-law decay (with power-law index $p> - 5/3$) beginning near the peak of the flare, while for all the lower-mass black holes, the late-time emission exceeds the expected value from the early-time extrapolation with best-fit $p$ (Figs.~\ref{fig:mass-PLpredict} \& \ref{fig:mass-PLearly}). In other words, the light curves of TDFs from low-mass black holes show two distinct phases: first, a steep decay, and second, a shallow late-time evolution.  In contrast, TDFs from high-mass black holes can be described with a single, relatively shallow power law.  For both mass ranges, fitted decay times $t_0$ are comparable to the theoretical $t_{\rm fb}$, and seem to follow the predicted $t_{\rm fb} \propto M_\bullet^{1/2}$ scaling (Fig. \ref{fig:mass-PLearly}). 

We stress that the observed flattening of the light curve cannot be explained merely by a change of the photospheric temperature with time. First of all, we corrected all observations to the luminosity at 150~nm in the rest frame. This matches the pivot frequency of the F150LP bandpass at $z\approx 0.1$, hence the k-corrected luminosity of the late-time {\it HST} FUV observations is nearly independent of temperature. The light curve model we use to predict the late-time 150~nm luminosity from the early-time observations includes the mean temperature as a free parameter; hence the uncertainty on the mean blackbody temperature has been propagated into this prediction. An increase of the temperature is sometimes observed in TDFs \citep{Holoien16b, Mockler18,vanVelzen18_NedStark}. Indeed, for some sources in our sample (e.g. ASASSN-14ae) we see some frequency dependence in the residuals to our single-temperature light curves (also the small increase of the 150~nm luminosity based on the late-time NUV and FUV observations of \mbox{PS1-10jh} is likely due our single-temperature k-correction). However, the residuals are very small, a few tens of percent at most. 
To conclude, the existing multi-band observations exclude a temperature change that would lead to a dramatic flatting of the FUV light curve.

\subsection{A stellar origin for the late-time UV detections is very unlikely}\label{sec:fuv_compare}
For five sources in our sample (PS1-10jh, ASASSN-14ae, ASASSN-14li, ASASSN-15oi, iPTF-16fnl), a stellar origin for late-time UV detections is conclusively ruled out because the late-time emission exceeds the upper limits (or detections) obtained before the flare. In the case of the remaining sources, the FUV emission is far more compact than the optical emission, and therefore both cannot come from a single stellar distribution (see section \ref{sec:HSTphot} for a source-by-source discussion). Compact FUV emission could be explained by a two-component stellar population, but only if {\it all} of the younger, UV-bright stars are concentrated within our HST aperture of 0.1" or $\sim 0.1$~kpc. We can use {\it HST} imaging at optical/IR wavelengths to rule out such a compact stellar population for two host galaxies in our sample (PTF-09ge, and GALEX-D3-13, see section \ref{sec:HSTphot}). 

We are thus left with three TDF host galaxies (SDSS-TDE2, GALEX-D1-9, and PTF-09djl) for which we cannot directly rule out a central concentration of young stars as an explanation for the late-time FUV emission. For these three sources, the late-time FUV luminosity is $L_{\rm FUV} \ge 10^{42.2}\rm \,erg\,s^{-1}$. In this subsection, we show that this combination of high UV luminosity and compact emitting region is unprecedented for the typical host galaxies of TDFs.


First of all, nuclear star clusters (NSCs) are not consistent with the observed late-time FUV emission because the implied mass of the cluster is too high. For the typical FUV-$r$ colors of blue galaxies \citep{Wyder07}, the optical luminosity implied would be $\sim 10^{43}~{\rm erg~s}^{-1}$, which is an order of magnitude higher than the most luminous NCS in the samples of \citet{Georgiev14,Carson15}. Likewise, even if we generously assume a large NSC radius of 10~pc, the projected mass density of the cluster would be $\sim 10^7 M_\odot\, \rm pc^{-2}$, which is an order of magnitude larger than the densest known NSCs \citep{Walcher05}. 

Our FUV observations could, however, potentially be explained by a young population of stars that extends beyond the typical sizes of NSCs. In this scenario, the surface density within our aperture of $\approx 0.3$~kpc would be $\sim 10^{4}~M_\odot \rm pc^{-2}$. This would also be an extreme conclusion, as none of the galaxy bulges studied by \citet{Walcher05} reach this density. 

While the observed nuclear UV emission seems to be very atypical, we note that a large fraction of TDF hosts are classified as rare post-starburst galaxies \citep{Arcavi14,French16}, which may often have ``blue cores". These core colors could be due to a mix of young stellar populations, or AGN \citep{Yang06}. Quite often, however, strong color gradients are seen on $\sim 1$~kpc scales \citep{Pracy12}. This an order of magnitude larger than the aperture of our late-time {\it HST} observation, implying that for most post-starburst galaxies, our observations should resolve the population of young stars that could explain the FUV emission. 

A direct measurement of the FUV flux on sub-kpc scales is possible for the very low redshift ($z=0.004$) post-starburst galaxy NGC~3156 \citep{StonevanVelzen16}. For this source, GALEX FUV images can probe emission on 0.3~kpc scales. While NGC~3156 is known for its extreme concentration of stars near the center, we use the GALEX data to measure a nuclear FUV luminosity of only $L_{\rm FUV} = 5\times 10^{40}\rm erg\,s^{-1}$, an order of magnitude lower than the post-peak FUV emission in TDF host galaxies. The observed mass density of NGC~3156 at 0.1~kpc is $\sim 10^3 M_\odot~\rm pc^{-2}$, which is also an order of magnitude lower than the mass density required if we wish to invoke a stellar origin of the late-time FUV detections in these TDF host galaxies.

However, NGC 3156 is only one post-starburst galaxy.  To search for UV-bright cores in a larger population, we used the catalog of quiescent Balmer-strong galaxies that \cite{French18} selected using SDSS spectroscopy. By construction, the lowest redshift in the \citet{French18} sample is $z=0.01$. We selected a comparison sample for TDF host galaxies by requiring a total galaxy mass in the range $10^{9.0}<M_*/M_\odot <10^{10.5}$ and maximum redshift $z<0.03$, leaving 759 galaxies. At the median redshift of these comparison galaxies, the GALEX PSF \citep{Morrissey07} corresponds to $\approx 2$~kpc. To estimate the PSF flux we use an aperture with a radius 3.75" (\verb flux_aper_3  listed as in the GALEX GR6plus7 catalog) and apply an aperture correction of 0.2~mag. Of the 759 quiescent Balmer-strong galaxies in the comparison sample, we find GALEX FUV detections for 373 sources. For the detected galaxies, the median FUV luminosity is $2\times 10^{41}\rm\,{erg}~{s}^{-1}$ with a dispersion of 0.5~dex. Of all the galaxies in our parent sample of 759 low-redshift quiescent Balmer-strong galaxies, less than 3\% have a FUV luminosity greater than $10^{42}~ {\rm erg}~ s^{-1}$. If we repeat this exercise using post-starburst galaxies alone (which are a smaller subset of the quiescent Balmer-strong population), this fraction increases to $\approx$5\%. 

Under the conservative assumption that the TDF host galaxies are similar to quiescent Balmer-strong galaxies, the probability to detect 3 sources with $L_{\rm FUV}>10^{42}~\rm~erg~s^{-1}$ is negligible ($\sim 0.03^3)$. This estimate is conservative because the GALEX measurements probe the flux on $\sim$~kpc scales, while our HST observations probe regions an order of magnitude smaller. To estimate how this difference in physical scale affects our estimate of $L_{\rm FUV}$ in the comparison sample, we use {\it HST} NUV (F275W) observations that have been obtained for four post-starburst galaxies with $z\approx 0.025$ (GO program 14785: PI van~Velzen). We find that for these galaxies, the NUV flux on 0.1~kpc scales is between 3 and 30 times lower than the NUV flux seen through the GALEX aperture, implying that our measurements of $L_{\rm FUV}$ in the comparison sample of quiescent Balmer-strong galaxies overestimates the true nuclear FUV flux by at least this amount. 

We thus conclude that unresolved FUV emission at the level observed in the sample of TDF host galaxies is very rare, even in post-starburst galaxies. Typical levels of FUV emission are at least an order of magnitude lower than those we observe in TDF host galaxies.


\subsection{Late-time UV detections provide conclusive evidence against a SNe scenario}
Before interpreting the nature of the late-time FUV luminosity, it is important to stress that these detections further substantiate that these sources are not due to stellar explosions. Some supernovae can be UV-bright near peak \citep{Gezari09b}, but observations across all SNe subtypes show the post-peak NUV luminosity decreasing on timescales of months \citep{Brown09,Pritchard14,Lunnan18}.  
Given that our sample contains a mix of TDFs, selected either based on spectroscopic properties or optical/UV properties, it is encouraging that all show late-time UV detections (while SDSS-TDE1 was not detected in our {\it HST} observations, it was detected one year post-peak by GALEX).

\begin{figure*}[t!]
\begin{center}
\includegraphics[width=400pt, trim=5mm 0mm 4mm 2mm, clip]{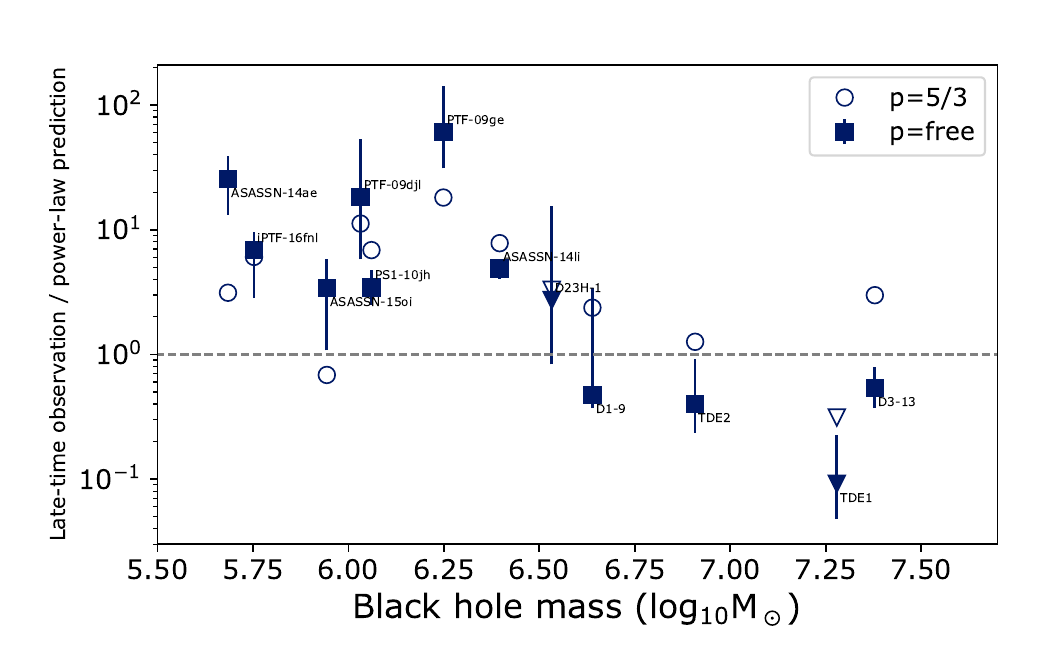} 
\end{center}
\caption{The excess of late-time emission compared to the expected value obtained from extrapolating the early-time light curve. For this extrapolation we used a single power law (Eq.~\ref{eq:lc}), either with a fixed index $p=-5/3$ (open circles) or with $p$ as a free parameter (filled squares). The uncertainty due to the extrapolation has been propagated into the 90\% credible intervals shown here. TDFs from the lower end of the mass distribution ($M_\bullet<10^{6.5}\,M_\odot$) show a significant excess compared to the expected late-time emission for a single power law. In other words, their light curves show a significant flattening. The TDFs from higher-mass black holes, on the other hand, are broadly consistent with a single power law that decays more slowly (Fig.~\ref{fig:mass-PLearly}). 
}\label{fig:mass-PLpredict}
\end{figure*}

\begin{figure*}
\begin{center}
\includegraphics[width=250pt, trim=5mm 3mm 4mm 2mm, clip]{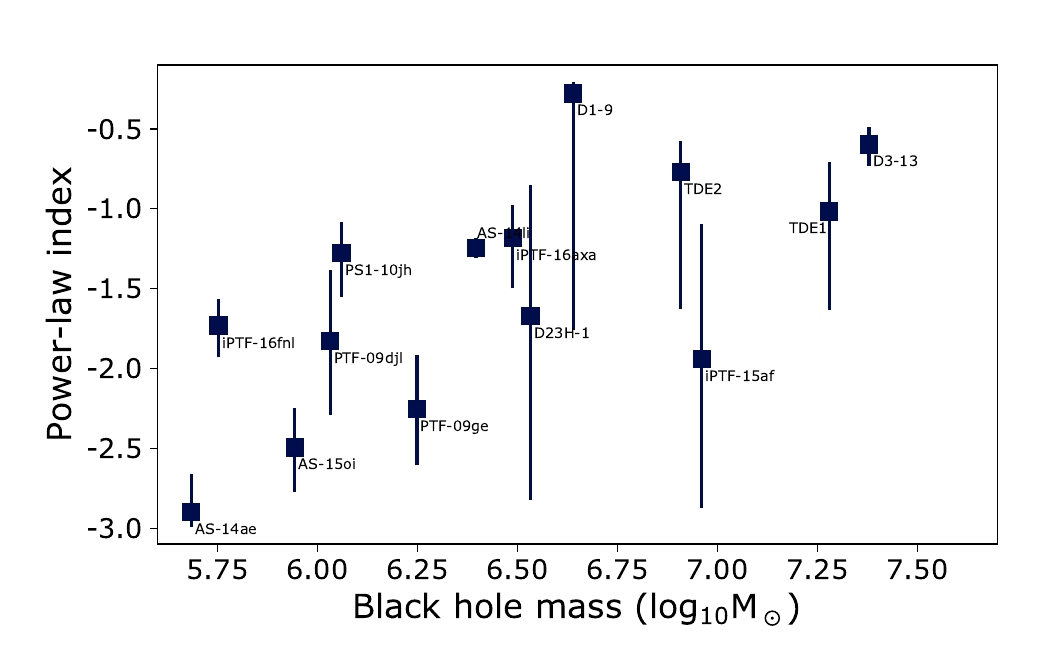} \quad
\includegraphics[width=250pt, trim=5mm 3mm 4mm 2mm, clip]{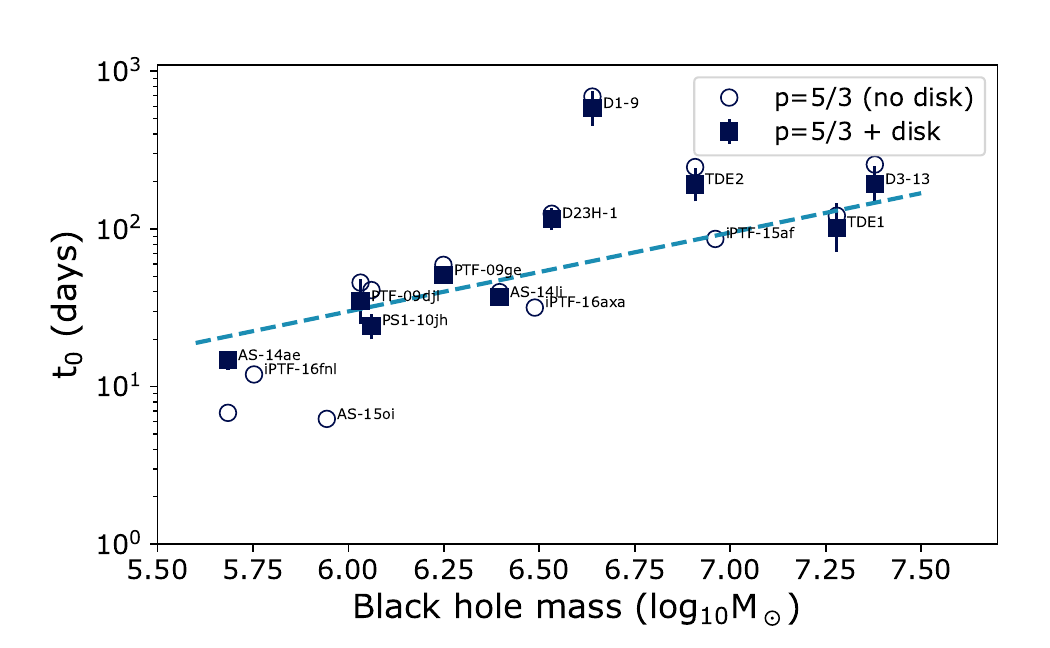} 
\end{center}
\caption{Best-fit parameters for a single power law, $L\propto (t/t_0)^p$, applied to the early-time observations of TDFs. We show the result as a function of the black hole mass obtained from the $M$--$\sigma$ relation. We define ``early-time'' as $t<10t_{\rm fb}$. {\it Left:} the power-law index is observed to increase with black hole mass. {\it Right:} the characteristic decay time $t_0$ compared to the expected scaling of the fallback time with mass, $t_{\rm fb}\propto M_\bullet^{1/2}$ (Eq. \ref{eq:fb}, shown as a blue dashed line). We show the result obtained by fitting a fixed power law to the early-time data, with (filled squares) and without (open circles) the addition of nearly constant flux from the disk model that fits to the late-time observations. In both panels, the errorbars span the 90\% credible interval. }\label{fig:mass-PLearly}
\end{figure*}

\subsection{The case for late-time disk dominance}
The flattening seen in the light curves of TDFs from low-mass SMBHs suggests that a different emission mechanism may be operating at late times.  We now discuss the challenges that previously proposed early-time emission mechanisms (reprocessing and stream self-intersection shocks) would face in explaining the late-time FUV observations. 

To explain---within the reprocessing paradigm---two phases in the light curves of TDFs with low-mass hosts, we would have to conclude either that accretion power becomes roughly constant at late times (i.e. $t>10t_{\rm fb}$), or that an initially low reprocessing efficiency steadily increases.  The former possibility is at odds with what we know of TDF theory: early in the flare, if the disk viscous time $t_{\rm visc} \ll t_{\rm fb}$, the bolometric luminosity of the disk would decline roughly as $L \propto t^{-5/3}$.  At late times, $t_{\rm visc} \gg t_{\rm fb}$ and the accretion rate will be controlled by internal processes in the disk, with $L \propto t^{-1.2}$ \citep{Cannizzo90}.

The second possibility is challenged by the decreasing opacity of most plausible reprocessing layers, which should lead to a {\it decreasing} reprocessing efficiency.  This is clearly the case for wind reprocessing, where the monotonically declining column density in the reprocessing outflow eventually becomes too low to absorb a significant fraction of the X-ray/EUV photons from the disk. The resulting X-ray ionization breakout will rapidly decrease the optical/NUV/FUV emission, while the observed X-ray flux should increase. \citet{Metzger16} compute that for a disruption with $M_\bullet = 10^6\, M_\odot$, a reprocessing shell produced in a wind or outflow can remain optically thick to X-rays for at most a few fallback times. This requirement is clearly at odds with our detections of FUV emission at $t>10t_{\rm fb}$.  Since the maximum time for significant reprocessing scales with $M_\bullet^{-5/6}$ \citep{Metzger16}, we would expect the highest-mass black holes to exhibit almost no late-time UV emission from a reprocessing layer. Instead we observed the opposite: TDFs from higher-mass black holes have more shallow light curves (Fig.~\ref{fig:mass-PLearly}), with detections up to 10 years post peak.  Gravitationally bound reprocessing layers are, in this respect, more plausible explanations for late-time FUV emission.  Such layers could consist of poorly circularized debris from the disrupted star \citep[e.g.][]{Guillochon14,Coughlin14}.  Mass would steadily drain from such a layer due to continued dissipation of orbital energy in shocks (at stream-stream and stream-disk intersections), but even if enough mass remains at large scales to be optically thick, it is unclear why the reprocessing efficiency should increase.  

A second, and more decisive, challenge for the reprocessing picture is the size of the observed photosphere radii. We find that the late-time photosphere radius, $R_{\rm late}$, is typically of the same order as the tidal radius (Table~\ref{tab:prop}). If the late-time accretion disk is fully circularized, it must extend to at least $R_{\rm c} = 2R_{\rm t}/\beta$, where the tidal radius $R_{\rm t} = R_*(M_\odot / M_*)^{1/3}$, and the penetration parameter $\beta$ is a dimensionless inverse of the stellar pericenter $R_{\rm p}$ (defined as $\beta = R_{\rm t}/R_{\rm p}$).  Viscous spreading can increase the disk size by a factor of a few (Section \ref{sec:disk}).  It seems implausible that the reprocessing layer has the same physical size as the power source responsible for heating the layer (i.e. the accretion disk).  A hypothetical hydrodynamic structure subtending a large solid angle on scales $\sim R_{\rm late}$ would dissipatively interact with the late-time accretion disk, and likely merge into it.

Further evidence against late-time reprocessing follows from the observations of ASASSN-14li. For this source, the X-ray luminosity decreases \citep{Bright18} while the UV luminosity flattens off to a plateau (Fig.~\ref{fig:lcfitswift}). The exponential tail of the X-ray light curve of ASASSN-14li \citep{Bright18} can be explained by a decrease in the disk temperature \citep[see][Table 1]{PashamvanVelzen17} and hence the accretion rate for this source may still follow a power-law decay. Even if the accretion rate is tracking the fallback rate, the observed decrease of the X-ray luminosity translates to a decrease of the power available for reprocessing. The observed disconnect between the UV and X-ray light curves thus provides evidence against a reprocessing layer as the origin of late-time UV emission in ASASSN-14li. 

Circularization-powered emission is likewise an unsatisfactory explanation for the late-time UV detections.  In the circularization paradigm, the optical/UV luminosity is due to stream self-intersection shocks at large distances from the SMBH.  While the stream self-intersection point $R_{\rm SI}$ may be close to the tidal radius for highly relativistic pericenters ($R_{\rm p} \lesssim 20R_{\rm g}$, where the gravitational radius $R_{\rm g}=GM_\bullet / c^2$; \citealt{Hayasaki13, Bonnerot16}), it is located at least an order of magnitude farther away for less relativistic pericenters, which should be more common, particularly around low-mass SMBHs \citep{Dai15}.  A large self-intersection radius is a key feature of the circularization paradigm's appeal, as it naturally explains the large blackbody radii (typically $\sim {\rm few}~\times 10^{14}~{\rm cm}$) fitted to early-time TDF emission \citep{Piran15}.  Shock dissipation on these scales produces a low radiative efficiency, which we can bound from above assuming (i) complete thermalization of stream kinetic energy and (ii) no adiabatic losses of thermalized energy: thus, $\eta_{\rm c} \le GM_\bullet / (R_{\rm SI} c^2) $.  The late-time mass fallback rate is \citep[e.g.][]{Lodato09}
\begin{equation}
    \dot{M}_{\rm fb} = \frac{M_\star}{3t_{\rm fb}}\left( \frac{t}{t_{\rm fb}} \right)^{-5/3}, \label{eq:fb}
\end{equation}
producing a late-time circularization luminosity $L_{\rm c}=\eta_{\rm c} \dot{M}_{\rm fb} c^2$.  If we use early-time blackbody radii to estimate $R_{\rm SI}$ and $\eta_{\rm c}$, we would be unable to reproduce the observed late-time FUV emission of PS1-10jh, PTF-09ge, PTF-09djl, TDE2, and D1-9.  If we instead set the smaller late-time blackbody radii $R_{\rm late}=R_{\rm SI}$, and use very generous assumptions ($\eta_{\rm c}$ equals its upper limit; $100\%$ of radiated energy is in the FUV bands; the disrupted star has an unusually large $M_\star = M_\odot$), we still have observed late-time FUV luminosities $\nu L_{\nu} > L_{\rm c}$ for PS1-10jh, PTF-09ge, PTF-09djl, and TDE2.  While D3-13 and perhaps D1-9 may evade the energetic constraints above, both exhibit shallow intermediate- and late-time declines that are inconsistent with the $\propto t^{-5/3}$ predictions of the circularization paradigm.  Circularization shocks alone are thus an unlikely power source for most of our observed late-time FUV detections.

These challenges for the reprocessing and circularization paradigms lead us to consider a different power source for late-time FUV emission. An unobscured accretion disk, the scenario first envisaged for TDFs \citep{Rees88, Cannizzo90}, is the most natural candidate.  As we explain in the next section, simple accretion disk models (that generally fail to reproduce early-time optical/UV TDF emission) are easily able to accommodate our late-time FUV detections\footnote{A late-time transition to disk-dominated emission is analogous to the \citet{Leloudas16} interpretation of the TDF candidate ASASSN-15lh, where circularization shocks were invoked as the power source for the early-time ($\lesssim 75~{\rm d}$) optical/UV light curve, and accretion power was invoked to explain later ($\gtrsim 75~{\rm d}$) observations.  The analogy is inexact, however, as the fitted blackbody radius for ASASSN-15lh remained much larger than the plausible size of a quasi-circular accretion disk, requiring a reprocessing layer.}.

\subsection{Disk models}
\label{sec:disk}
To better quantify the arguments of the previous subsection, we have constructed a library of simple, time-dependent disk models that can be compared to our FUV observations of each TDF.  The quantitative details of these models are given in Appendix \ref{app:disk}, but their key features and assumptions are:
\begin{enumerate}
    \item TDF accretion disks are well-described as planar and azimuthally symmetric; furthermore, we vertically average their fluid properties so as to model them in 1D.  These assumptions are probably quite incorrect at early times, when the process of circularization will imprint strong non-axisymmetries on the inner accretion flow \citep{Hayasaki13, Guillochon14, Shiokawa15}.  Furthermore, TDF disks form with large generic tilts, and differential nodal precession (due to Lense-Thirring frame-dragging) can create warps within the disk or lead to global precession \citep{Fragile+07, Stone12}.  However idealized these assumptions may be at early times, they are likely robust at the late times we are concerned with in this paper.  All of our TDFs have been observed after $\gtrsim 10$ fallback times; at which point the returning $\dot{M}_{\rm fb}$ is low enough to have little impact on the circularized inner disk.  Likewise, internal torques within an initially tilted disk are likely to align it with the SMBH equatorial plane on timescales of $\sim 10^{1-4}~{\rm d}$ \citep{Franchini+16}.  Even for PS1-10jh, the ``youngest'' TDF with {\it HST} observations ($\Delta t=1877~{\rm d}$), alignment will only be prevented if both the SMBH spin $a_\bullet$ and the dimensionless viscosity parameter $\alpha$ are extremely low (e.g. $a_\bullet \lesssim 0.2$ and $\alpha \lesssim 0.03$; \citealt{Franchini+16}).
   
    \item We evolve TDF accretion disks in time in a quasi-viscous way, i.e. according to the standard viscous diffusion equation of 1D accretion theory \citep[e.g.][]{Abramowicz13}.  In real accretion disks, angular momentum transport is generally due to magnetized turbulence, but we employ an effective viscosity similar to the Shakura-Sunyaev prescription \citep{Shakura73}.  However, rather than implementing the usual $\alpha$-viscosity, we employ a ``$\beta$-viscosity'' where kinematic viscosity $\nu \propto P_{\rm gas}$, the gas pressure alone \citep{Sakimoto&Coroniti81}.  This approximation is chosen for simplicity, as it prevents the onset of thermal and viscous instabilities that would occur at late times in $\alpha$-disks. We discuss the implications of these instabilities (and their absence) in detail below (Sec.~\ref{sec:alpha}).  We also neglect the time-dependent addition of mass to the disk from returning debris streams; our initial conditions place the bound half of the star into a Gaussian surface density profile centered on the circularization radius $R_{\rm c}=2R_{\rm t}/\beta$.  This assumption is crude at early times, but since most of the stellar mass returns after $t_{\rm fb}$, it is reasonably accurate at the late times we are concerned with, as the disks enter a self-similar spreading phase \citep{Cannizzo90}.
    
    \item We assume that TDF disks emit as multicolor blackbodies and neglect coronal contributions or any type of reprocessing.  The midplane temperature in the disk is set by a local energetic balance between viscous heating, radiative cooling, and advective cooling.  The local radiative flux is determined by combining this energy equation with an opacity model; we consider both electron scattering opacity and a Kramer's law for bound-free opacity.  We neglect the effects of gravitational lensing and redshift (this is a reasonable approximation for the FUV, which is sourced primarily from the outermost disk annuli at $\sim 50 R_{\rm g}$, but is less justifiable for soft X-ray or EUV predictions).
\end{enumerate}

Our disk models have the following free parameters: SMBH mass $M_\bullet$; dimensionless SMBH spin $a_\bullet<1$, which determines the disk inner edge; stellar mass $M_\star$; stellar radius $R_\star$; stellar penetration parameter $\beta \equiv R_{\rm t}/R_{\rm p}$; effective viscosity parameter $\alpha$, and time since disruption $\Delta t$.  Our fiducial models assume $M_\bullet$ values determined for each TDF (Table~\ref{tab:prop}); SMBH spins $a_\bullet = 0.9$; a ZAMS stellar radius appropriate for stars of slightly super-solar metallicity ($Z=1.4Z_\odot$; \citealt{Girardi+00}); grazing, $\beta=1$ disruptions; a time since disruption $\Delta t+t_0$ as obtained from the observed light curve (Table \ref{tab:prop}). We then construct, for each TDF under study, a grid of models that samples $0.01 \le \alpha \le 1$ and $0.1 \le M_\star \le 3$.  The model grid samples both variables in a log-flat way; we consider 9 different values of $\alpha$, and 10 of $M_\star$.

Our assumed value of $a_\bullet$ sets the inner disk boundary by determining the innermost stable circular orbit (ISCO)\footnote{Our model grid uses $a_\bullet=0.9$ to avoid complications from the Hills mass \citep{Hills75}, which would affect high-mass TDFs like TDE1, TDE2, and D3-13.}.  This choice significantly affects the bolometric emission of the disk, which is dominated by unobservable EUV bands.  The choice of $a_\bullet$ has an even larger effect on disk predictions for thermal soft X-ray flux, which is always on the Wien tail of even the inner disk annuli.  However, the value of $a_\bullet$ has little effect on the observed FUV flux, which is (at the time of observation) dominated by the outermost annuli of the spreading disk.  Our assumed value of $\beta$ determines the initial conditions of the disk (the initial surface density profile $\Sigma(R)$ is peaked at a radius $R=2R_{\rm p}$), but a viscously spreading disk asymptotes to a self-similar expansion that loses most memory of its initial mass distribution (though not its total initial mass and angular momentum), so this assumption is also reasonable.

\begin{deluxetable}{l c c}
\tablewidth{200pt}
\tablecaption{Disk parameter inference}
\tablehead{name & stellar mass~& $\alpha$\\
                & ($\log M_\odot$) & (log)}
\startdata
PS1-10jh        & $  \phantom{-}0.0_{0.1}^{0.1}$ &$ -0.3_{0.2}^{0.3}$ \\
PTF-09ge        & $ -0.9_{0.2}^{0.1}$ &$ -1.1_{0.2}^{0.3}$ \\
PTF-09djl       & $  \phantom{-}0.1_{0.1}^{0.8}$ &$ -0.4_{0.3}^{0.4}$ \\
TDE2            & $ -0.6_{0.2}^{0.3}$ &$ -0.4_{0.3}^{0.4}$ \\
D1-9            & $ -0.2_{0.1}^{0.7}$ &$ -0.5_{0.4}^{0.5}$ \\
D23H-1          & $ -0.9_{0.1}^{0.1}$ &$ -0.7_{0.4}^{0.4}$ \\
D3-13           & $ -0.8_{0.2}^{0.1}$ &$ -0.4_{0.3}^{0.5}$ \\
TDE1            &   --   & $<-1.7$ \\
ASASSN-14ae     & $ -0.6_{0.2}^{0.2}$ &$ -0.3_{0.2}^{0.4}$ \\
ASASSN-14li     & $ -0.4_{0.2}^{0.1}$ &$ -0.2_{0.2}^{0.3}$ \\
iPTF-16fnl      & $ -0.7_{0.2}^{0.2}$ &$ -0.3_{0.2}^{0.4}$ \\
\label{tab:params}
\enddata
\tablecomments{Uncertainties listed correspond to the 90\% credible interval. Contour plots with samples from the posterior distribution of these parameters are shown in Fig.~\ref{fig:diskchi2}.}
\end{deluxetable}

\begin{figure*}
\includegraphics[width=145pt, trim=0mm 5mm 5mm 5mm, clip]{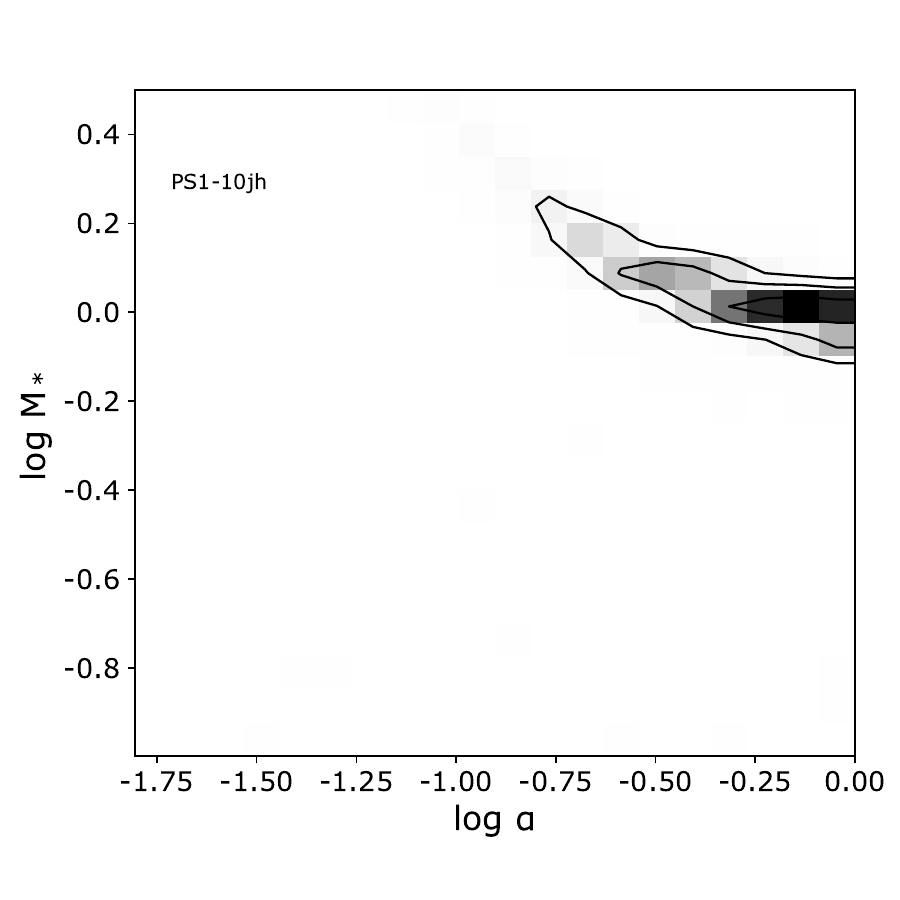} 
\includegraphics[width=145pt, trim=0mm 5mm 5mm 5mm, clip]{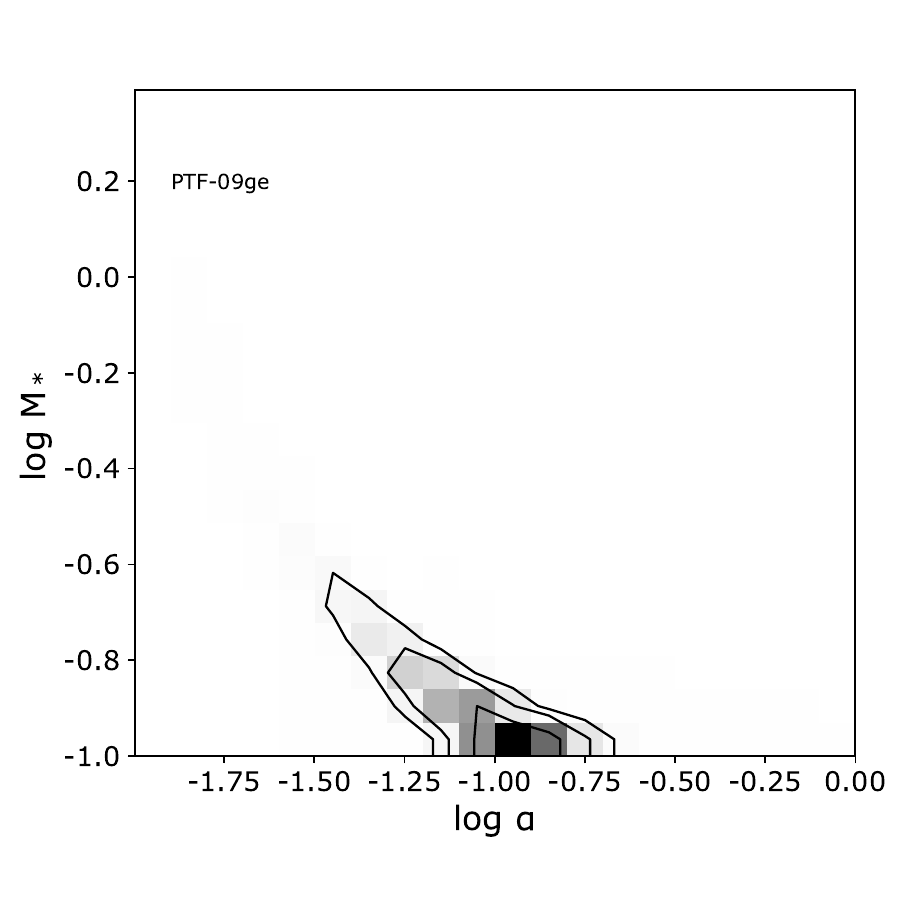} 
\includegraphics[width=145pt, trim=0mm 5mm 5mm 5mm, clip]{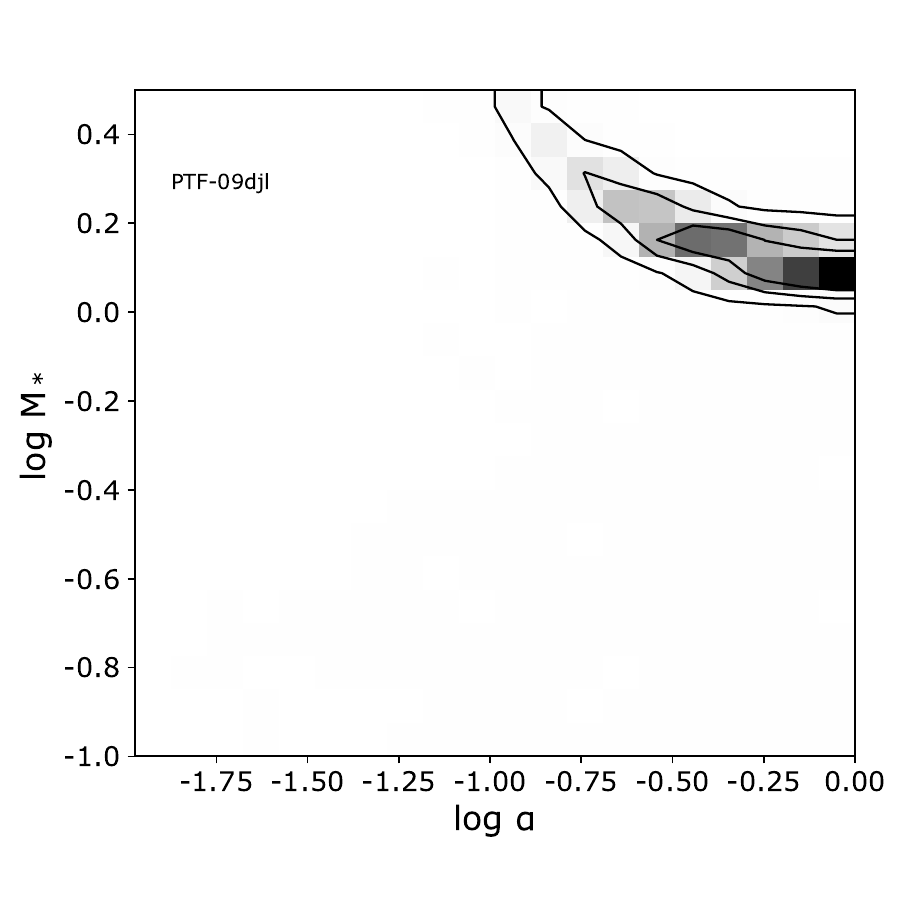}

\includegraphics[width=145pt, trim=0mm 5mm 5mm 5mm, clip]{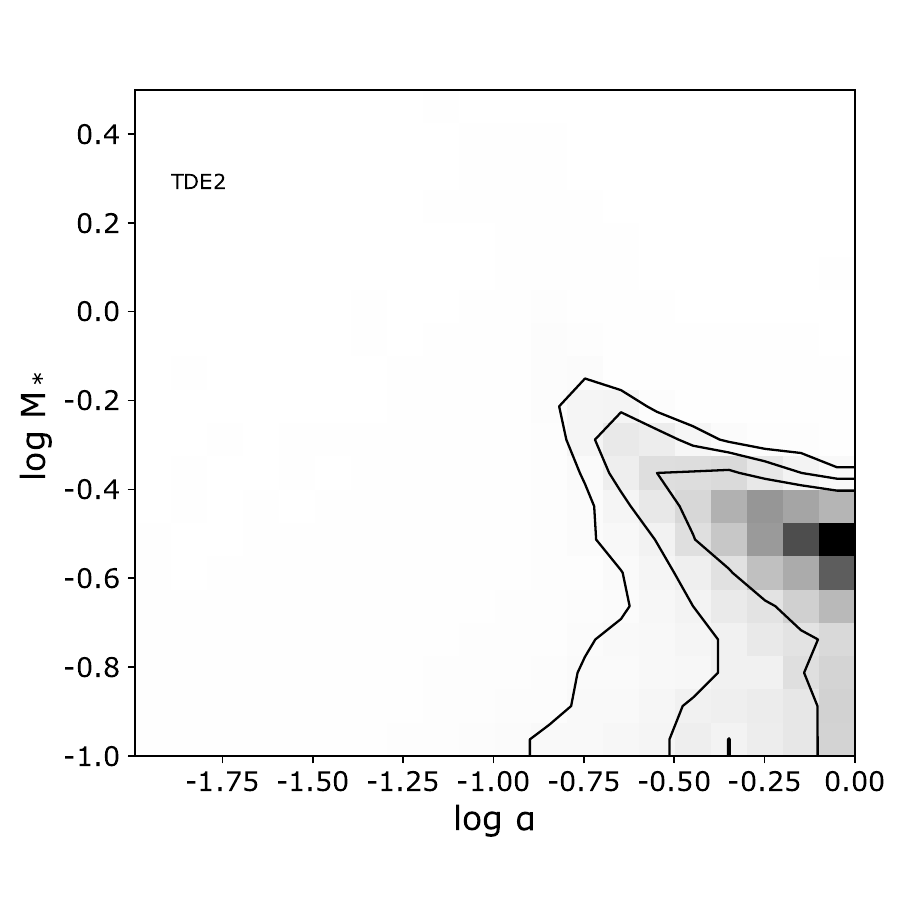}
\includegraphics[width=145pt, trim=0mm 5mm 5mm 5mm, clip]{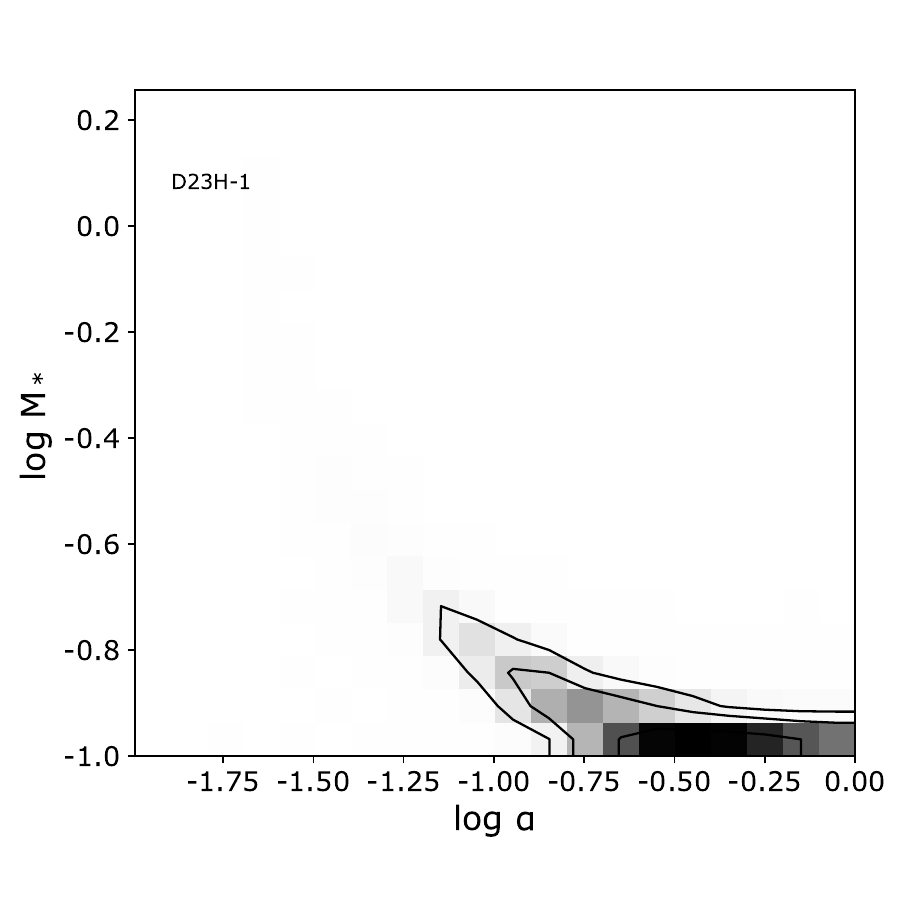} 
\includegraphics[width=145pt, trim=0mm 5mm 5mm 5mm, clip]{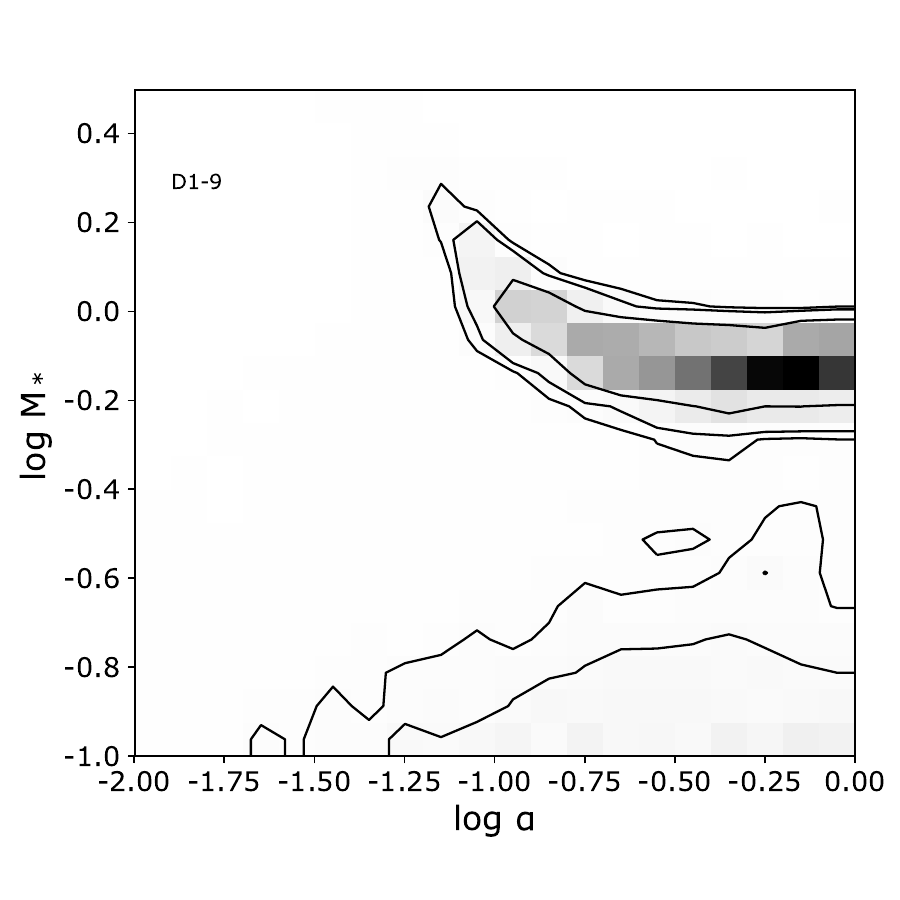} 

\includegraphics[width=145pt, trim=0mm 5mm 5mm 5mm, clip]{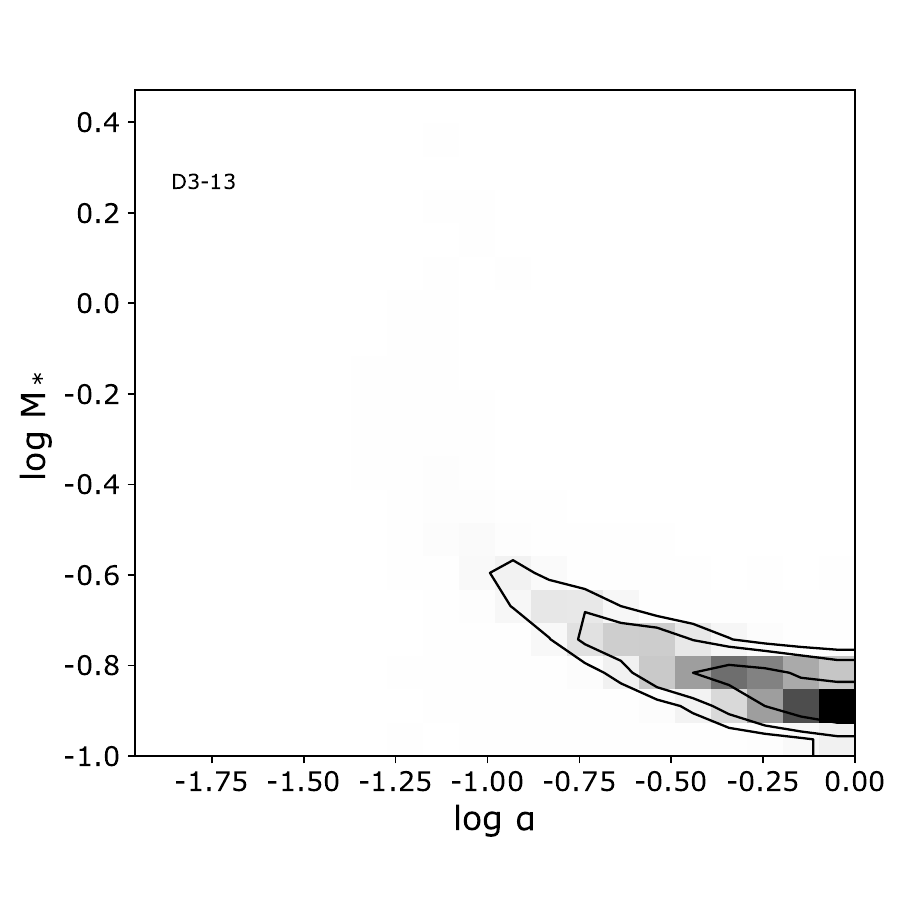} 
\includegraphics[width=145pt, trim=0mm 5mm 5mm 5mm, clip]{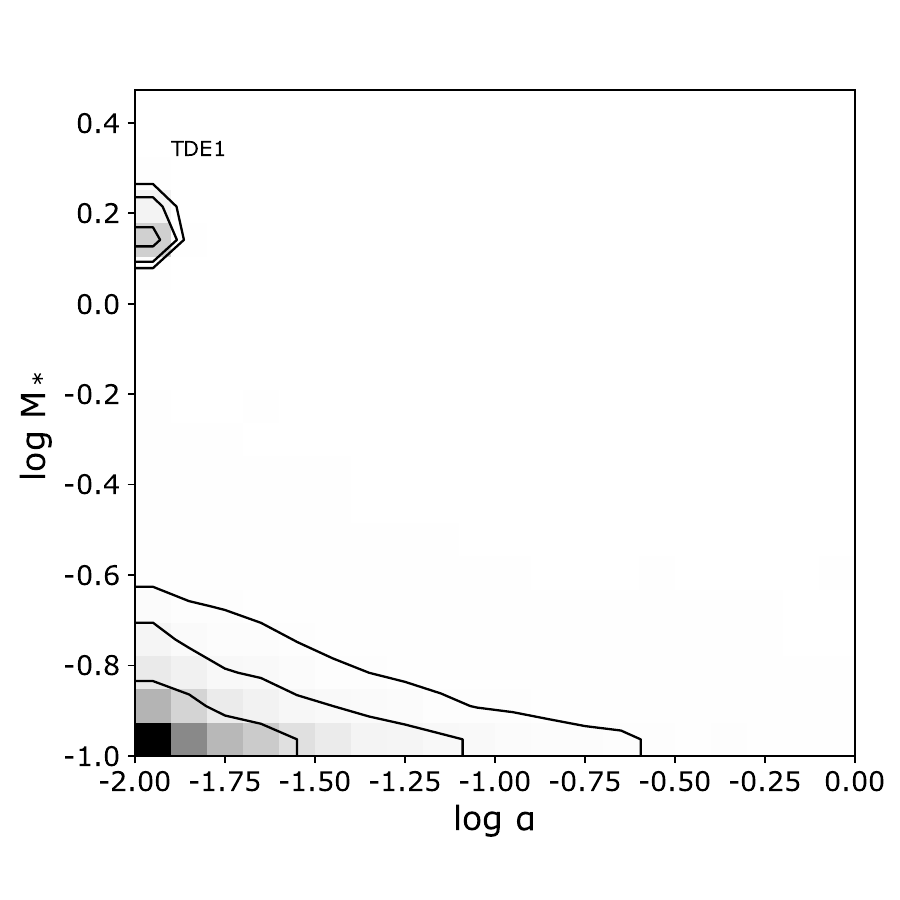}
\includegraphics[width=145pt, trim=0mm 5mm 5mm 5mm, clip]{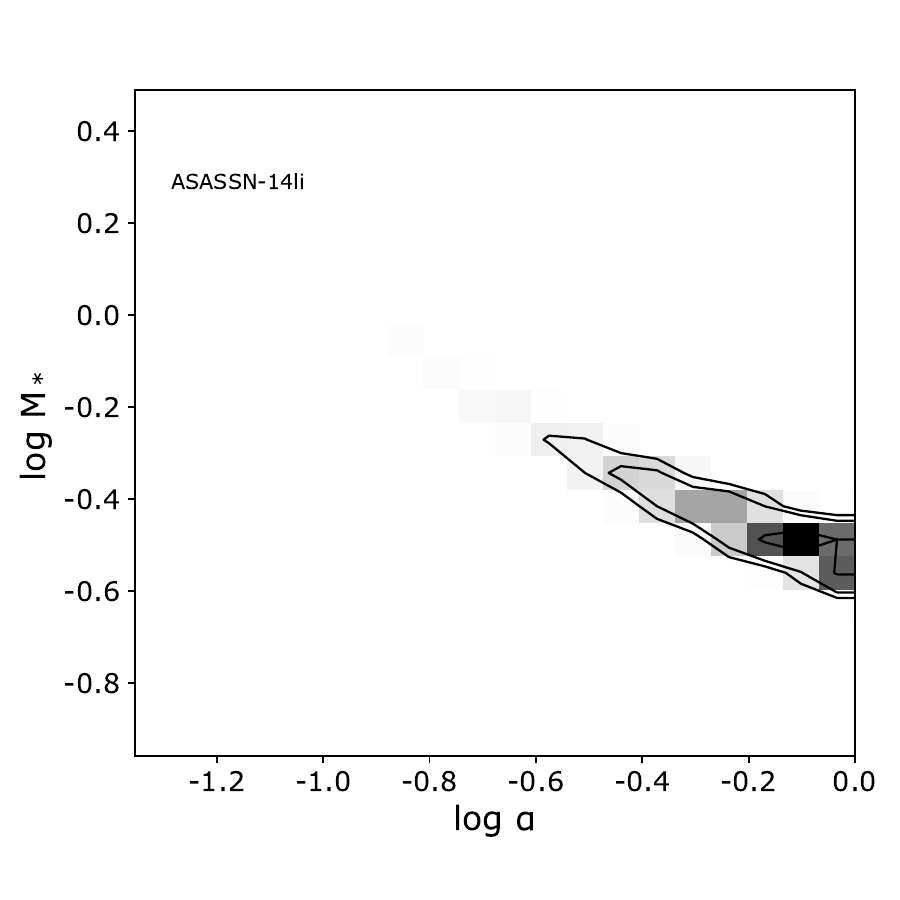} 

\includegraphics[width=145pt, trim=0mm 5mm 5mm 5mm, clip]{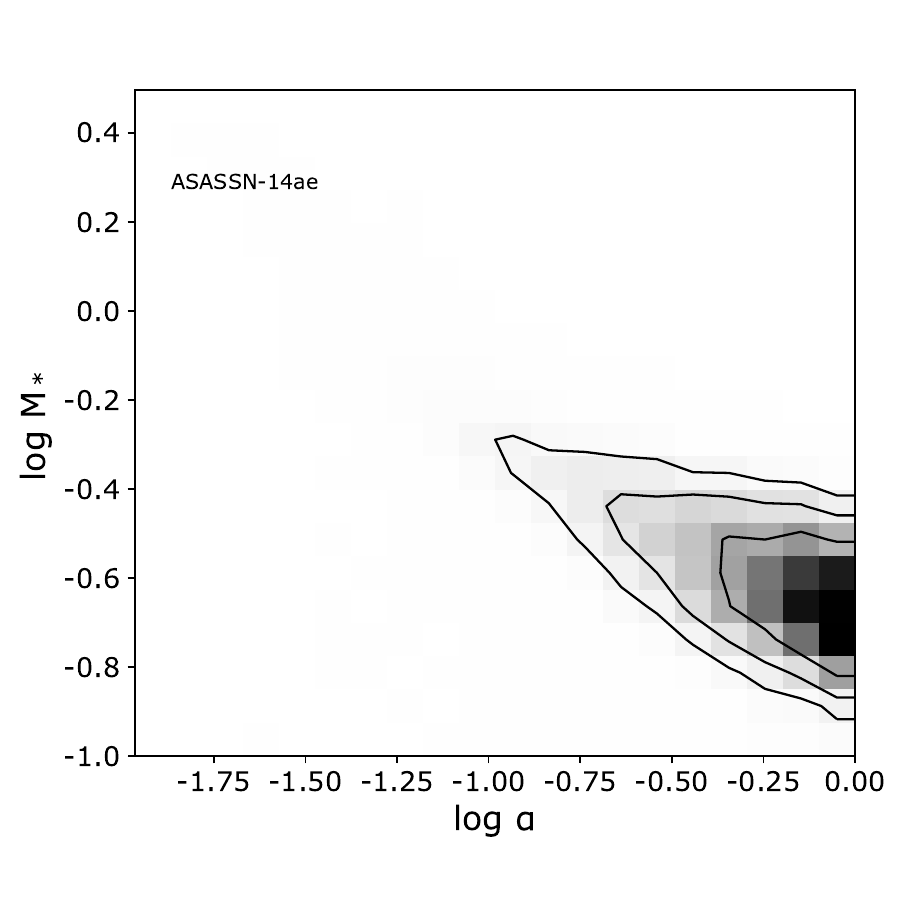}
\includegraphics[width=145pt, trim=0mm 5mm 5mm 5mm, clip]{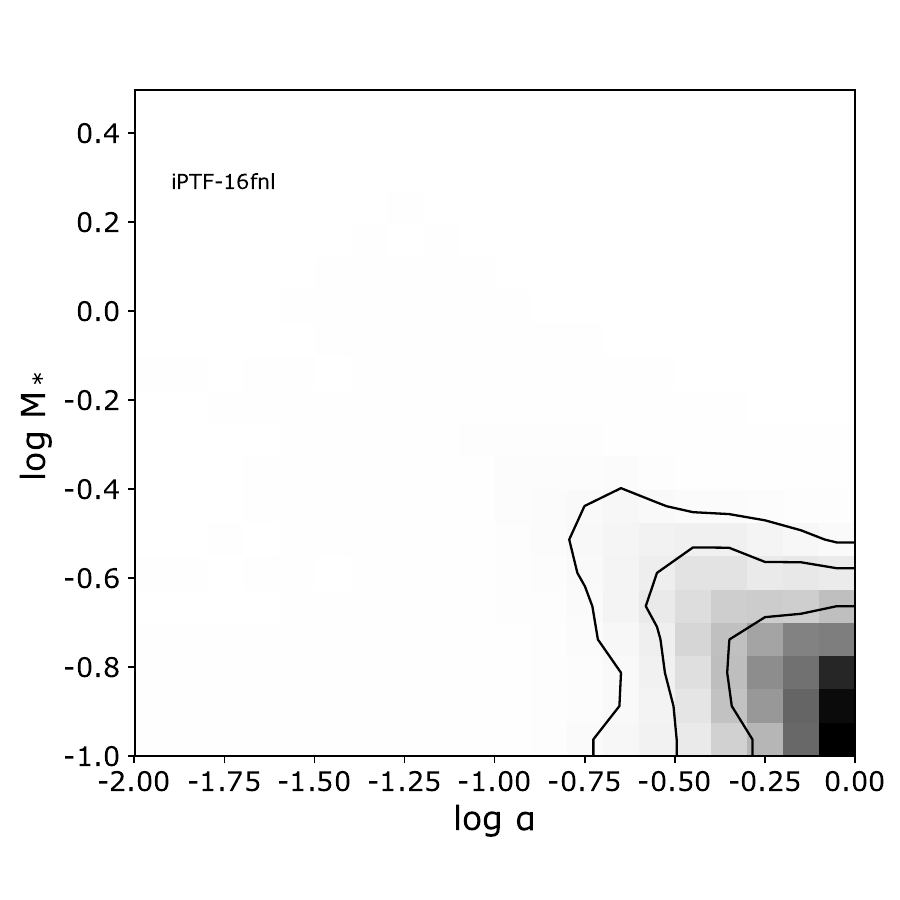} 

\caption{Contour plots showing the density of samples from the posterior distribution of stellar mass ($M_\star$) and disk viscosity ($\alpha$), obtained by comparing the late-time FUV observations to viscously-spreading accretion disk models (see Sec. \ref{sec:disk} and Appendix \ref{app:disk} for more details). Black lines enclose the credible intervals with a probability corresponding to 1$\sigma$, 2$\sigma$, and 3$\sigma$ from a Gaussian distribution.  For the six TDFs with unambiguous late-time {\it HST} detections, we can fit observed FUV luminosities with reasonable parameter choices.  The deep upper limits on late-time FUV flux from TDE1 cannot easily be explained by our model, suggesting either the accretion of little mass ($\ll 0.1 M_\odot$) in a partial disruption, or a state change in the accretion disk.}
\label{fig:diskchi2}
\end{figure*}

We are able to obtain satisfactory fits to all our late-time (i.e. $t>10t_{\rm fb}$) detections by using these simple time-dependent disk models. Similar to the approach applied to the light curves (Sec.~\ref{sec:lightcurves}), we apply a MCMC method to infer the two parameters of the disk model  (again we also include a factor $f$ that accounts for any underestimation of the measurement uncertainty). We use a flat prior for the viscosity, $-2<\log \alpha<0$. The \citet{Salpeter55} mass function is used to yield the prior for the stellar mass, $P(M_\star)\propto M_\star^{-2.3}$, with upper and lower bounds of 0.1 and 3~$M_\odot$, respectively. 

Contour plots with samples from the posterior distribution for each TDF  are shown in Fig. \ref{fig:diskchi2}, and the best-fit parameters are written explicitly in Table~\ref{tab:params}.  Our six unambiguous detections can be explained as follows.  PS1-10jh, D1-9, and PTF-09djl can be adequately fitted with roughly Solar-type stars ($M_\star\approx M_\odot$) and a high effective viscosity, $\alpha \gtrsim 0.3$.  TDE2, and D3-13 favor similar $\alpha$ parameters, but lower main sequence stars ($M_\star \approx 0.15-0.25 M_\odot$). PTF-09ge is a mild outlier among the late-time detections; it requires a significantly lower effective viscosity ($\alpha \approx 0.1$), and favors the lowest stellar mass we consider.

\begin{figure}[t!]
\begin{center}
\includegraphics[width=240pt, trim=1pt 2pt 0pt 2pt, clip]{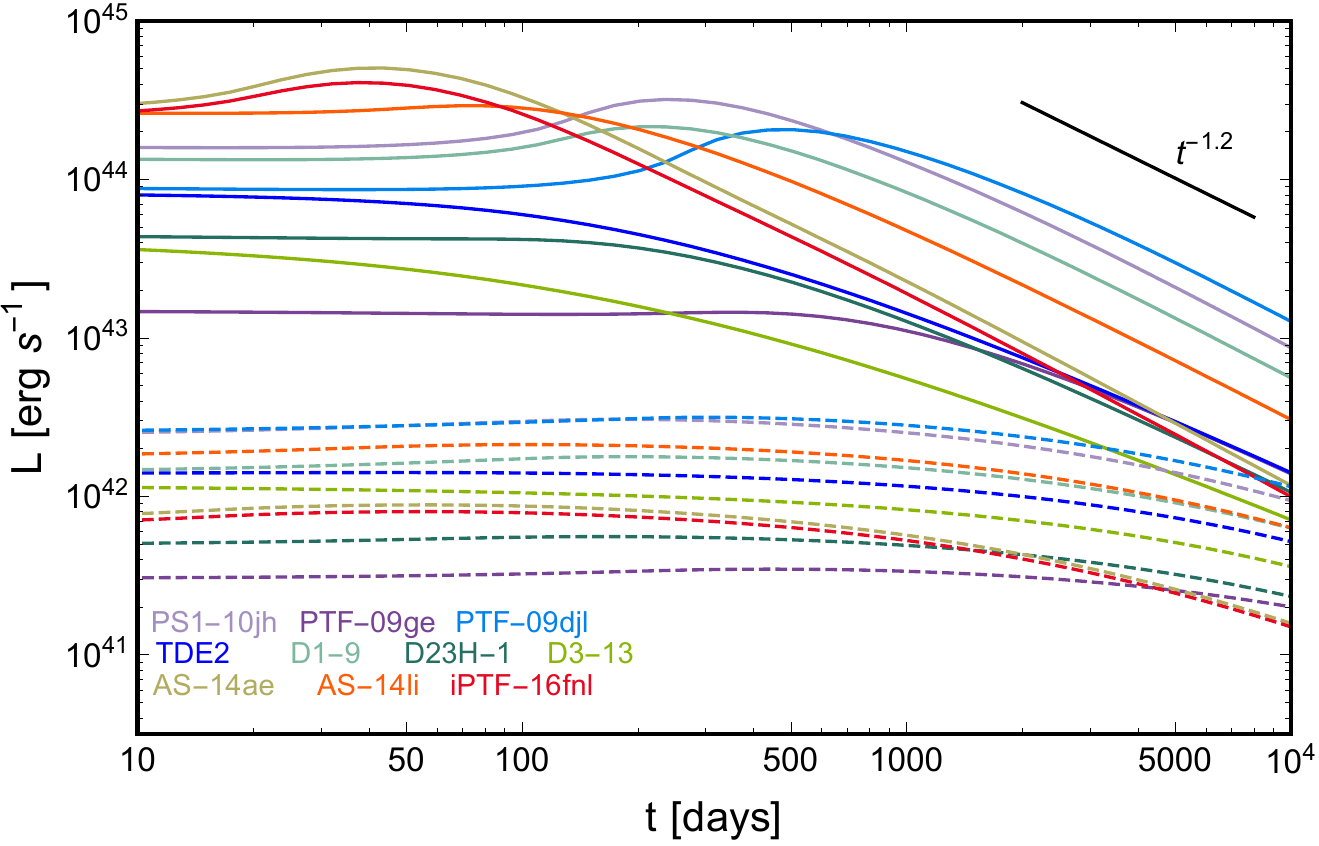} 
\end{center}
\caption{Light curves corresponding to the best-fit models for each TDF in our {\it HST} sample, as well as for three other TDFs seen to undergo late-time flattening in {\it Swift}/UVOT data.  Bolometric light curves have been recalculated with $a_\bullet=0$ (with the exception of D3-13; in this plot, the D3-13 bolometric light curve is shown for $a_\bullet = 0.69$, the minimum value that supports disk formation outside the ISCO) and are shown as solid lines, while FUV light curves are shown as dashed.}\label{fig:modelLs}
\end{figure}

Light curves for all best-fit models, both bolometric and in the FUV band, are presented in Fig. \ref{fig:modelLs}.  We recalculate our best-fit models with $a_\bullet = 0$ to estimate these bolometric light curves (this has little effect on the FUV luminosity).  After a viscous time has passed, the bolometric light curves follow the $\propto t^{-1.2}$ power law, as predicted in \citet{Cannizzo90}.  The FUV light curves evolve in a more shallow way.  One reason for this is that the FUV bands are near the Rayleigh-Jeans tail of emission, and so for a fixed disk size, they would be expected to evolve $\propto t^{-0.3}$.  However, the disks are viscously spreading, increasing the emitting area for Rayleigh-Jeans radiation and causing the FUV light curves to decay even more slowly than $t^{-0.3}$.

If we interpret our FUV observations of D23H-1 as transient (rather than stellar) in origin, the inferred TDF parameters are similar to those of PTF-09ge.  The nondetection of TDE1 is not easy to fit within our grid of disk models, and would require either $\alpha \ll 0.01$ or the accretion of $\ll 0.1 M_\odot$. A low accreted mass might occur in a partial disruption, but we disfavor this explanation for the late-time properties of TDE1 because its luminosity at peak is similar to the rest of the sample (Table~\ref{tab:prop}).  More speculatively, this nondetection could indicate that the original TDE1 transient was an exotic TDF impostor, such as a mildly relativistic stellar collision \citep{Metzger17}.  A third explanation is discussed in the following section.

Interestingly, we are also able to fit the most recent observations of ASASSN-14li, ASASSN-14ae, and iPTF-16fnl within the context of our model, suggesting that these more recent TDFs may also have transitioned to a ``bare'' accretion disk, as is suggested by the flattening of their light curves (Fig. \ref{fig:lcfitswift}). 

\begin{figure*}
\begin{center}
\includegraphics[width=250pt, trim=6mm 11mm 4mm 2mm, clip]{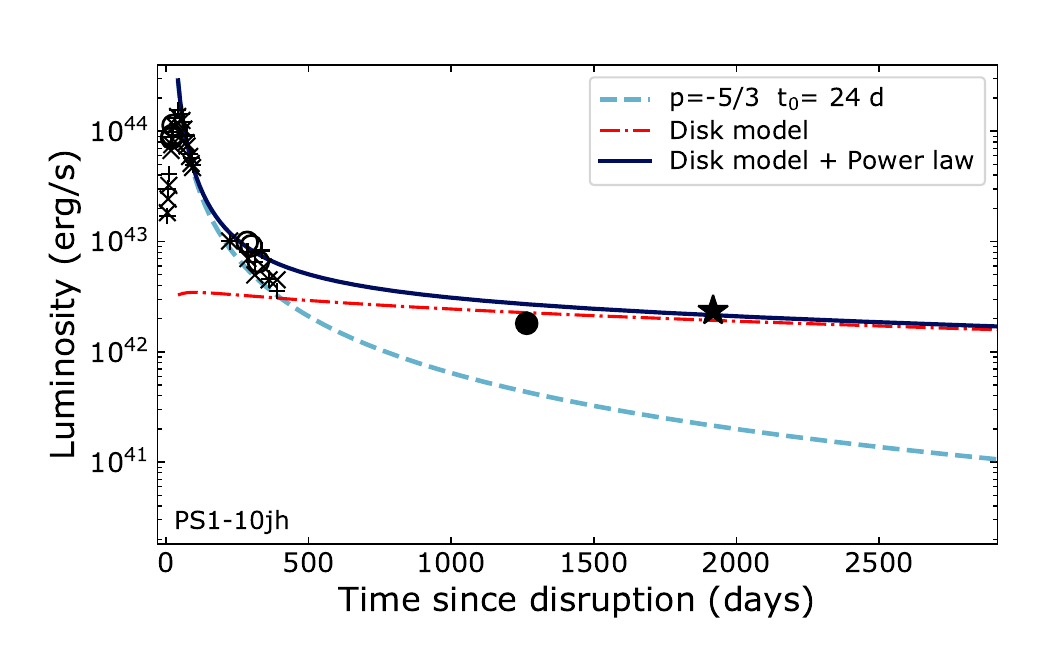}
\includegraphics[width=250pt, trim=6mm 11mm 4mm 2mm, clip]{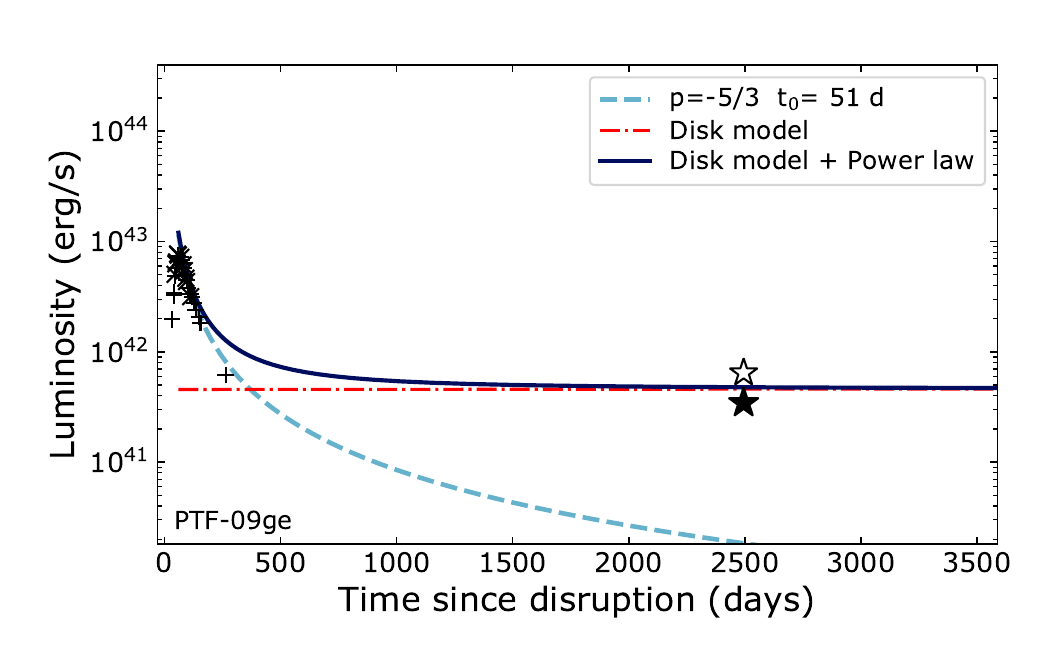}
\includegraphics[width=250pt, trim=6mm 11mm 4mm 2mm, clip]{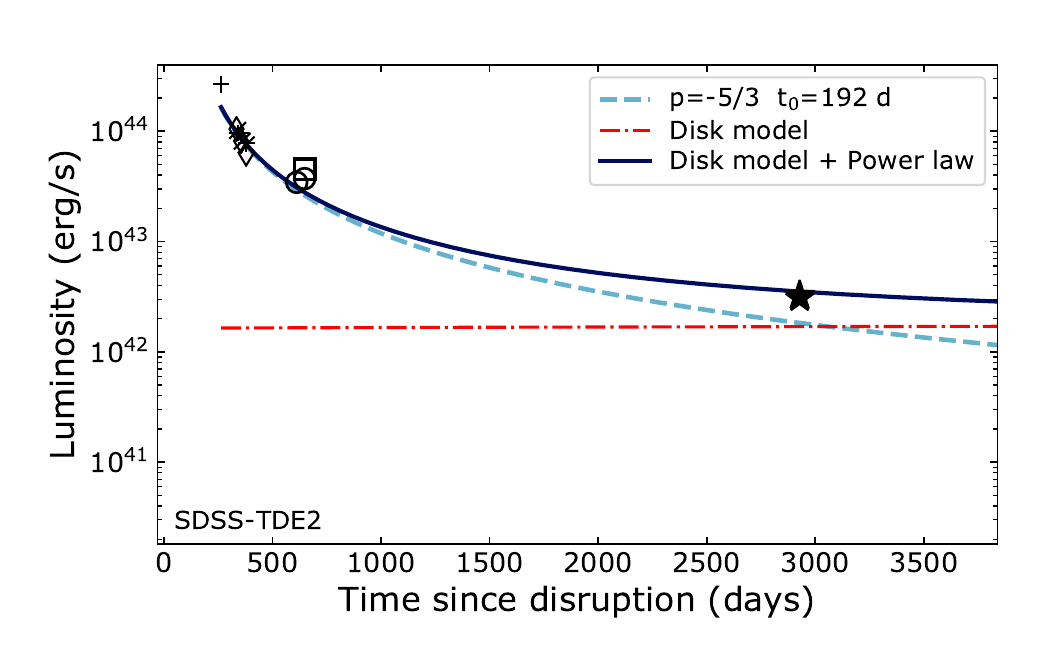}
\includegraphics[width=250pt, trim=6mm 11mm 4mm 2mm, clip]{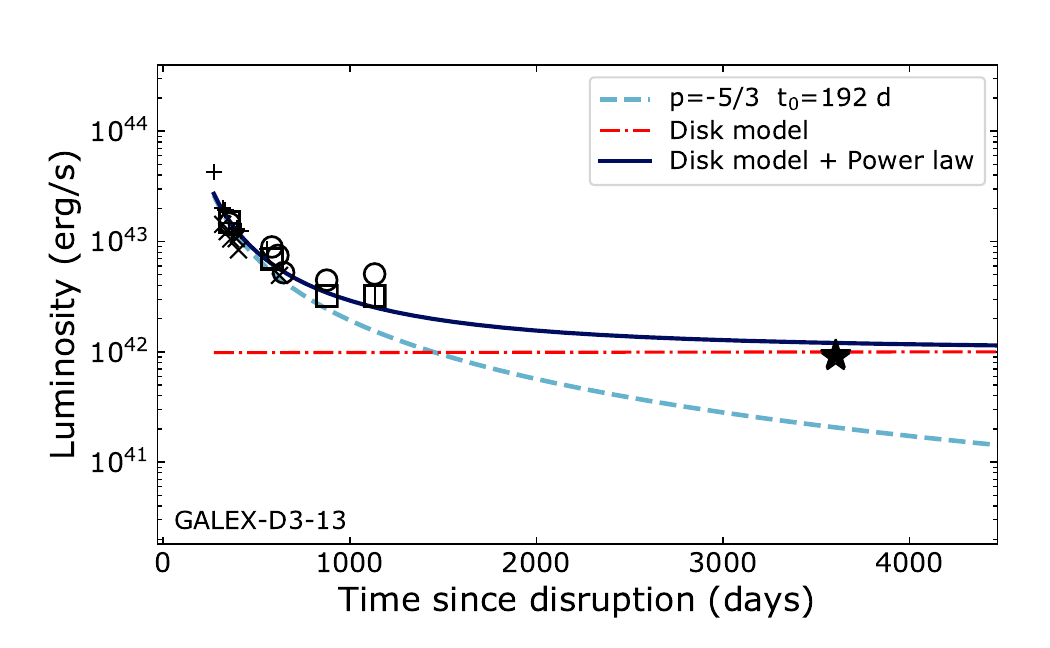}
\includegraphics[width=250pt, trim=6mm 0mm 4mm 2mm, clip]{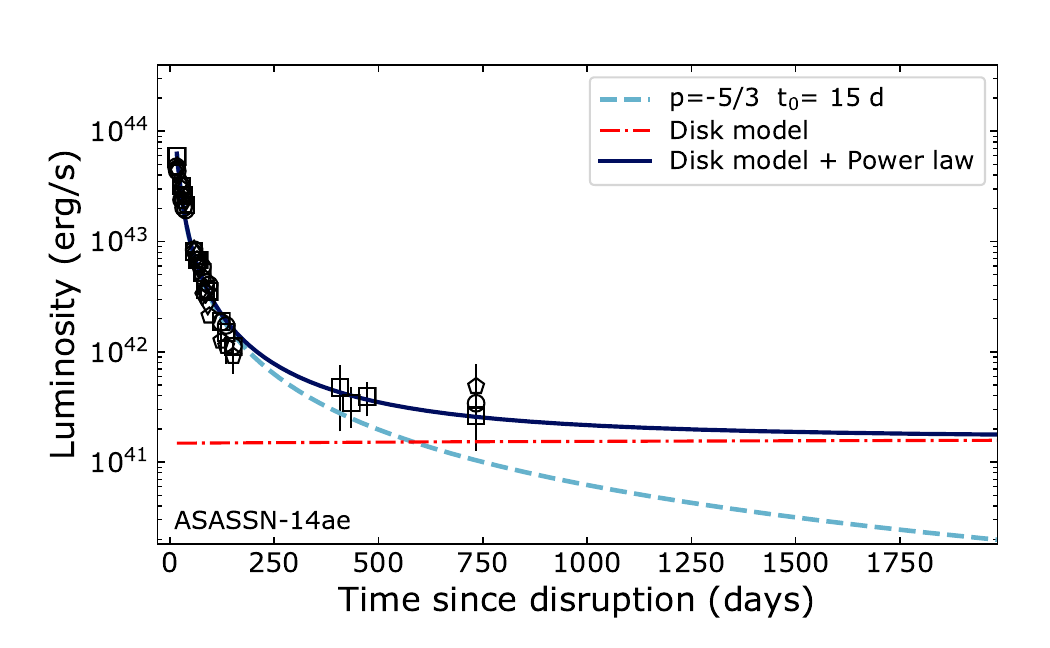} 
\includegraphics[width=250pt, trim=6mm 0mm 4mm 2mm, clip]{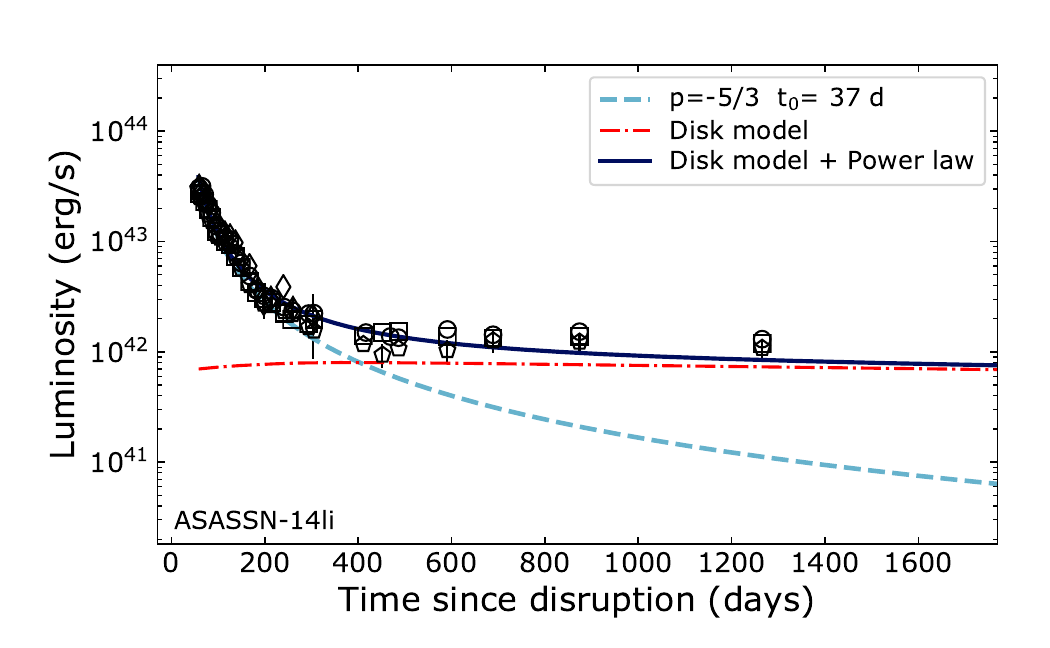} 
\caption{Two-component (disk and power law) fits for our best-sampled flares. Data identical to Figs.~\ref{fig:lcfit} \& \ref{fig:lcfitswift}, but showing the result for a power-law with an index fixed at $p=-5/3$ plus our disk model. The slowly evolving FUV emission from this disk model is not important for the early time observations, but provides a good explanation for the late-time observations. }
\label{fig:lc_flat}
\end{center}
\end{figure*}

We summarize our light curve modeling in Fig. \ref{fig:lc_flat}, where we show two-component light curve models for six well-sampled TDFs.  We see that this simple two-component model does a good job of describing both the older flares observed with {\it HST} (where we have a long temporal baseline but sparse sampling), and younger flares observed by {\it Swift} (where the flattening of the FUV light curve has only begun recently, but is well-resolved in time).

\subsection{Accretion physics implications}\label{sec:alpha}

The simple 1D theory of accretion disks has long predicted viscous \citep{Lightman&Eardley74} and thermal \citep{Shakura&Sunyaev76} instabilities in moderately sub-Eddington regimes, when the disk is radiation pressure-dominated and cooling radiatively.  These instabilities emerge when when the effective diffusion coefficient of Eq. \ref{eq:PDE} becomes negative, and when cooling rates depend more sensitively on temperature than do heating rates, respectively.  In the context of the $\alpha$-viscosity model, these effects occur only at intermediate accretion rates.  They are absent in advectively-cooled super-Eddington disks, and also in colder disks dominated by gas pressure.  Some alternative ad hoc forms for an effective kinematic viscosity, such as the $\beta$-disk we employ \citep{Sakimoto&Coroniti81} and its more complicated generalizations \citep{Taam&Lin84}, are always viscously and thermally stable.

While some examples of these instabilities may have been observed \citep[see e.g.][for a possible thermal instability in the GRS 1915+105 microquasar]{Taam+97, Fender&Belloni04}, their physical reality has frequently been debated due to the many simplifying assumptions that enter into quasi-viscous 1D models.  Angular momentum transport in real accretion disks is generally governed by 3D magnetohydrodynamic (MHD) stresses related to the magnetorotational instability \citep{Balbus&Hawley91}, and MHD simulations have sometimes found that 3D effects suppress thermal \citep{Hirose+09} or viscous instability.  However, the debate is by no means settled: more recent radiation MHD simulations find thermal instability \citep{Jiang+13, Mishra+16} persists in three dimensions, although the instability may be suppressed in disks with sufficiently high iron opacity \citep{Jiang+16}.  The longer linear growth time of the viscous instability makes it more challenging to study, but recent global simulations have found evidence for a clumping \citep{Mishra+16, Fragile+18} analogous to the predictions of \citet{Lightman&Eardley74}.  Both viscous and thermal instabilities may, however, be suppressed in nature by large-scale toroidal fields \citep{Begelman&Pringle07, Oda+09, Sadowski16}.  

This long-standing debate in accretion theory matters for TDF disks because a simple implementation of the $\alpha$-disk model predicts the onset of thermal instability on timescales $\lesssim 1~{\rm yr}$ after disruption \citep{Shen14}.  The 1D development of this instability would lead to a dramatic drop in the temperature of the disk, reducing the FUV luminosity far below the levels predicted in Section \ref{sec:disk}, and making it impossible to reproduce the observed FUV emission.  

More specifically, the $\alpha$-models of \citet{Shen14} are predicted to undergo a thermal instability once the local accretion rate falls below the advection-dominated threshold, i.e. $\dot{M}(R)/\dot{M}_{\rm Edd} = 3^{-1/2} (R/R_{\rm g})$.  In these models, disks accreting below this threshold rapidly transition to a gas-pressure dominated, radiatively-cooled state.  In this cold state, the initial viscous time is extremely long, and the characteristic luminosity is $L/L_{\rm Edd} \approx 0.5 \times 10^{-4} \eta (\alpha/0.1)^{1/8} (M_\bullet / 10^6 M_\odot)^{-1.71}.$  Such low luminosities are at least two orders of magnitude dimmer than the faintest of our late-time detections, and thus we can rule out this type of model for the outcome of thermal instabilities in most TDF disks.  We note that the model of \citet{Shen14} predicts limit-cycle behavior, with intermittent episodes of runaway heating that cause late-time TDF disks to expand back to a highly luminous advective state.  However, these episodes are so brief that it is unlikely for an individual observing epoch to catch any late-time TDF disk, let alone six, in them.

However, the low luminosities predicted after the state changes of \citet{Shen14} offer a speculative way to explain the late-time nondetection of TDE1.  If thermal instabilities are suppressed in nature by iron opacity bumps \citep{Jiang+16}, then TDF disks should often be thermally stable.  On occasion, however, a SMBH will disrupt a low-metallicity star that may lack substantial quantities of iron-group elements, leading to a state change and collapse of the TDF disk to unobservably low FUV luminosities.  A piece of circumstantial evidence supporting this hypothesis may be found in Fig. \ref{fig:popsyn}.  Of all the TDF host galaxies we modeled, the host of TDE1 has the oldest stellar population, which is best fit by a single burst of star formation $12.8~{\rm Gyr}$ ago, hinting at a low metallicity.  

A final possible explanation for the nondetection of TDE1 is a different type of disk state change, from a thermal, radiatively efficient state to a ``low-hard,'' radiatively inefficient accretion flow.  In X-ray binaries, this type of transition is seen to occur at accretion rates below $\approx 2\%$ of the Eddington rate \citep{Maccarone03}.  This type of state change produces a large drop in disk thermal emission.  The $M_\bullet-\sigma$ relationship suggests that TDE1 has one of the highest black hole masses in our sample, making it plausible that this TDF could have been the first to undergo such a state transition.  Radiatively inefficient accretion flows are often associated with efficient jet launching, and past work has argued that radio-dim TDFs should become radio-bright at late times as jets turn on at a low Eddington fraction \citep{vanVelzen11, Tchekhovskoy13}.  In our fiducial, $a_\bullet = 0.9$ grid of models, the present-day accretion rates in all observed TDFs are well above $2\%$ of the Eddington rate, with the exceptions of TDE2 ($\dot{M} = 0.032\dot{M}_{\rm Edd}$, for best-fit $M_\star$ and $\alpha$) and D3-13 ($\dot{M} = 0.0024\dot{M}_{\rm Edd}$, for best-fit $M_\star$ and $\alpha$).  The persistence of thermal emission in D3-13 suggests that either one of our model parameters for this TDF is significantly incorrect, or some aspect of TDF accretion disks prevents (or delays) the transition to a radiatively inefficient state.  Late-time radio observations of TDE1, TDE2, and D3-13 could better constrain this hypothesis.

While we do not claim that the late-time observation of FUV emission in TDFs validates the \citet{Sakimoto&Coroniti81} $\beta$-disk---indeed, detailed MHD simulations typically find that stress scales proportionally to total pressure \citep{Mishra+16} - our models seem to demonstrate two features of late-time TDF disk hydrodynamics.  First, these disks must have long initial viscous times.  This is achieved naturally in the $\beta$-disk prescription, but could also arise in $\alpha$-disks with small $\alpha$ parameters.  Second, it seems clear that the classic thermal and viscous instabilities of 1D $\alpha$-theory are not developing in TDF disks we observe, with the possible exception of TDE1.  This result is in agreement with the observation that observed black hole X-ray binaries generally fail to display the limit cycle behavior that would result from nonlinear development of the \citet{Lightman&Eardley74} viscous instability \citep{Cannizzo96, Done+07}.

\subsection{Solution to the missing energy problem}
The observed radiated energy of TDFs, $E_{\rm bol}^{\rm early}$, is obtained by integrating fitted early-time bolometric light curves (typically estimated from single-temperature blackbody fits).  Empirically, $E_{\rm bol}^{\rm early} \sim 10^{49-51}~{\rm erg}$, which is usually one to two orders of magnitude lower than $E_{\rm bol}^{\rm tot} = 0.05 M_\star c^2 \sim 10^{52-53}~{\rm erg}$, the energy that should be released during radiatively efficient accretion of the bound stellar debris. This ``missing energy problem'' \citep{Piran15,Stone16b, Lu18} may be explained in a variety of ways: for example, TDFs may emit most of their energy in unobservable EUV wavelengths, as is suggested by infrared dust echoes from some TDFs \citep{vanVelzen16b}; alternatively, TDFs may have unexpectedly small energy budgets due to severe mass loss during the circularization process \citep{Metzger16}, a preference for partial disruptions \citep{Guillochon13}, or the prevalence of low-energy TDF impostors \citep{Metzger17}.  Some of the missing energy can also be explained by reddening of the TDF spectrum due to the dust in the host galaxy. This reddening will decrease the observed blackbody temperature, leading to an underestimation of the blackbody luminosity by up to an order of magnitude \citep{Gezari09}.  The missing energy problem may also be resolved if the flare's  radiative efficiency $\eta \ll 0.1$.  This explanation is realized in the circularization paradigm \citep{Piran15}, where the early-time light curve is powered by shocks at the stream self-intersection radius $R_{\rm SI}$.  As mentioned previously, the typical radiative efficiency of bulk kinetic energy thermalized at these shocks is $\eta_{\rm c} \approx GM_\bullet/ (R_{\rm SI}c^2) \ll 0.1$.  If disk formation is delayed, then a much greater amount of energy can be emitted, at lower luminosities, long after the peak of the flare.  

In contrast, our slowly evolving accretion disk models address the missing energy problem in two of the ways discussed above: by displacing the TDF's energy both in wavelength and in time.  The SEDs of our theoretical models typically peak near $\sim 0.1~{\rm keV}$, and a majority of their power comes out in the EUV.  However, the long viscous times of our models can, in most cases, prevent early emission of this radiation, even under the assumption of efficient circularization.  Our models, which are based on \citet{Cannizzo90}, differ strongly from ``delayed circularization'' explanations for the missing energy problem in the large early-time EUV (and often X-ray) luminosities we predict.  There are also likely to be more subtle temporal differences at longer wavelengths as well: delayed accumulation of bound material into a disk that efficiently transports angular momentum will produce a steeper late-time light curve than do our models.

\begin{deluxetable*}{l c c c c c c c c c}
\tablewidth{0pt}
\tablecolumns{9}
\tablecaption{Tidal disruption flare energetics}
\tablehead{name & \tablenotemark{a}$E_{\rm bol}^{\rm early}$ & \tablenotemark{a}$E_{\rm bol}^{\rm early}$ & \tablenotemark{b}$E_{\rm FUV}^{\rm late}$ & \tablenotemark{b}$E_{\rm FUV}^{\rm late}$ & \tablenotemark{c}$E_{\rm bol}^{\rm late}$ & \tablenotemark{c}$E_{\rm bol}^{\rm late}$ & \tablenotemark{d}$C^{\rm late}$ & \tablenotemark{e}$L_{\rm X}^{\rm late}$ & \tablenotemark{f}$E_{\rm bol}^{\rm tot}$\\
                    & ($\log$\,erg) & ($\%$)  & ($\log$\,erg)  & ($\%$)  &($\log$\,erg) & ($\%$) & & ($\log$\,erg$~{\rm s}^{-1}$) & ($\log$\,erg)}
\startdata
PS1-10jh   & $51.3^{+0.1}_{-0.2}$ & $3.88\%$ & 50.6 & $0.792\%$ & 52.4 & $49.7\%$ & 26.4 & $41.3^{+1.7}_{-1.2}$ & 52.7 \\
PTF-09ge   & $50.5^{+0.2}_{-0.2}$ & $4.89\%$ & 49.8 & $1.03\%$ & 51.3 & $33.2\%$ & 18.3 & $37.0^{+3.6}_{-2.6}$ & 51.8 \\
PTF-09djl  & $51.0^{+0.1}_{-0.2}$ &  $1.55\%$ & 50.8 & $0.905\%$ & 52.4 & $37.8\%$ & 24.1 & $41.5^{+1.6}_{-1.1}$ & 52.8  \\
TDE2       & $51.6^{+0.1}_{-0.0}$ & $30.8\%$ & 50.9 & $5.64\%$ & 51.6 & $30.7\%$ & 1.78 & $30.3^{+8.0}_{-7.6}$ & 52.1  \\
D1-9       & $50.7^{+0.0}_{-0.0}$ & $1.55\%$ & 50.7 & $1.39\%$ & 52.3 & $63.8\%$ & 13.9 & $39.8^{+2.4}_{-1.5}$ & 52.5  \\
D23H-1     & $50.7^{+0.1}_{-0.1}$ & $7.74\%$ & $50.1$ & $1.86\%$ & 51.5 & $51.1\%$ & 9.00 & $35.1^{+4.9}_{-3.7}$ & 51.8  \\
D3-13      & $51.4^{+0.1}_{-0.1}$ & $17.6\%$ & 50.4 & $1.60\%$ & 51.2 & $11.5\%$ & 2.50 & $22.7^{+7.9}_{-0.0}$ & 52.2 \\
TDE1       & $50.7^{+0.2}_{-0.2}$ & -- & $<49.0$ & -- & -- & -- & -- & -- & -- \\
ASASSN-14ae & $50.2^{+0.0}_{-0.0}$ & $1.23\%$ & 49.3 & $0.152\%$ & 51.9 & $66.1\%$ & 107 & $41.7^{+1.3}_{-0.9}$ & 52.1 \\
ASASSN-14li & $50.5^{+0.0}_{-0.0}$ & $1.54\%$ & 50.1 & $0.642\%$ & 52.1 & $57.6\%$ & 30.6 & $39.2^{+2.9}_{-2.1}$ & 52.3 \\
iPTF-16fnl & $49.8^{+0.0}_{-0.0}$ & $0.615\%$ & 49.2 & $0.157\%$ & 51.7 & $51.0\%$ & 127 & $42.5^{+1.1}_{-0.8}$ & 52.0 \\
\enddata

\tablenotetext{a}{The observed bolometric energy release from early-time ($t<10t_{\rm fb}$) light curves, as computed from our posterior distributions of power-law light curves for single-temperature blackbodies (\S \ref{sec:lightcurves}).}
\tablenotetext{b}{The approximate energy radiated to the present day in FUV wavelengths, which we estimate as $E_{\rm FUV}^{\rm late} \equiv t_{\rm late}\nu L_\nu$, using quantities in Table \ref{tab:prop}.}
\tablenotetext{c}{The total bolometric energy radiated to the present day for our best-fit models for each TDF.}
\tablenotetext{d}{The predicted late-time bolometric correction $C^{\rm late} \equiv L_{\rm bol} / (\nu L_\nu)$.  The numerator of this correction factor uses our best-fit theoretical model, while the denominator uses the observed FUV luminosity.}
\tablenotetext{e}{The predicted late-time soft X-ray luminosity ($E\ge 0.3~{\rm keV}$) for our best-fit models.  This is an approximate estimate, as we have ignored host absorption and general relativistic effects.}
\tablenotetext{f}{The total bolometric energy available to each TDF for our best-fit models: $E_{\rm bol}^{\rm tot} \equiv 0.058 (M_\star / 2)c^2$. }

\tablecomments{For model-dependent quantities, the fiducial values quoted are for an SMBH spin $a_\bullet=0.0$.  The error range on $L_{\rm X}^{\rm late}$ covers a range of spins from $a_\bullet = -0.9$ to $a_\bullet = 0.9$, with the exception of D3-13.  For this TDF, retrograde and slow prograde spins would push the Hills mass below $M_\bullet$, and so the fiducial values here were computed with $a_\bullet=0.69$, the lowest value that permits disk formation outside the ISCO.  All quantities in parentheses represent percentages of $E_{\rm bol}^{\rm tot}$ in our best-fit models (for the fiducial, $a_\bullet = 0.0$, case).  Estimates and models for D23H-1 assume for the sake of argument that all observed FUV emission is of transient, rather than stellar, origin.}\label{tab:energy}
\end{deluxetable*}

The amount of energy radiated so far in the FUV can be crudely estimated as $E_{\rm FUV}^{\rm late} \equiv t_{\rm late} L_{\rm FUV}^{\rm late}$.  Typically, $E_{\rm FUV}^{\rm late} \sim 0.1-1 E_{\rm bol}^{\rm early}$, indicating that the late-time emission we have directly observed can be as energetically important as the better-studied early-time light curves of TDFs.  For every TDF in our sample, $E_{\rm FUV}^{\rm late} \ll E_{\rm bol}^{\rm tot}$, so in order to radiate the expected gravitational energy of $E_{\rm bol}^{\rm tot} \sim 2\times 10^{52}$~erg in the FUV, the observed late-time plateau of $10^{42.5}~{\rm erg}\,{\rm s}^{-1}$ would have to persist for $\sim 300$~yr. However, the FUV bands are on the Rayleigh-Jeans side of each blackbody we have fit to the data, implying a substantial bolometric correction, $C^{\rm late}\equiv L_{\rm bol}^{\rm late} / (\nu L_{\rm FUV}^{\rm late})$, should exist for each TDF detected at late times.  We estimate these bolometric corrections using our best-fit accretion disk models, and also estimate $E_{\rm bol}^{\rm late} \equiv \int_{t_0}^{t_{\rm late}}L_{\rm bol}(t){\rm d}t$, the total bolometric energy outputted so far by the TDF.  

We quantify all observed and modeled energy scales in Table \ref{tab:energy}.  In this table, fiducial values\footnote{Even though our model grid was built with $a_\bullet = 0.9$, we have re-run best fit combinations of $\{\alpha, M_\star\}$ with different values of SMBH spin.  This usually does not impact the FUV luminosity substantially, since it is dominated by the outermost disk annuli.  The two exceptions to this are TDFs with large SMBH masses, TDE2 and D3-13 (where the ISCO is not far from the disk outer edge).  For TDE2, switching to $a_\bullet \lesssim 0$ will decrease the FUV luminosity substantially, making the best-fit $\{M_\star, \alpha\}$ less trustworthy.  The same caveat applies to D3-13, but here the best-fit $M_\star$ and large SMBH mass $M_\bullet$ puts this TDF dangerously close to the Hills mass.  The minimum SMBH spin value that can produce a disk outside the ISCO is $a_\bullet = 0.69$, so for D3-13's entry in Table \ref{tab:energy}, we take this value as both the fiducial and lowest one.} for modeled energies use $a_\bullet =0.0$, and quoted error ranges go from $a_\bullet = - 0.9$ to $a_\bullet=0.9$.  In general, we find substantial late-time bolometric corrections, with $8 \lesssim C^{\rm late} \lesssim 300$.  Bolometric corrections are smallest for the oldest flares in our sample, (TDE2, D1-9, and D3-13), and are largest for the youngest flares in the sample (ASASSN-14ae, ASASSN-14li, iPTF-16fnl).  In all cases, the disks have radiated an order unity fraction of their total available energy budget ($E_{\rm bol}^{\rm late} \sim E_{\rm bol}^{\rm tot}$), though a majority of $E_{\rm bol}^{\rm tot}$ remains to be radiated.  

One testable prediction of our models is the late-time soft X-ray luminosity at energies $\ge 0.3~{\rm keV}$, which can be substantial for many of the flares in our sample.  In Table \ref{tab:energy}, we present our predictions for the late-time thermal soft X-ray luminosity, $L_{\rm X}^{\rm late}$, for each best-fit model.  However, we note that these predictions are likely optimistic, as we have neglected photoelectric absorption in the host galaxy, and have assumed $a_\bullet =0.9$ (the predicted $L_{\rm FUV}^{\rm late}$ values are insensitive to these assumptions, but $L_{\rm X}^{\rm late}$ is quite sensitive to both).  Many of these TDFs have constraining upper limits on their early-time X-ray luminosities, which may indicate that disks had not yet circularized at early times, or alternatively, that obscuration due to tidal debris or outflows imparted a strong viewing angle-dependence on inner disk annuli \citep{Dai18}.  However, our FUV observations suggest that at sufficiently late times, compact TDF disks exist and are not obscured by a larger-scale photosphere.  For most flares, constraining late-time X-ray observations do not yet exist.  However, our prediction for $L_{\rm X}^{\rm late}$ in ASASSN-14li is quite close to this TDF's observed late-time X-ray luminosity \citep{Bright18}.

If the disk-powered plateau indeed lasts for $\sim 10$~yr ($\sim 10^2$~yr) and the TDF rate is $10^{-4}\,$galaxy$^{-1}\,$yr$^{-1}$ \citep[e.g.][]{Stone16b, vanVelzen18,Hung18}, then $\sim 0.1$\% ($\sim 1\%$) of galaxies are currently hosting a TDF disk with a considerable FUV luminosity. However, recent observations suggest that the tidal disruption rate in post-starburst galaxies is enhanced by a factor of $\sim 10$ \citep[e.g.][]{French16,Law-Smith17}, implying that the fraction of this galaxy class with elevated central UV emission could be as high as 10\%.  As we have shown in this work, distinguishing this long-lived TDF emission from UV produced by stellar sources is possible with high-resolution {\it HST} observations.

\section{Summary}
\label{sec:conclusions}

We have obtained late-time {\it HST} FUV imaging of eight galaxies hosting TDF candidates.  Each of these TDF candidates were originally identified by thermal emission in the optical/UV between 5 and 9 years prior to our observations.  In six of the eight targets (PS1-10jh, PTF-09ge, PTF-09djl, SDSS-TDE2, GALEX-D1-9, and GALEX-D3-13) we unambiguously detect unresolved nuclear point sources with FUV luminosities $\nu L_\nu \sim 10^{41.5 - 42.5}~{\rm erg~s}^{-1}$. 
For PS1-10jh and GALEX-D3-13, we are able to unambiguously conclude that the observed FUV flux is of transient origin (as a result of a pre-flare {\it GALEX} upper limit and {\it HST} optical imaging observations).  For the other four of these detections, a stellar origin for the unresolved {\it HST} FUV emission is unlikely but cannot be ruled out. Late-time nuclear FUV emission of similar magnitude is seen in D23H-1, but here the FUV luminosity is spatially extended, and it is unclear what fraction of the nuclear component, if any, is due to a TDF. Only for one target, TDE1, do we obtain a nondetection and a deep upper limit ($\nu L_\nu < 10^{40.6}~{\rm erg~s}^{-1}$) on the late-time flare luminosity. 

We also analyzed all the public {\it Swift} UVOT observations of TDFs, finding detections up to 2 years post-peak, with a luminosity similar to that observed at late times with {\it HST}, for three other TDFs: ASASSN-14ae, ASASSN-14li, iPTF-16fnl. 
The {\it HST} and {\it Swift} observations represent the largest sample of late-time TDF emission.

Compact accretion disks on scales comparable to the tidal radius are the simplest theoretical models for TDF thermal emission, but these models are severely discrepant with early-time optical/UV observations.  This discrepancy has sparked much debate about the origins of early-time TDF emission (e.g. reprocessing of disk X-rays versus shock-powered flares), but appears to be absent for the late-time detections in this paper, all of which are consistent with simple analytic/semi-analytic models of viscously spreading, compact accretion disks \citep{Cannizzo90}.  This has interesting implications for TDF physics and possibly disk accretion physics more generally, which we list below.
\begin{enumerate}
    \item The six late-time FUV detections in our sample are inconsistent with a canonical $\propto t^{-5/3}$ power law decay; if such a power law is fit to the early-time flare photometry, the late-time emission represents a significant {\it flattening} of the light curve. 
   
    \item The  {\it Swift} UVOT  photometry shows a similar late-time NUV flattening for ASASSN-14ae, ASASSN-14li, and iPTF-16fnl, indicating that the time of this transition may be typically $\sim 1-2~{\rm yr}$ after peak.  
    
    \item By using two {\it HST} FUV bands (or, for the more recent sample, multiple {\it Swift}/UVOT bands), we fit single-color blackbodies to observed late-time emission.  The best-fit blackbody temperatures are relatively cool, with $T_{\rm late} \approx 10^{4.5}~{\rm K}$, and the blackbody radii are compact, with $R_{\rm late} \approx 10^{13.2-13.7}~{\rm cm}$ (one outlier, SDSS-TDE2, has $R_{\rm late}=10^{14.2}~{\rm cm}$).  In all cases, these blackbody radii are at most a small multiple of the circularization radius $2R_{\rm t}$.  This obviates the need for a reprocessing layer to explain late-time emission, and we show that the other early-time power source (circularization shocks from returning debris) is energetically incapable of explaining high late-time luminosities.  The sustained late-time UV luminosities we observe represent further evidence against alternative supernovae interpretations of these flares.
    
    \item For all 10 TDFs with late-time detections, we are able to fit the observed UV luminosity with simple 1D models for time-dependent, viscously spreading accretion disks.  The detailed models we employ use an effective viscosity that is proportional only to gas pressure (the so-called ``beta-viscosity'' of \citealt{Sakimoto&Coroniti81}) as in \citet{Cannizzo90}.  This choice of viscosity prescription is somewhat artificial, but its success suggests that late-time tidal disruption disks (i) retain a fairly high gas mass, and (ii) have not undergone a thermal instability.  The luminosities we observe are significantly higher than those predicted by simple 1D alpha-viscosity models that are permitted to collapse to a dim, gas pressure-dominated state due to thermal instability \citep{Shen14}.  Thus, if the alpha-viscosity picture is a reasonable approximation to real, magnetohydrodynamic TDF disks, there must be additional effects \citep[e.g. iron opacity;][]{Jiang+16}
    that suppress the onset of thermal instabilities.
    
    \item We have identified an interesting, mass-dependent trend in the properties of the TDF sample considered here.  As shown in Fig.~\ref{fig:mass-PLearly}, the best-fit power-law indices for early-time light curves are systematically steeper in TDFs sourced by low-mass SMBHs ($M_\bullet \lesssim 10^{6.5} M_\odot$) and shallower in those from high-mass SMBHs ($M_\bullet \gtrsim 10^{6.5} M_\odot$).  
    Early-time power laws for the high-mass subsample are shallow enough that our late-time observations are consistent with little-to-no flattening in the light curve (Fig.~\ref{fig:mass-PLpredict}). A delay in the disk formation for the lower-mass black holes could explain why, for these TDFs, the late-time excess is larger with respect to the early-time power law. In this scenario, the early-time light curves of low-mass TDFs have a larger fractional emission contribution from stream self-intersection shocks  \citep{Piran15}. However, our current disk models cannot be used to quantitatively test if this speculation works (see  caveats below).
    
    \item The late-time FUV emission is consistent with compact, quasi-circular accretion disks possessing long viscous times.  If this general model is correct, then the TDF ``missing energy problem'' is solved by TDFs radiating most of their available energy budget, $E_{\rm bol}^{\rm tot}$, in EUV bands at late times ($\sim 10-100~{\rm yr}$ post-peak).  Stated another way, the initial disk masses inferred from our viscously-spreading models are a sizable fraction of the debris that remains dynamically bound following the tidal disruption of a low-mass star.  
    
\end{enumerate}
There are several caveats to the above conclusions that should be clarified with future work.  First, it would be useful to confirm the transient nature of the late-time emission in the four sources lacking stringent constraints from pre-flare FUV upper limits or high-resolution {\it HST} imaging. For the TDF host galaxies, the observed late-time FUV luminosity on $\sim 0.1$ kpc scales is at least an order of magnitude larger than in similar galaxies (see section~\ref{sec:fuv_compare}). Yet it is theoretically possible that very compact and young nuclear stellar populations could produce the observed FUV emission.  This possibility could be tested directly with follow-up UV observations. Temporal or spectral variability in UV emission would firmly establish the transient nature of the late-time signal, although our theoretical models predict very slow evolution of TDF accretion disks at this stage.  However, high-resolution NUV imaging would supply a decisive test, since stellar population models predict {\it more} NUV than FUV flux, while TDF disks are expected to emit both these wavelengths from the Rayleigh-Jeans tail of a multicolor blackbody (an expectation borne out by our two-filter FUV observations), meaning that {\it less} NUV is expected than FUV from disk emission.

Second, it is important to test a wider range of time-dependent accretion models than those explored in this paper.  For simplicity, we have focused only on very simple ``alpha-disk'' and ``beta-disk'' types of effective viscosity.  While the former appears inconsistent, and the latter consistent, with the limited late-time observations so far obtained, these models are only approximations for the actual magnetohydrodynamic processes governing the evolution of astrophysical accretion disks.  The models employed in this paper are fundamentally Newtonian, and eventually, late-time TDF light curves should be compared to general relativistic spreading disk models (e.g. the work of \citealt{Balbus&Mummery18}, which was published while this paper was under review).  It would be useful to explore time-dependent disk models with more realistic microphysics and initial conditions.  For example, the addition of time-dependent source terms reflecting the slow accumulation of bound debris into the accretion disk could probe the extent to which the late-time luminosities we observed can be explained by inefficient circularization (e.g. \citealt{Shiokawa15}).

Third, soft X-ray follow-up observations will offer a complementary probe of late-time TDF disk physics.  Many TDF candidates detected by thermal optical/UV emission appear X-ray dim at early times, while others show a diversity of thermal soft X-ray luminosities.  The wide range of early-time behavior seen at X-ray wavelengths may be due to viewing angle effects, or perhaps a wide range in circularization efficiencies.  However, the small best-fit blackbody radii $R_{\rm late}$ of our observations suggest that if X-rays were once being absorbed and reprocessed by an extended, optically thick shroud, that reprocessing layer no longer exists; conversely, if X-rays were absent at early times because no inner disk had yet circularized, such a disk is now present.  The magnitude and spectral properties of X-ray emission will aid in testing the different disk models discussed above.

The apparent near-ubiquity of late-time disk emission in TDF candidates raises exciting opportunities for studying both TDF and accretion physics.  In stark contrast to the uncertain and contested early-time hydrodynamics of a tidal disruption event, it is natural to expect that at late times, the stellar debris should settle down into a more axisymmetric configuration resembling a traditional accretion disk.  The simple 1D disk models used in this paper are, indeed, able to match observed late-time FUV emission.  If these or analogous disk models continue to match future late-time TDF observations, we would possess a powerful new tool for measuring the intrinsic parameters of TDFs (e.g. SMBH mass, or mass of the disrupted star), independent of the uncertainties of early-time emission mechanisms.  In the meantime, however, late-time TDF observations can be used to narrow down the parameter space of viable disk models, which can probe the nature of effective viscosity and associated instabilities in astrophysical accretion disks.



\section*{Acknowledgments}
We dedicate this manuscript to the memory of the late John Cannizzo, whose pioneering early work on TDF disks played a central role in our interpretation of these observations.  We gratefully acknowledge useful discussions and  assistance from John Cannizzo, James Guillochon, Glennys Farrar, Decker French, Chris Fragile, Julian Krolik, and Ann Zabludoff. We thank the Aspen Center for Physics for its hospitality during the completion of this work.  SVV received financial support from NASA through HST-HF2-51350. NCS received financial support from NASA through Einstein Postdoctoral Fellowship Award Number PF5-160145.  BDM acknowledges support from NASA through the Astrophysics Theory Program (grant number NNX17AK43G). SG is supported in part by NSF CAREER grant 1454816 and NSF AAG grant 1616566. 

This research made use of Astropy, a community-developed core Python package for Astronomy \citep{Astropy-Collaboration18} and corner.py  \citep{Foreman-Mackey16}. Support for program GO-14255 was provided by NASA through a grant from the Space Telescope Science Institute (STScI), which is operated by the Association of Universities for Research in Astronomy Inc., under NASA contract NAS 5-26555.

\software{astropy \citep{Astropy-Collaboration18}, 
PhytonPhot \citep{Jones15}, 
pysynphot \citep{Lim15}, 
gPhoton \citep{Million16}, 
HEAsoft \citep{Arnaud96}, 
FSPS \citep{Conroy09,Conroy10} with Python binding  \citep{Foreman-Mackey14}, 
emcee \citep{Foreman-Mackey13}, 
corner.py  \citep{Foreman-Mackey16}.
}


\bibliographystyle{aasjournal}
\bibliography{general_desk,Nick.bib}

\begin{thebibliography}{}
\expandafter\ifx\csname natexlab\endcsname\relax\def\natexlab#1{#1}\fi
\providecommand{\url}[1]{\href{#1}{#1}}

\bibitem[{{Abramowicz} \& {Fragile}(2013)}]{Abramowicz13}
{Abramowicz}, M.~A., \& {Fragile}, P.~C. 2013, Living Reviews in Relativity,
  16, 1

\bibitem[{{Arcavi} {et~al.}(2014){Arcavi}, {Gal-Yam}, {Sullivan}, {Pan},
  {Cenko}, {Horesh}, {Ofek}, {De Cia}, {Yan}, {Yang}, {Howell}, {Tal},
  {Kulkarni}, {Tendulkar}, {Tang}, {Xu}, {Sternberg}, {Cohen}, {Bloom},
  {Nugent}, {Kasliwal}, {Perley}, {Quimby}, {Miller}, {Theissen}, \&
  {Laher}}]{Arcavi14}
{Arcavi}, I., {Gal-Yam}, A., {Sullivan}, M., {et~al.} 2014, \apj, 793, 38

\bibitem[{{Arnaud}(1996)}]{Arnaud96}
{Arnaud}, K.~A. 1996, in Astronomical Society of the Pacific Conference Series,
  Vol. 101, Astronomical Data Analysis Software and Systems V, ed. G.~H.
  {Jacoby} \& J.~{Barnes}, 17

\bibitem[{{Astropy Collaboration} {et~al.}(2018){Astropy Collaboration},
  {Price-Whelan}, {Sip{\'{o}}cz}, {G{\"u}nther}, {Lim}, {Crawford}, {Conseil},
  {Shupe}, {Craig}, {Dencheva}, {Ginsburg}, {VanderPlas}, {Bradley},
  {P{\'e}rez-Su{\'a}rez}, {de Val- Borro}, {Aldcroft}, {Cruz}, {Robitaille},
  {Tollerud}, {Ardelean}, {Babej}, {Bach}, {Bachetti}, {Bakanov}, {Bamford},
  {Barentsen}, {Barmby}, {Baumbach}, {Berry}, {Biscani}, {Boquien}, {Bostroem},
  {Bouma}, {Brammer}, {Bray}, {Breytenbach}, {Buddelmeijer}, {Burke},
  {Calderone}, {Cano Rodr{\'\i}guez}, {Cara}, {Cardoso}, {Cheedella}, {Copin},
  {Corrales}, {Crichton}, {D'Avella}, {Deil}, {Depagne}, {Dietrich}, {Donath},
  {Droettboom}, {Earl}, {Erben}, {Fabbro}, {Ferreira}, {Finethy}, {Fox},
  {Garrison}, {Gibbons}, {Goldstein}, {Gommers}, {Greco}, {Greenfield},
  {Groener}, {Grollier}, {Hagen}, {Hirst}, {Homeier}, {Horton}, {Hosseinzadeh},
  {Hu}, {Hunkeler}, {Ivezi{\'c}}, {Jain}, {Jenness}, {Kanarek}, {Kendrew},
  {Kern}, {Kerzendorf}, {Khvalko}, {King}, {Kirkby}, {Kulkarni}, {Kumar},
  {Lee}, {Lenz}, {Littlefair}, {Ma}, {Macleod}, {Mastropietro}, {McCully},
  {Montagnac}, {Morris}, {Mueller}, {Mumford}, {Muna}, {Murphy}, {Nelson},
  {Nguyen}, {Ninan}, {N{\"o}the}, {Ogaz}, {Oh}, {Parejko}, {Parley}, {Pascual},
  {Patil}, {Patil}, {Plunkett}, {Prochaska}, {Rastogi}, {Reddy Janga},
  {Sabater}, {Sakurikar}, {Seifert}, {Sherbert}, {Sherwood-Taylor}, {Shih},
  {Sick}, {Silbiger}, {Singanamalla}, {Singer}, {Sladen}, {Sooley},
  {Sornarajah}, {Streicher}, {Teuben}, {Thomas}, {Tremblay}, {Turner},
  {Terr{\'o}n}, {van Kerkwijk}, {de la Vega}, {Watkins}, {Weaver}, {Whitmore},
  {Woillez}, {Zabalza}, \& {Astropy Contributors}}]{Astropy-Collaboration18}
{Astropy Collaboration}, {Price-Whelan}, A.~M., {Sip{\'{o}}cz}, B.~M., {et~al.}
  2018, \aj, 156, 123

\bibitem[{{Auchettl} {et~al.}(2017){Auchettl}, {Guillochon}, \&
  {Ramirez-Ruiz}}]{Auchettl16}
{Auchettl}, K., {Guillochon}, J., \& {Ramirez-Ruiz}, E. 2017, \apj, 838, 149

\bibitem[{{Balbus} \& {Hawley}(1991)}]{Balbus&Hawley91}
{Balbus}, S.~A., \& {Hawley}, J.~F. 1991, \apj, 376, 214

\bibitem[{{Balbus} \& {Mummery}(2018)}]{Balbus&Mummery18}
{Balbus}, S.~A., \& {Mummery}, A. 2018, \mnras, 481, 3348

\bibitem[{{Begelman} \& {Pringle}(2007)}]{Begelman&Pringle07}
{Begelman}, M.~C., \& {Pringle}, J.~E. 2007, \mnras, 375, 1070

\bibitem[{{Blagorodnova} {et~al.}(2017){Blagorodnova}, {Gezari}, {Hung},
  {Kulkarni}, {Cenko}, {Pasham}, {Yan}, {Arcavi}, {Ben-Ami}, {Bue}, {Cantwell},
  {Cao}, {Castro-Tirado}, {Fender}, {Fremling}, {Gal-Yam}, {Ho}, {Horesh},
  {Hosseinzadeh}, {Kasliwal}, {Kong}, {Laher}, {Leloudas}, {Lunnan}, {Masci},
  {Mooley}, {Neill}, {Nugent}, {Powell}, {Valeev}, {Vreeswijk}, {Walters}, \&
  {Wozniak}}]{Blagorodnova17}
{Blagorodnova}, N., {Gezari}, S., {Hung}, T., {et~al.} 2017, \apj, 844, 46

\bibitem[{{Blagorodnova} {et~al.}(2019){Blagorodnova}, {Cenko}, {Kulkarni},
  {Arcavi}, {Bloom}, {Duggan}, {Filippenko}, {Fremling}, {Horesh},
  {Hosseinzadeh}, {Karamehmetoglu}, {Levan}, {Masci}, {Nugent}, {Pasham},
  {Veilleux}, {Walters}, {Yan}, \& {Zheng}}]{Blagorodnova19}
{Blagorodnova}, N., {Cenko}, S.~B., {Kulkarni}, S.~R., {et~al.} 2019, \apj,
  873, 92

\bibitem[{{Bonnerot} {et~al.}(2016){Bonnerot}, {Rossi}, {Lodato}, \&
  {Price}}]{Bonnerot16}
{Bonnerot}, C., {Rossi}, E.~M., {Lodato}, G., \& {Price}, D.~J. 2016, \mnras,
  455, 2253

\bibitem[{{Bright} {et~al.}(2018){Bright}, {Fender}, {Motta}, {Mooley},
  {Perrott}, {van Velzen}, {Carey}, {Hickish}, {Razavi-Ghods}, {Titterington},
  {Scott}, {Grainge}, {Scaife}, {Cantwell}, \& {Rumsey}}]{Bright18}
{Bright}, J.~S., {Fender}, R.~P., {Motta}, S.~E., {et~al.} 2018, \mnras,
  arXiv:1801.03094

\bibitem[{{Brown} {et~al.}(2017){Brown}, {Holoien}, {Auchettl}, {Stanek},
  {Kochanek}, {Shappee}, {Prieto}, \& {Grupe}}]{Brown16b}
{Brown}, J.~S., {Holoien}, T.~W.-S., {Auchettl}, K., {et~al.} 2017, \mnras,
  466, 4904

\bibitem[{{Brown} {et~al.}(2009){Brown}, {Holland}, {Immler}, {Milne},
  {Roming}, {Gehrels}, {Nousek}, {Panagia}, {Still}, \& {Vanden
  Berk}}]{Brown09}
{Brown}, P.~J., {Holland}, S.~T., {Immler}, S., {et~al.} 2009, \aj, 137, 4517

\bibitem[{{Calzetti} {et~al.}(2000){Calzetti}, {Armus}, {Bohlin}, {Kinney},
  {Koornneef}, \& {Storchi-Bergmann}}]{Calzetti00}
{Calzetti}, D., {Armus}, L., {Bohlin}, R.~C., {et~al.} 2000, \apj, 533, 682

\bibitem[{{Cannizzo}(1996)}]{Cannizzo96}
{Cannizzo}, J.~K. 1996, \apjl, 466, L31

\bibitem[{{Cannizzo} {et~al.}(1990){Cannizzo}, {Lee}, \&
  {Goodman}}]{Cannizzo90}
{Cannizzo}, J.~K., {Lee}, H.~M., \& {Goodman}, J. 1990, \apj, 351, 38

\bibitem[{{Cardelli} {et~al.}(1989){Cardelli}, {Clayton}, \&
  {Mathis}}]{cardelli89}
{Cardelli}, J.~A., {Clayton}, G.~C., \& {Mathis}, J.~S. 1989, \apj, 345, 245

\bibitem[{{Carson} {et~al.}(2015){Carson}, {Barth}, {Seth}, {den Brok},
  {Cappellari}, {Greene}, {Ho}, \& {Neumayer}}]{Carson15}
{Carson}, D.~J., {Barth}, A.~J., {Seth}, A.~C., {et~al.} 2015, \aj, 149, 170

\bibitem[{{Chambers} {et~al.}(2016){Chambers}, {Magnier}, {Metcalfe},
  {Flewelling}, {Huber}, {Waters}, {Denneau}, {Draper}, {Farrow}, {Finkbeiner},
  {Holmberg}, {Koppenhoefer}, {Price}, {Saglia}, {Schlafly}, {Smartt},
  {Sweeney}, {Wainscoat}, {Burgett}, {Grav}, {Heasley}, {Hodapp}, {Jedicke},
  {Kaiser}, {Kudritzki}, {Luppino}, {Lupton}, {Monet}, {Morgan}, {Onaka},
  {Stubbs}, {Tonry}, {Banados}, {Bell}, {Bender}, {Bernard}, {Botticella},
  {Casertano}, {Chastel}, {Chen}, {Chen}, {Cole}, {Deacon}, {Frenk},
  {Fitzsimmons}, {Gezari}, {Goessl}, {Goggia}, {Goldman}, {Grebel}, {Hambly},
  {Hasinger}, {Heavens}, {Heckman}, {Henderson}, {Henning}, {Holman}, {Hopp},
  {Ip}, {Isani}, {Keyes}, {Koekemoer}, {Kotak}, {Long}, {Lucey}, {Liu},
  {Martin}, {McLean}, {Morganson}, {Murphy}, {Nieto-Santisteban}, {Norberg},
  {Peacock}, {Pier}, {Postman}, {Primak}, {Rae}, {Rest}, {Riess}, {Riffeser},
  {Rix}, {Roser}, {Schilbach}, {Schultz}, {Scolnic}, {Szalay}, {Seitz},
  {Shiao}, {Small}, {Smith}, {Soderblom}, {Taylor}, {Thakar}, {Thiel},
  {Thilker}, {Urata}, {Valenti}, {Walter}, {Watters}, {Werner}, {White},
  {Wood-Vasey}, \& {Wyse}}]{Chambers16}
{Chambers}, K.~C., {Magnier}, E.~A., {Metcalfe}, N., {et~al.} 2016, ArXiv
  e-prints, arXiv:1612.05560

\bibitem[{{Chornock} {et~al.}(2014){Chornock}, {Berger}, {Gezari}, {Zauderer},
  {Rest}, {Chomiuk}, {Kamble}, {Soderberg}, {Czekala}, {Dittmann}, {Drout},
  {Foley}, {Fong}, {Huber}, {Kirshner}, {Lawrence}, {Lunnan}, {Marion},
  {Narayan}, {Riess}, {Roth}, {Sanders}, {Scolnic}, {Smartt}, {Smith},
  {Stubbs}, {Tonry}, {Burgett}, {Chambers}, {Flewelling}, {Hodapp}, {Kaiser},
  {Magnier}, {Martin}, {Neill}, {Price}, \& {Wainscoat}}]{Chornock14}
{Chornock}, R., {Berger}, E., {Gezari}, S., {et~al.} 2014, \apj, 780, 44

\bibitem[{{Conroy} \& {Gunn}(2010)}]{Conroy10}
{Conroy}, C., \& {Gunn}, J.~E. 2010, \apj, 712, 833

\bibitem[{{Conroy} {et~al.}(2009){Conroy}, {Gunn}, \& {White}}]{Conroy09}
{Conroy}, C., {Gunn}, J.~E., \& {White}, M. 2009, \apj, 699, 486

\bibitem[{{Coughlin} \& {Begelman}(2014)}]{Coughlin14}
{Coughlin}, E.~R., \& {Begelman}, M.~C. 2014, \apj, 781, 82

\bibitem[{{Dai} {et~al.}(2015){Dai}, {McKinney}, \& {Miller}}]{Dai15}
{Dai}, L., {McKinney}, J.~C., \& {Miller}, M.~C. 2015, \apjl, 812, L39

\bibitem[{{Dai} {et~al.}(2018){Dai}, {McKinney}, {Roth}, {Ramirez-Ruiz}, \&
  {Miller}}]{Dai18}
{Dai}, L., {McKinney}, J.~C., {Roth}, N., {Ramirez-Ruiz}, E., \& {Miller},
  M.~C. 2018, \apj, 859, L20

\bibitem[{{Done} {et~al.}(2007){Done}, {Gierli{\'n}ski}, \& {Kubota}}]{Done+07}
{Done}, C., {Gierli{\'n}ski}, M., \& {Kubota}, A. 2007, \aapr, 15, 1

\bibitem[{{Fender} \& {Belloni}(2004)}]{Fender&Belloni04}
{Fender}, R., \& {Belloni}, T. 2004, \araa, 42, 317

\bibitem[{{Foreman-Mackey}(2016)}]{Foreman-Mackey16}
{Foreman-Mackey}, D. 2016, The Journal of Open Source Software, 1,
  doi:10.21105/joss.00024

\bibitem[{{Foreman-Mackey} {et~al.}(2013){Foreman-Mackey}, {Hogg}, {Lang}, \&
  {Goodman}}]{Foreman-Mackey13}
{Foreman-Mackey}, D., {Hogg}, D.~W., {Lang}, D., \& {Goodman}, J. 2013, \pasp,
  125, 306

\bibitem[{Foreman-Mackey {et~al.}(2014)Foreman-Mackey, Sick, \&
  Johnson}]{Foreman-Mackey14}
Foreman-Mackey, D., Sick, J., \& Johnson, B. 2014, python-fsps: Python bindings
  to FSPS (v0.1.1), doi:10.5281/zenodo.12157.
\newblock \url{https://doi.org/10.5281/zenodo.12157}

\bibitem[{{Fragile} {et~al.}(2007){Fragile}, {Blaes}, {Anninos}, \&
  {Salmonson}}]{Fragile+07}
{Fragile}, P.~C., {Blaes}, O.~M., {Anninos}, P., \& {Salmonson}, J.~D. 2007,
  \apj, 668, 417

\bibitem[{{Fragile} {et~al.}(2018){Fragile}, {Etheridge}, {Anninos}, {Mishra},
  \& {Klu{\'z}niak}}]{Fragile+18}
{Fragile}, P.~C., {Etheridge}, S.~M., {Anninos}, P., {Mishra}, B., \&
  {Klu{\'z}niak}, W. 2018, \apj, 857, 1

\bibitem[{{Franchini} {et~al.}(2016){Franchini}, {Lodato}, \&
  {Facchini}}]{Franchini+16}
{Franchini}, A., {Lodato}, G., \& {Facchini}, S. 2016, \mnras, 455, 1946

\bibitem[{{French} {et~al.}(2016){French}, {Arcavi}, \& {Zabludoff}}]{French16}
{French}, K.~D., {Arcavi}, I., \& {Zabludoff}, A. 2016, \apjl, 818, L21

\bibitem[{{French} \& {Zabludoff}(2018)}]{French18}
{French}, K.~D., \& {Zabludoff}, A.~I. 2018, \apj, 868, 99

\bibitem[{{Fruchter} \& {Hook}(2002)}]{Fruchter02}
{Fruchter}, A.~S., \& {Hook}, R.~N. 2002, \pasp, 114, 144

\bibitem[{{Fukugita} {et~al.}(1996){Fukugita}, {Ichikawa}, {Gunn}, {Doi},
  {Shimasaku}, \& {Schneider}}]{fukugita96}
{Fukugita}, M., {Ichikawa}, T., {Gunn}, J.~E., {et~al.} 1996, \aj, 111, 1748

\bibitem[{{Gehrels} {et~al.}(2004){Gehrels}, {Chincarini}, {Giommi}, {Mason},
  {Nousek}, {Wells}, {White}, {Barthelmy}, {Burrows}, {Cominsky}, {Hurley},
  {Marshall}, {M{\'e}sz{\'a}ros}, {Roming}, {Angelini}, {Barbier}, {Belloni},
  {Campana}, {Caraveo}, {Chester}, {Citterio}, {Cline}, {Cropper}, {Cummings},
  {Dean}, {Feigelson}, {Fenimore}, {Frail}, {Fruchter}, {Garmire}, {Gendreau},
  {Ghisellini}, {Greiner}, {Hill}, {Hunsberger}, {Krimm}, {Kulkarni}, {Kumar},
  {Lebrun}, {Lloyd-Ronning}, {Markwardt}, {Mattson}, {Mushotzky}, {Norris},
  {Osborne}, {Paczynski}, {Palmer}, {Park}, {Parsons}, {Paul}, {Rees},
  {Reynolds}, {Rhoads}, {Sasseen}, {Schaefer}, {Short}, {Smale}, {Smith},
  {Stella}, {Tagliaferri}, {Takahashi}, {Tashiro}, {Townsley}, {Tueller},
  {Turner}, {Vietri}, {Voges}, {Ward}, {Willingale}, {Zerbi}, \&
  {Zhang}}]{Gehrels04}
{Gehrels}, N., {Chincarini}, G., {Giommi}, P., {et~al.} 2004, \apj, 611, 1005

\bibitem[{{Georgiev} \& {B{\"o}ker}(2014)}]{Georgiev14}
{Georgiev}, I.~Y., \& {B{\"o}ker}, T. 2014, \mnras, 441, 3570

\bibitem[{{Gezari} {et~al.}(2015){Gezari}, {Chornock}, {Lawrence}, {Rest},
  {Jones}, {Berger}, {Challis}, \& {Narayan}}]{Gezari15}
{Gezari}, S., {Chornock}, R., {Lawrence}, A., {et~al.} 2015, \apjl, 815, L5

\bibitem[{{Gezari} {et~al.}(2006){Gezari}, {Martin}, {Milliard}, {Basa},
  {Halpern}, {Forster}, {Friedman}, {Morrissey}, {Neff}, {Schiminovich},
  {Seibert}, {Small}, \& {Wyder}}]{Gezari06}
{Gezari}, S., {Martin}, D.~C., {Milliard}, B., {et~al.} 2006, \apjl, 653, L25

\bibitem[{{Gezari} {et~al.}(2008){Gezari}, {Basa}, {Martin}, {Bazin},
  {Forster}, {Milliard}, {Halpern}, {Friedman}, {Morrissey}, {Neff},
  {Schiminovich}, {Seibert}, {Small}, \& {Wyder}}]{Gezari08}
{Gezari}, S., {Basa}, S., {Martin}, D.~C., {et~al.} 2008, \apj, 676, 944

\bibitem[{{Gezari} {et~al.}(2009{\natexlab{a}}){Gezari}, {Heckman}, {Cenko},
  {Eracleous}, {Forster}, {Gon{\c c}alves}, {Martin}, {Morrissey}, {Neff},
  {Seibert}, {Schiminovich}, \& {Wyder}}]{Gezari09}
{Gezari}, S., {Heckman}, T., {Cenko}, S.~B., {et~al.} 2009{\natexlab{a}}, \apj,
  698, 1367

\bibitem[{{Gezari} {et~al.}(2009{\natexlab{b}}){Gezari}, {Halpern}, {Grupe},
  {Yuan}, {Quimby}, {McKay}, {Chamarro}, {Sisson}, {Akerlof}, {Wheeler},
  {Brown}, {Cenko}, {Rau}, {Djordjevic}, \& {Terndrup}}]{Gezari09b}
{Gezari}, S., {Halpern}, J.~P., {Grupe}, D., {et~al.} 2009{\natexlab{b}}, \apj,
  690, 1313

\bibitem[{{Gezari} {et~al.}(2012){Gezari}, {Chornock}, {Rest}, {Huber},
  {Forster}, {Berger}, {Challis}, {Neill}, {Martin}, {Heckman}, {Lawrence},
  {Norman}, {Narayan}, {Foley}, {Marion}, {Scolnic}, {Chomiuk}, {Soderberg},
  {Smith}, {Kirshner}, {Riess}, {Smartt}, {Stubbs}, {Tonry}, {Wood-Vasey},
  {Burgett}, {Chambers}, {Grav}, {Heasley}, {Kaiser}, {Kudritzki}, {Magnier},
  {Morgan}, \& {Price}}]{Gezari12}
{Gezari}, S., {Chornock}, R., {Rest}, A., {et~al.} 2012, \nat, 485, 217

\bibitem[{{Girardi} {et~al.}(2000){Girardi}, {Bressan}, {Bertelli}, \&
  {Chiosi}}]{Girardi+00}
{Girardi}, L., {Bressan}, A., {Bertelli}, G., \& {Chiosi}, C. 2000, \aaps, 141,
  371

\bibitem[{{Guillochon} {et~al.}(2014){Guillochon}, {Manukian}, \&
  {Ramirez-Ruiz}}]{Guillochon14}
{Guillochon}, J., {Manukian}, H., \& {Ramirez-Ruiz}, E. 2014, \apj, 783, 23

\bibitem[{{Guillochon} {et~al.}(2018){Guillochon}, {Nicholl}, {Villar},
  {Mockler}, {Narayan}, {Mandel}, {Berger}, \& {Williams}}]{Guillochon18}
{Guillochon}, J., {Nicholl}, M., {Villar}, V.~A., {et~al.} 2018, The
  Astrophysical Journal Supplement Series, 236, 6

\bibitem[{{Guillochon} \& {Ramirez-Ruiz}(2013)}]{Guillochon13}
{Guillochon}, J., \& {Ramirez-Ruiz}, E. 2013, \apj, 767, 25

\bibitem[{{G{\"u}ltekin} {et~al.}(2009){G{\"u}ltekin}, {Richstone}, {Gebhardt},
  {Lauer}, {Tremaine}, {Aller}, {Bender}, {Dressler}, {Faber}, {Filippenko},
  {Green}, {Ho}, {Kormendy}, {Magorrian}, {Pinkney}, \& {Siopis}}]{Gultekin09}
{G{\"u}ltekin}, K., {Richstone}, D.~O., {Gebhardt}, K., {et~al.} 2009, \apj,
  698, 198

\bibitem[{{Hayasaki} {et~al.}(2013){Hayasaki}, {Stone}, \& {Loeb}}]{Hayasaki13}
{Hayasaki}, K., {Stone}, N., \& {Loeb}, A. 2013, \mnras, 434, 909

\bibitem[{{Hills}(1975)}]{Hills75}
{Hills}, J.~G. 1975, \nat, 254, 295

\bibitem[{{Hirose} {et~al.}(2009){Hirose}, {Krolik}, \& {Blaes}}]{Hirose+09}
{Hirose}, S., {Krolik}, J.~H., \& {Blaes}, O. 2009, \apj, 691, 16

\bibitem[{{Holoien} {et~al.}(2014){Holoien}, {Prieto}, {Bersier}, {Kochanek},
  {Stanek}, {Shappee}, {Grupe}, {Basu}, {Beacom}, {Brimacombe}, {Brown},
  {Davis}, {Jencson}, {Pojmanski}, \& {Szczygie{\l}}}]{Holoien14}
{Holoien}, T.~W.-S., {Prieto}, J.~L., {Bersier}, D., {et~al.} 2014, \mnras,
  445, 3263

\bibitem[{{Holoien} {et~al.}(2016{\natexlab{a}}){Holoien}, {Kochanek},
  {Prieto}, {Stanek}, {Dong}, {Shappee}, {Grupe}, {Brown}, {Basu}, {Beacom},
  {Bersier}, {Brimacombe}, {Danilet}, {Falco}, {Guo}, {Jose}, {Herczeg},
  {Long}, {Pojmanski}, {Simonian}, {Szczygie{\l}}, {Thompson}, {Thorstensen},
  {Wagner}, \& {Wo{\'z}niak}}]{Holoien16}
{Holoien}, T.~W.-S., {Kochanek}, C.~S., {Prieto}, J.~L., {et~al.}
  2016{\natexlab{a}}, \mnras, 455, 2918

\bibitem[{{Holoien} {et~al.}(2016{\natexlab{b}}){Holoien}, {Kochanek},
  {Prieto}, {Grupe}, {Chen}, {Godoy-Rivera}, {Stanek}, {Shappee}, {Dong},
  {Brown}, {Basu}, {Beacom}, {Bersier}, {Brimacombe}, {Carlson}, {Falco},
  {Johnston}, {Madore}, {Pojmanski}, \& {Seibert}}]{Holoien16b}
---. 2016{\natexlab{b}}, \mnras, 463, 3813

\bibitem[{{Hung} {et~al.}(2017){Hung}, {Gezari}, {Blagorodnova}, {Roth},
  {Cenko}, {Kulkarni}, {Horesh}, {Arcavi}, {McCully}, {Yan}, {Lunnan},
  {Fremling}, {Cao}, {Nugent}, \& {Wozniak}}]{Hung17}
{Hung}, T., {Gezari}, S., {Blagorodnova}, N., {et~al.} 2017, \apj, 842, 29

\bibitem[{{Hung} {et~al.}(2018){Hung}, {Gezari}, {Cenko}, {van Velzen},
  {Blagorodnova}, {Yan}, {Kulkarni}, {Lunnan}, {Kupfer}, {Leloudas}, {Kong},
  {Nugent}, {Fremling}, {Laher}, {Masci}, {Cao}, {Roy}, \&
  {Petrushevska}}]{Hung18}
{Hung}, T., {Gezari}, S., {Cenko}, S.~B., {et~al.} 2018, The Astrophysical
  Journal Supplement Series, 238, 15

\bibitem[{{Jiang} {et~al.}(2016){Jiang}, {Davis}, \& {Stone}}]{Jiang+16}
{Jiang}, Y.-F., {Davis}, S.~W., \& {Stone}, J.~M. 2016, \apj, 827, 10

\bibitem[{{Jiang} {et~al.}(2013){Jiang}, {Stone}, \& {Davis}}]{Jiang+13}
{Jiang}, Y.-F., {Stone}, J.~M., \& {Davis}, S.~W. 2013, \apj, 778, 65

\bibitem[{{Jones} {et~al.}(2015){Jones}, {Scolnic}, \& {Rodney}}]{Jones15}
{Jones}, D.~O., {Scolnic}, D.~M., \& {Rodney}, S.~A. 2015, {PythonPhot: Simple
  DAOPHOT-type photometry in Python}, ascl:1501.010

\bibitem[{{Krolik} {et~al.}(2016){Krolik}, {Piran}, {Svirski}, \&
  {Cheng}}]{Krolik16}
{Krolik}, J., {Piran}, T., {Svirski}, G., \& {Cheng}, R.~M. 2016, \apj, 827,
  127

\bibitem[{{Kroupa}(2001)}]{Kroupa01}
{Kroupa}, P. 2001, \mnras, 322, 231

\bibitem[{{Law-Smith} {et~al.}(2017){Law-Smith}, {Ramirez-Ruiz}, {Ellison}, \&
  {Foley}}]{Law-Smith17}
{Law-Smith}, J., {Ramirez-Ruiz}, E., {Ellison}, S.~L., \& {Foley}, R.~J. 2017,
  \apj, 850, 22

\bibitem[{{Leloudas} {et~al.}(2016){Leloudas}, {Fraser}, {Stone}, {van Velzen},
  {Jonker}, {Arcavi}, {Fremling}, {Maund}, {Smartt}, {Kr{\`\i}hler},
  {Miller-Jones}, {Vreeswijk}, {Gal-Yam}, {Mazzali}, {De Cia}, {Howell},
  {Inserra}, {Patat}, {de Ugarte Postigo}, {Yaron}, {Ashall}, {Bar},
  {Campbell}, {Chen}, {Childress}, {Elias-Rosa}, {Harmanen}, {Hosseinzadeh},
  {Johansson}, {Kangas}, {Kankare}, {Kim}, {Kuncarayakti}, {Lyman}, {Magee},
  {Maguire}, {Malesani}, {Mattila}, {McCully}, {Nicholl}, {Prentice},
  {Romero-Ca{\~n}izales}, {Schulze}, {Smith}, {Sollerman}, {Sullivan},
  {Tucker}, {Valenti}, {Wheeler}, \& {Young}}]{Leloudas16}
{Leloudas}, G., {Fraser}, M., {Stone}, N.~C., {et~al.} 2016, Nature Astronomy,
  1, 0002

\bibitem[{{Lightman} \& {Eardley}(1974)}]{Lightman&Eardley74}
{Lightman}, A.~P., \& {Eardley}, D.~M. 1974, \apjl, 187, L1

\bibitem[{{Lim} {et~al.}(2015){Lim}, {Diaz}, \& {Laidler}}]{Lim15}
{Lim}, P.~L., {Diaz}, R.~I., \& {Laidler}, V. 2015, pySynphot User's Guide,
  ascl:1303.023

\bibitem[{{Lodato}(2012)}]{Lodato12}
{Lodato}, G. 2012, in European Physical Journal Web of Conferences, Vol.~39,
  European Physical Journal Web of Conferences, ed. R.~{Saxton} \&
  S.~{Komossa}, 01001

\bibitem[{{Lodato} {et~al.}(2009){Lodato}, {King}, \& {Pringle}}]{Lodato09}
{Lodato}, G., {King}, A.~R., \& {Pringle}, J.~E. 2009, \mnras, 392, 332

\bibitem[{{Lodato} \& {Rossi}(2011)}]{Lodato11}
{Lodato}, G., \& {Rossi}, E.~M. 2011, \mnras, 410, 359

\bibitem[{{Loeb} \& {Ulmer}(1997)}]{LoebUlmer97}
{Loeb}, A., \& {Ulmer}, A. 1997, \apj, 489, 573

\bibitem[{{Lu} \& {Kumar}(2018)}]{Lu18}
{Lu}, W., \& {Kumar}, P. 2018, ArXiv e-prints, arXiv:1802.02151

\bibitem[{{Lunnan} {et~al.}(2018){Lunnan}, {Chornock}, {Berger}, {Jones},
  {Rest}, {Czekala}, {Dittmann}, {Drout}, {Foley}, {Fong}, {Kirshner},
  {Laskar}, {Leibler}, {Margutti}, {Milisavljevic}, {Narayan}, {Pan}, {Riess},
  {Roth}, {Sanders}, {Scolnic}, {Smartt}, {Smith}, {Chambers}, {Draper},
  {Flewelling}, {Huber}, {Kaiser}, {Kudritzki}, {Magnier}, {Metcalfe},
  {Wainscoat}, {Waters}, \& {Willman}}]{Lunnan18}
{Lunnan}, R., {Chornock}, R., {Berger}, E., {et~al.} 2018, \apj, 852, 81

\bibitem[{{Maccarone}(2003)}]{Maccarone03}
{Maccarone}, T.~J. 2003, \aap, 409, 697

\bibitem[{{Magnier} {et~al.}(2016){Magnier}, {Sweeney}, {Chambers},
  {Flewelling}, {Huber}, {Price}, {Waters}, {Denneau}, {Draper}, {Jedicke},
  {Hodapp}, {Kaiser}, {Kudritzki}, {Metcalfe}, {Stubbs}, \&
  {Wainscoast}}]{Magnier16}
{Magnier}, E.~A., {Sweeney}, W.~E., {Chambers}, K.~C., {et~al.} 2016, ArXiv
  e-prints, arXiv:1612.05244

\bibitem[{{Martin} {et~al.}(2005){Martin}, {Fanson}, {Schiminovich},
  {Morrissey}, {Friedman}, {Barlow}, {Conrow}, {Grange}, {Jelinsky},
  {Milliard}, {Siegmund}, {Bianchi}, {Byun}, {Donas}, {Forster}, {Heckman},
  {Lee}, {Madore}, {Malina}, {Neff}, {Rich}, {Small}, {Surber}, {Szalay},
  {Welsh}, \& {Wyder}}]{Martin05}
{Martin}, D.~C., {Fanson}, J., {Schiminovich}, D., {et~al.} 2005, \apjl, 619,
  L1

\bibitem[{{Metzger} \& {Stone}(2016)}]{Metzger16}
{Metzger}, B.~D., \& {Stone}, N.~C. 2016, \mnras, 461, 948

\bibitem[{{Metzger} \& {Stone}(2017)}]{Metzger17}
---. 2017, \apj, 844, 75

\bibitem[{{Miller}(2015)}]{Miller15}
{Miller}, M.~C. 2015, \apj, 805, 83

\bibitem[{{Million} {et~al.}(2016){Million}, {Fleming}, {Shiao}, {Seibert},
  {Loyd}, {Tucker}, {Smith}, {Thompson}, \& {White}}]{Million16}
{Million}, C., {Fleming}, S.~W., {Shiao}, B., {et~al.} 2016, \apj, 833, 292

\bibitem[{{Mishra} {et~al.}(2016){Mishra}, {Fragile}, {Johnson}, \&
  {Klu{\'z}niak}}]{Mishra+16}
{Mishra}, B., {Fragile}, P.~C., {Johnson}, L.~C., \& {Klu{\'z}niak}, W. 2016,
  \mnras, 463, 3437

\bibitem[{{Mockler} {et~al.}(2019){Mockler}, {Guillochon}, \&
  {Ramirez-Ruiz}}]{Mockler18}
{Mockler}, B., {Guillochon}, J., \& {Ramirez-Ruiz}, E. 2019, \apj, 872, 151

\bibitem[{{Morrissey} {et~al.}(2007){Morrissey}, {Conrow}, {Barlow}, {Small},
  {Seibert}, {Wyder}, {Budav{\'a}ri}, {Arnouts}, {Friedman}, {Forster},
  {Martin}, {Neff}, {Schiminovich}, {Bianchi}, {Donas}, {Heckman}, {Lee},
  {Madore}, {Milliard}, {Rich}, {Szalay}, {Welsh}, \& {Yi}}]{Morrissey07}
{Morrissey}, P., {Conrow}, T., {Barlow}, T.~A., {et~al.} 2007, \apjs, 173, 682

\bibitem[{{Nealon} {et~al.}(2018){Nealon}, {Price}, {Bonnerot}, \&
  {Lodato}}]{Nealon+18}
{Nealon}, R., {Price}, D.~J., {Bonnerot}, C., \& {Lodato}, G. 2018, \mnras,
  474, 1737

\bibitem[{{Oda} {et~al.}(2009){Oda}, {Machida}, {Nakamura}, \&
  {Matsumoto}}]{Oda+09}
{Oda}, H., {Machida}, M., {Nakamura}, K.~E., \& {Matsumoto}, R. 2009, \apj,
  697, 16

\bibitem[{{Oke}(1974)}]{oke74}
{Oke}, J.~B. 1974, \apjs, 27, 21

\bibitem[{{Pasham} \& {van Velzen}(2018)}]{PashamvanVelzen17}
{Pasham}, D.~R., \& {van Velzen}, S. 2018, ApJ, 856, 14

\bibitem[{{Piran} {et~al.}(2015){Piran}, {Svirski}, {Krolik}, {Cheng}, \&
  {Shiokawa}}]{Piran15}
{Piran}, T., {Svirski}, G., {Krolik}, J., {Cheng}, R.~M., \& {Shiokawa}, H.
  2015, \apj, 806, 164

\bibitem[{{Pracy} {et~al.}(2012){Pracy}, {Owers}, {Couch}, {Kuntschner},
  {Bekki}, {Briggs}, {Lah}, \& {Zwaan}}]{Pracy12}
{Pracy}, M.~B., {Owers}, M.~S., {Couch}, W.~J., {et~al.} 2012, \mnras, 420,
  2232

\bibitem[{{Pritchard} {et~al.}(2014){Pritchard}, {Roming}, {Brown}, {Bayless},
  \& {Frey}}]{Pritchard14}
{Pritchard}, T.~A., {Roming}, P.~W.~A., {Brown}, P.~J., {Bayless}, A.~J., \&
  {Frey}, L.~H. 2014, \apj, 787, 157

\bibitem[{{Rees}(1988)}]{Rees88}
{Rees}, M.~J. 1988, \nat, 333, 523

\bibitem[{{Roming} {et~al.}(2005){Roming}, {Kennedy}, {Mason}, {Nousek}, {Ahr},
  {Bingham}, {Broos}, {Carter}, {Hancock}, {Huckle}, {Hunsberger}, {Kawakami},
  {Killough}, {Koch}, {McLelland}, {Smith}, {Smith}, {Soto}, {Boyd},
  {Breeveld}, {Holland}, {Ivanushkina}, {Pryzby}, {Still}, \&
  {Stock}}]{Roming05}
{Roming}, P.~W.~A., {Kennedy}, T.~E., {Mason}, K.~O., {et~al.} 2005, \ssr, 120,
  95

\bibitem[{{Sakimoto} \& {Coroniti}(1981)}]{Sakimoto&Coroniti81}
{Sakimoto}, P.~J., \& {Coroniti}, F.~V. 1981, \apj, 247, 19

\bibitem[{{Salpeter}(1955)}]{Salpeter55}
{Salpeter}, E.~E. 1955, \apj, 121, 161

\bibitem[{{S{\c a}dowski}(2016)}]{Sadowski16}
{S{\c a}dowski}, A. 2016, \mnras, 459, 4397

\bibitem[{{Schlegel} {et~al.}(1998){Schlegel}, {Finkbeiner}, \&
  {Davis}}]{Schlegel98}
{Schlegel}, D.~J., {Finkbeiner}, D.~P., \& {Davis}, M. 1998, \apj, 500, 525

\bibitem[{{Shakura} \& {Sunyaev}(1973)}]{Shakura73}
{Shakura}, N.~I., \& {Sunyaev}, R.~A. 1973, \aap, 24, 337

\bibitem[{{Shakura} \& {Sunyaev}(1976)}]{Shakura&Sunyaev76}
---. 1976, \mnras, 175, 613

\bibitem[{{Shen} \& {Matzner}(2014)}]{Shen14}
{Shen}, R.-F., \& {Matzner}, C.~D. 2014, \apj, 784, 87

\bibitem[{{Shiokawa} {et~al.}(2015){Shiokawa}, {Krolik}, {Cheng}, {Piran}, \&
  {Noble}}]{Shiokawa15}
{Shiokawa}, H., {Krolik}, J.~H., {Cheng}, R.~M., {Piran}, T., \& {Noble}, S.~C.
  2015, \apj, 804, 85

\bibitem[{{Stone} \& {Loeb}(2012)}]{Stone12}
{Stone}, N., \& {Loeb}, A. 2012, Physical Review Letters, 108, 061302

\bibitem[{{Stone} {et~al.}(2013){Stone}, {Sari}, \& {Loeb}}]{Stone13}
{Stone}, N., {Sari}, R., \& {Loeb}, A. 2013, \mnras, 435, 1809

\bibitem[{{Stone} \& {Metzger}(2016)}]{Stone16b}
{Stone}, N.~C., \& {Metzger}, B.~D. 2016, \mnras, 455, 859

\bibitem[{{Stone} \& {van Velzen}(2016)}]{StonevanVelzen16}
{Stone}, N.~C., \& {van Velzen}, S. 2016, \apjl, 825, L14

\bibitem[{{Stoughton} {et~al.}(2002){Stoughton}, {Lupton}, {Bernardi},
  {Blanton}, {Burles}, {Castander}, {Connolly}, {Eisenstein}, {Frieman},
  {Hennessy}, {Hindsley}, {Ivezi{\'c}}, {Kent}, {Kunszt}, {Lee}, {Meiksin},
  {Munn}, {Newberg}, {Nichol}, {Nicinski}, {Pier}, {Richards}, {Richmond},
  {Schlegel}, {Smith}, {Strauss}, {SubbaRao}, {Szalay}, {Thakar}, {Tucker},
  {Vanden Berk}, {Yanny}, {Adelman}, {Anderson}, {Anderson}, {Annis},
  {Bahcall}, {Bakken}, {Bartelmann}, {Bastian}, {Bauer}, {Berman},
  {B{\"o}hringer}, {Boroski}, {Bracker}, {Briegel}, {Briggs}, {Brinkmann},
  {Brunner}, {Carey}, {Carr}, {Chen}, {Christian}, {Colestock}, {Crocker},
  {Csabai}, {Czarapata}, {Dalcanton}, {Davidsen}, {Davis}, {Dehnen},
  {Dodelson}, {Doi}, {Dombeck}, {Donahue}, {Ellman}, {Elms}, {Evans}, {Eyer},
  {Fan}, {Federwitz}, {Friedman}, {Fukugita}, {Gal}, {Gillespie}, {Glazebrook},
  {Gray}, {Grebel}, {Greenawalt}, {Greene}, {Gunn}, {de Haas}, {Haiman},
  {Haldeman}, {Hall}, {Hamabe}, {Hansen}, {Harris}, {Harris}, {Harvanek},
  {Hawley}, {Hayes}, {Heckman}, {Helmi}, {Henden}, {Hogan}, {Hogg}, {Holmgren},
  {Holtzman}, {Huang}, {Hull}, {Ichikawa}, {Ichikawa}, {Johnston}, {Kauffmann},
  {Kim}, {Kimball}, {Kinney}, {Klaene}, {Kleinman}, {Klypin}, {Knapp},
  {Korienek}, {Krolik}, {Kron}, {Krzesi{\'n}ski}, {Lamb}, {Leger},
  {Limmongkol}, {Lindenmeyer}, {Long}, {Loomis}, {Loveday}, {MacKinnon},
  {Mannery}, {Mantsch}, {Margon}, {McGehee}, {McKay}, {McLean}, {Menou},
  {Merelli}, {Mo}, {Monet}, {Nakamura}, {Narayanan}, {Nash}, {Neilsen},
  {Newman}, {Nitta}, {Odenkirchen}, {Okada}, {Okamura}, {Ostriker}, {Owen},
  {Pauls}, {Peoples}, {Peterson}, {Petravick}, {Pope}, {Pordes}, {Postman},
  {Prosapio}, {Quinn}, {Rechenmacher}, {Rivetta}, {Rix}, {Rockosi}, {Rosner},
  {Ruthmansdorfer}, {Sandford}, {Schneider}, {Scranton}, {Sekiguchi}, {Sergey},
  {Sheth}, {Shimasaku}, {Smee}, {Snedden}, {Stebbins}, {Stubbs}, {Szapudi},
  {Szkody}, {Szokoly}, {Tabachnik}, {Tsvetanov}, {Uomoto}, {Vogeley}, {Voges},
  {Waddell}, {Walterbos}, {Wang}, {Watanabe}, {Weinberg}, {White}, {White},
  {Wilhite}, {Wolfe}, {Yasuda}, {York}, {Zehavi}, \& {Zheng}}]{stoughton02}
{Stoughton}, C., {Lupton}, R.~H., {Bernardi}, M., {et~al.} 2002, \aj, 123, 485

\bibitem[{{Taam} {et~al.}(1997){Taam}, {Chen}, \& {Swank}}]{Taam+97}
{Taam}, R.~E., {Chen}, X., \& {Swank}, J.~H. 1997, \apjl, 485, L83

\bibitem[{{Taam} \& {Lin}(1984)}]{Taam&Lin84}
{Taam}, R.~E., \& {Lin}, D.~N.~C. 1984, \apj, 287, 761

\bibitem[{{Tchekhovskoy} {et~al.}(2014){Tchekhovskoy}, {Metzger}, {Giannios},
  \& {Kelley}}]{Tchekhovskoy13}
{Tchekhovskoy}, A., {Metzger}, B.~D., {Giannios}, D., \& {Kelley}, L.~Z. 2014,
  \mnras, 437, 2744

\bibitem[{{van Velzen}(2018)}]{vanVelzen18}
{van Velzen}, S. 2018, \apj, 852, 72

\bibitem[{van Velzen {et~al.}(2011)van Velzen, K\"ording, \&
  Falcke}]{vanVelzen11}
van Velzen, S., K\"ording, E., \& Falcke, H. 2011, \mnras, 417, L51

\bibitem[{{van Velzen} {et~al.}(2016){van Velzen}, {Mendez}, {Krolik}, \&
  {Gorjian}}]{vanVelzen16b}
{van Velzen}, S., {Mendez}, A.~J., {Krolik}, J.~H., \& {Gorjian}, V. 2016,
  \apj, 829, 19

\bibitem[{{van Velzen} {et~al.}(2011){van Velzen}, {Farrar}, {Gezari},
  {Morrell}, {Zaritsky}, {{\"O}stman}, {Smith}, {Gelfand}, \&
  {Drake}}]{vanVelzen10}
{van Velzen}, S., {Farrar}, G.~R., {Gezari}, S., {et~al.} 2011, \apj, 741, 73

\bibitem[{{van Velzen} {et~al.}(2019){van Velzen}, {Gezari}, {Cenko}, {Kara},
  {Miller-Jones}, {Hung}, {Bright}, {Roth}, {Blagorodnova}, {Huppenkothen},
  {Yan}, {Ofek}, {Sollerman}, {Frederick}, {Ward}, {Graham}, {Fender},
  {Kasliwal}, {Canella}, {Stein}, {Giomi}, {Brinnel}, {van Santen}, {Nordin},
  {Bellm}, {Dekany}, {Fremling}, {Golkhou}, {Kupfer}, {Kulkarni}, {Laher},
  {Mahabal}, {Masci}, {Miller}, {Neill}, {Riddle}, {Rigault}, {Rusholme},
  {Soumagnac}, \& {Tachibana}}]{vanVelzen18_NedStark}
{van Velzen}, S., {Gezari}, S., {Cenko}, S.~B., {et~al.} 2019, \apj, 872, 198

\bibitem[{{Vazdekis} {et~al.}(2010){Vazdekis}, {S{\'a}nchez-Bl{\'a}zquez},
  {Falc{\'o}n-Barroso}, {Cenarro}, {Beasley}, {Cardiel}, {Gorgas}, \&
  {Peletier}}]{Vazdekis10}
{Vazdekis}, A., {S{\'a}nchez-Bl{\'a}zquez}, P., {Falc{\'o}n-Barroso}, J.,
  {et~al.} 2010, \mnras, 404, 1639

\bibitem[{{Walcher} {et~al.}(2005){Walcher}, {van der Marel}, {McLaughlin},
  {Rix}, {B{\"o}ker}, {H{\"a}ring}, {Ho}, {Sarzi}, \& {Shields}}]{Walcher05}
{Walcher}, C.~J., {van der Marel}, R.~P., {McLaughlin}, D., {et~al.} 2005,
  \apj, 618, 237

\bibitem[{{Wevers} {et~al.}(2019){Wevers}, {Stone}, {van Velzen}, {Jonker},
  {Hung}, {Auchettl}, {Gezari}, \& {Onori}}]{Wevers19}
{Wevers}, T., {Stone}, N.~C., {van Velzen}, S., {et~al.} 2019, arXiv e-prints,
  arXiv:1902.04077

\bibitem[{{Wevers} {et~al.}(2017){Wevers}, {van Velzen}, {Jonker}, {Stone},
  {Hung}, {Onori}, {Gezari}, \& {Blagorodnova}}]{Wevers17}
{Wevers}, T., {van Velzen}, S., {Jonker}, P.~G., {et~al.} 2017, \mnras, 471,
  1694

\bibitem[{{Wyder} {et~al.}(2007){Wyder}, {Martin}, {Schiminovich}, {Seibert},
  {Budav{\'a}ri}, {Treyer}, {Barlow}, {Forster}, {Friedman}, {Morrissey},
  {Neff}, {Small}, {Bianchi}, {Donas}, {Heckman}, {Lee}, {Madore}, {Milliard},
  {Rich}, {Szalay}, {Welsh}, \& {Yi}}]{Wyder07}
{Wyder}, T.~K., {Martin}, D.~C., {Schiminovich}, D., {et~al.} 2007, \apjs, 173,
  293

\bibitem[{Yang {et~al.}(2006)Yang, Tremonti, Zabludoff, \& Zaritsky}]{Yang06}
Yang, Y., Tremonti, C.~A., Zabludoff, A.~I., \& Zaritsky, D. 2006, \apjl, 646,
  L33

\bibitem[{{York} {et~al.}(2000){York}, {Adelman}, {Anderson}, {Anderson},
  {Annis}, {Bahcall}, {Bakken}, {Barkhouser}, {Bastian}, {Berman}, {Boroski},
  {Bracker}, {Briegel}, {Briggs}, {Brinkmann}, {Brunner}, {Burles}, {Carey},
  {Carr}, {Castander}, {Chen}, {Colestock}, {Connolly}, {Crocker}, {Csabai},
  {Czarapata}, {Davis}, {Doi}, {Dombeck}, {Eisenstein}, {Ellman}, {Elms},
  {Evans}, {Fan}, {Federwitz}, {Fiscelli}, {Friedman}, {Frieman}, {Fukugita},
  {Gillespie}, {Gunn}, {Gurbani}, {de Haas}, {Haldeman}, {Harris}, {Hayes},
  {Heckman}, {Hennessy}, {Hindsley}, {Holm}, {Holmgren}, {Huang}, {Hull},
  {Husby}, {Ichikawa}, {Ichikawa}, {Ivezi{\'c}}, {Kent}, {Kim}, {Kinney},
  {Klaene}, {Kleinman}, {Kleinman}, {Knapp}, {Korienek}, {Kron}, {Kunszt},
  {Lamb}, {Lee}, {Leger}, {Limmongkol}, {Lindenmeyer}, {Long}, {Loomis},
  {Loveday}, {Lucinio}, {Lupton}, {MacKinnon}, {Mannery}, {Mantsch}, {Margon},
  {McGehee}, {McKay}, {Meiksin}, {Merelli}, {Monet}, {Munn}, {Narayanan},
  {Nash}, {Neilsen}, {Neswold}, {Newberg}, {Nichol}, {Nicinski}, {Nonino},
  {Okada}, {Okamura}, {Ostriker}, {Owen}, {Pauls}, {Peoples}, {Peterson},
  {Petravick}, {Pier}, {Pope}, {Pordes}, {Prosapio}, {Rechenmacher}, {Quinn},
  {Richards}, {Richmond}, {Rivetta}, {Rockosi}, {Ruthmansdorfer}, {Sandford},
  {Schlegel}, {Schneider}, {Sekiguchi}, {Sergey}, {Shimasaku}, {Siegmund},
  {Smee}, {Smith}, {Snedden}, {Stone}, {Stoughton}, {Strauss}, {Stubbs},
  {SubbaRao}, {Szalay}, {Szapudi}, {Szokoly}, {Thakar}, {Tremonti}, {Tucker},
  {Uomoto}, {Vanden Berk}, {Vogeley}, {Waddell}, {Wang}, {Watanabe},
  {Weinberg}, {Yanny}, \& {Yasuda}}]{york02}
{York}, D.~G., {Adelman}, J., {Anderson}, Jr., J.~E., {et~al.} 2000, \aj, 120,
  1579

\end{thebibliography}


\appendix

\section{Bandpass throughput and blackbody temperature}
\label{app:bandpass}
\begin{figure}
\gridline{
		\fig{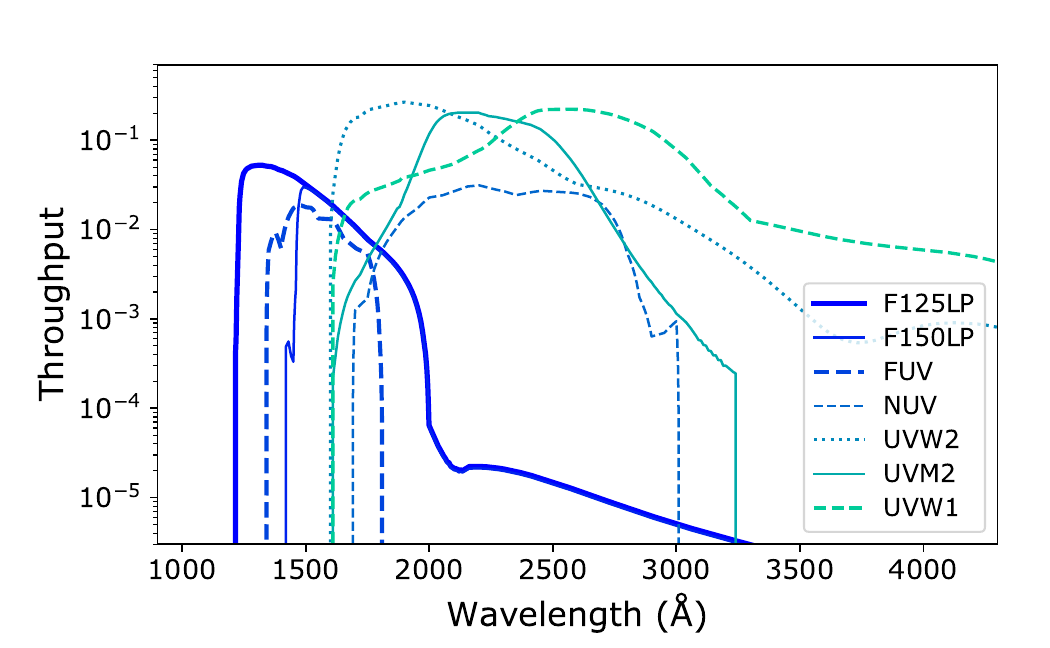}{0.49\textwidth}{(a)}
        \fig{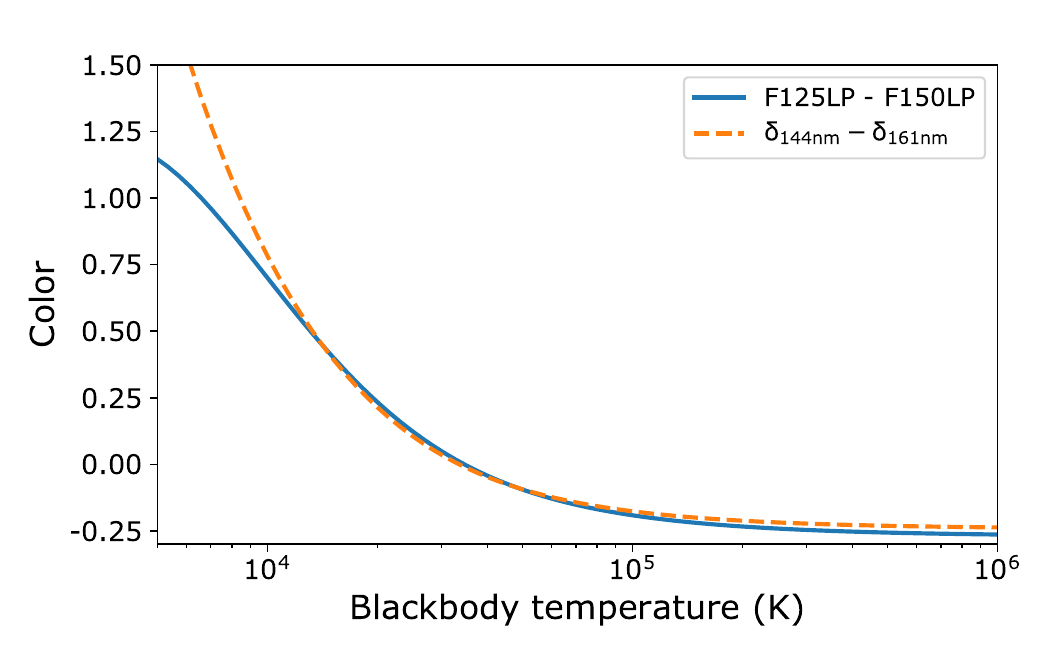}{0.49\textwidth}{(b)}
        }
\caption{(a). Throughput of the UV bandpasses used in this work. The red leak of the UVW1 and UVW2 filters is clearly visible. (b) Color as a function of blackbody temperature. While the HST/ACS F125LP and F150LP bandpass have some overlap, the color still contains sufficient information to estimate a blackbody temperature. For $T>2\times 10^4$~K, the inferred temperature is almost identical to the value obtained for a delta-function ($\delta$) centered at the pivot frequency of each bandpass.}\label{fig:bandpass}
\end{figure}

As shown in Fig.~\ref{fig:bandpass}, the HST/ACS F125LP and F150LP filters have a ``longpass'' throughput curve, which means they overlap at long wavelengths. Fortunately, this overlap does not significantly limit the inference of the blackbody temperature from the F125LP/F150LP color. We obtained the bandpass data for this figure using synphot\footnote{pysynphot.readthedocs.io}.

\section{Full MCMC output}
In Table~\ref{tab:mcmc} we list all free parameters of our light curve model (Eq.~\ref{eq:lc}) and the result obtained applying this model to the early time data.

\begin{deluxetable}{l c c c c c c c c c c}
\tablewidth{0pt}
\tablecolumns{11}
\tablecaption{Light curve (Eq.~\ref{eq:lc}) parameters inferred form early-time observations}
\tablehead{name &$L_{\rm peak}$\tablenotemark{a}  & $t_0$ & $t_{\rm peak}$\tablenotemark{b} & $p$ & $T$ & $\ln(f)$\\
 &$\log {\rm erg~s^{-1}}$  & $\log$~days & days &  & $\log$~K & }
\startdata
PS1-10jh         &$44.54_{-0.16}^{+0.21}$	 &$1.17_{-0.38}^{+0.28}$	 &0             	&$-1.28_{-0.28}^{+0.19}$	 &$4.59_{-0.09}^{+0.15}$	 &$-2.61_{-0.40}^{+0.38}$	 \\ 
PTF-09ge         &$42.97_{-0.15}^{+0.18}$	 &$1.97_{-0.11}^{+0.09}$	 &0             	&$-2.26_{-0.35}^{+0.34}$	 &$4.08_{-0.03}^{+0.03}$	 &$-2.76_{-0.16}^{+0.17}$	 \\ 
PTF-09djl        &$44.23_{-0.22}^{+0.18}$	 &$1.73_{-0.26}^{+0.18}$	 &0             	&$-1.83_{-0.46}^{+0.44}$	 &$4.41_{-0.08}^{+0.08}$	 &$-2.04_{-0.70}^{+0.65}$	 \\ 
SDSS-TDE2        &$44.62_{-0.37}^{+0.48}$	 &$1.13_{-0.58}^{+1.15}$	 &$14.48_{-12.80}^{+13.75}$	 &$-0.77_{-0.86}^{+0.19}$	 &$4.37_{-0.04}^{+0.05}$	 &$-2.28_{-0.28}^{+0.30}$	 \\ 
GALEX-D1-9       &$43.25_{-0.12}^{+0.17}$	 &$0.92_{-0.41}^{+1.92}$	 &$9.05_{-8.24}^{+17.02}$	 &$-0.28_{-1.48}^{+0.07}$	 &$4.59_{-0.05}^{+0.07}$	 &$-1.88_{-0.22}^{+0.25}$	 \\ 
GALEX-D23H-1     &$43.46_{-0.07}^{+0.09}$	 &$2.09_{-0.69}^{+0.37}$	 &0             	&$-1.67_{-1.15}^{+0.82}$	 &$4.70_{-0.09}^{+0.14}$	 &$-2.03_{-1.03}^{+0.65}$	 \\ 
GALEX-D3-13      &$43.83_{-0.21}^{+0.29}$	 &$0.81_{-0.30}^{+0.46}$	 &$7.77_{-6.86}^{+13.80}$	 &$-0.60_{-0.14}^{+0.11}$	 &$4.66_{-0.05}^{+0.06}$	 &$-1.78_{-0.22}^{+0.24}$	 \\ 
SDSS-TDE1        &$43.67_{-0.32}^{+0.50}$	 &$1.48_{-0.84}^{+0.54}$	 &$18.48_{-15.97}^{+10.54}$	 &$-1.02_{-0.62}^{+0.31}$	 &$4.42_{-0.10}^{+0.15}$	 &$-1.59_{-0.25}^{+0.26}$	 \\ 
ASASSN-14ae      &$44.42_{-0.61}^{+0.65}$	 &$1.44_{-0.27}^{+0.22}$	 &$19.31_{-16.98}^{+9.92}$	 &$-2.90_{-0.09}^{+0.24}$	 &$4.29_{-0.03}^{+0.03}$	 &$-2.13_{-0.23}^{+0.24}$	 \\ 
ASASSN-14li      &$44.15_{-0.51}^{+0.56}$	 &$0.96_{-0.44}^{+0.41}$	 &$16.45_{-14.06}^{+8.07}$	 &$-1.25_{-0.06}^{+0.06}$	 &$4.52_{-0.04}^{+0.04}$	 &$-1.74_{-0.10}^{+0.10}$	 \\ 
iPTF-16fnl       &$43.28_{-0.05}^{+0.06}$	 &$1.12_{-0.11}^{+0.11}$	 &0             	&$-1.73_{-0.20}^{+0.17}$	 &$4.47_{-0.02}^{+0.03}$	 &$-2.23_{-0.21}^{+0.21}$	 \\ 
ASASSN-15oi      &$45.49_{-1.00}^{+1.18}$	 &$1.07_{-0.52}^{+0.41}$	 &$19.42_{-16.88}^{+9.24}$	 &$-2.49_{-0.28}^{+0.25}$	 &$4.60_{-0.06}^{+0.08}$	 &$-1.98_{-0.16}^{+0.17}$	 \\ 
iPTF-16axa       &$44.29_{-0.42}^{+0.61}$	 &$1.02_{-0.49}^{+0.45}$	 &$15.21_{-12.99}^{+12.27}$	 &$-1.18_{-0.32}^{+0.21}$	 &$4.46_{-0.02}^{+0.02}$	 &$-2.32_{-0.27}^{+0.25}$	 \\ 
iPTF-15af        &$43.67_{-0.07}^{+0.05}$	 &$2.03_{-0.38}^{+0.23}$	 &0             	&$-1.94_{-0.93}^{+0.84}$	 &$4.85_{-0.19}^{+0.18}$	 &$-2.85_{-0.99}^{+0.73}$	 \\ 
\enddata
\label{tab:mcmc}
\tablenotetext{a}{The peak luminosity is reported at 150~nm in the rest frame of the TDF.}
\tablenotemark{b}{$t_{\rm peak}$ is a nuisance parameter that measures the time since peak, with respect to the first observation. When the peak is resolved by observations of the TDF, this parameter is not needed.} 
\end{deluxetable}

\section{Spreading Disks}
\label{app:disk}
In this appendix, we outline a simple time-dependent model for a viscously-spreading accretion disk, which we employ in the paper to compare to observations.  The viscous evolution of transient disks formed in tidal disruption was first considered by \citet{Cannizzo90}, and was explored in greater detail more recently by \citet{Shen14}.  We follow these works in studying the 1D evolution of an axisymmetric, vertically-averaged disk whose gas surface density $\Sigma(R)$ is governed by a diffusion equation
\begin{equation}
    \frac{\partial \Sigma}{\partial t} = \frac{3}{R} \frac{\partial}{\partial R}\left(R^{1/2} \frac{\partial}{\partial R}\left( \nu \Sigma r^{1/2} \right) \right),
    \label{eq:PDE}
\end{equation}
where $\nu$ is an effective kinematic viscosity.  Angular momentum transport in TDF disks is likely controlled by turbulent stresses produced via the magnetorotational instability (although see also \citealt{Nealon+18}), which we parametrize using the $\beta$-viscosity ansatz \citep{Sakimoto&Coroniti81}, i.e.
\begin{equation}
    \nu = \frac{2\alpha P_{\rm gas}}{3\Omega \rho},
\end{equation}
where the total midplane pressure $P = P_{\rm gas} + P_{\rm rad} = \rho k_{\rm B} T / (\mu m_{\rm p}) + a_{\rm rad}T^4/3$, $H(R)$ is the local scale-height, and we use midplane density $\rho$ and temperature $T$ ($\rho\equiv \frac{1}{2}\Sigma / H$).  Here $m_{\rm p}$, $k_{\rm B},$ and $a_{\rm rad}$ are the usual proton mass, Boltzmann constant, and radiation constant, respectively; $\mu \approx (2X + \frac{3}{4}Y + \frac{1}{2}Z)^{-1} \approx 0.60$ is the mean particle weight\footnote{Throughout this paper, we approximate the hydrogen mass fraction $X=0.7381$, the helium mass fraction $Y=0.2485$, the metal mass fraction $Z=0.0134$, and assume fully ionized gas.}, and $\alpha \le 1$ is the dimensionless Shakura-Sunyaev viscosity parameter \citep{Shakura73}.  Note that this approach differs from that of \citet{Shen14}, who used the more common ``$\alpha$-disk'' ansatz, i.e. $\nu \propto \alpha P$.

Disk temperature is governed by a local energy equation $Q^+ = Q^-_{\rm adv} + Q^-_{\rm rad}$ which balances viscous heating
\begin{equation}
    Q^+ = \frac{9}{4}\nu \Sigma \Omega^2
\end{equation}
against both radiative cooling
\begin{equation}
    Q^-_{\rm rad}=\frac{4a_{\rm rad}cT^4}{3\kappa \Sigma},
\end{equation}
and advective cooling
\begin{equation}
    Q^-_{\rm adv}= \frac{\dot{M}P}{2\pi R^2 \rho}.
\end{equation}
Here the local mass inflow rate $\dot{M}(R)=3\pi \nu \Sigma$, $c$ is the speed of light, and the local opacity $\kappa = \kappa_{\rm es} + \kappa_{\rm K}$, where electron scattering opacity $\kappa_{\rm es}=0.20(1+X)~{\rm cm}^2~{\rm g}^{-1}$ and Kramer's opacity $\kappa_{\rm K} = 4.0\times 10^{25} Z (1+X) (\rho)(T)^{-7/2}~{\rm cm}^2~{\rm g}^{-1}$.  The above $Q$ terms represent heating/cooling per unit area 

The vertical structure of the disk is computed assuming hydrostatic equilibrium,
\begin{equation}
    \frac{P}{\rho} = \Omega^2 H^2.
\end{equation}
Finally, at any snapshot in time $t$, we compute the spectral energy distribution of our disk models by treating the disk as a multicolor blackbody, where each annular ring has an effective temperature 
\begin{equation}
    \sigma_{\rm SB} T_{\rm eff}^4 = \frac{9}{4} \nu \Sigma \Omega^2,
\end{equation}
were $\sigma_{\rm SB}$ is the Stefan-Boltzmann constant.  Thus the bolometric luminosity is given by 
\begin{equation}
    L=\int 4\pi R \sigma_{\rm SB}T_{\rm eff}^4{\rm d R},
\end{equation}
the spectral energy density is
\begin{equation}
    L_\nu = \int 4 \pi^2 R B_\nu(T_{\rm eff}){\rm d}R,
\end{equation}
and $B_\nu(T)$ is the Planck function.

\end{document}